\documentclass[12pt,a4paper]{article}
\usepackage{jheppub}
\usepackage{graphicx} 
\usepackage{amsmath}
\usepackage{amssymb} 
\usepackage{dsfont}
\usepackage{xcolor}
\usepackage{cancel}
\usepackage{pgfplots}
\pgfplotsset{compat=1.18}
\usepackage{hyperref}
\usepackage[sep=5pt, offset=1.5em]{simpler-wick}
\usepackage[normalem]{ulem} 
\usepackage{amsthm} 

\usepackage{braket}
\usepackage{circuitikz}
\usepackage{fixcmex}
\usepackage{tikz}
\usetikzlibrary{shapes.geometric,positioning}
\usepackage{cleveref}
\usepackage{verbatim}

\preprint{CERN-TH-2025-206}
\title{Holography of K-complexity: Switchbacks and Shockwaves}

\author[a]{Marco Ambrosini,}

\author[b]{Eliezer Rabinovici} 

\author[a]{and Julian Sonner}

\affiliation[a]{Department of Theoretical Physics, University of Geneva, 24 quai Ernest-Ansermet, 1214 Gen\`eve 4, Switzerland} 

\affiliation[b]{Racah Institute of Physics, The Hebrew University, Jerusalem 9190401, Israel}
\affiliation[b]{CERN, Theoretical Physics Department, CH-1211 Geneva 23, Switzerland}

\emailAdd{Marco.Ambrosini@unige.ch}
\emailAdd{eliezer@mail.huji.ac.il}
\emailAdd{Julian.Sonner@unige.ch}

\abstract{In this paper we study Krylov complexity in the presence of single and multiple operators in the DSSYK model, where we can use the analytical techniques coming from chord diagrammatics. One of the results we obtain is that it showcases the switchback effect, when the appropriate ``triple-scaling limit'' is taken, under which the model becomes dual to semiclassical JT gravity.  We build on previous work, where it was shown that, in the continuum limit, Krylov complexity is defined as the sum of expectations value of right and left chord number operators. Here we argue that this property signals the emergence of the geometric nature of this notion of K-complexity. We show that in the regime where DSSYK is dual to semi-classical gravity, the light matter chord corresponds to a shockwave insertion in JT gravity. We identify the geodesic-length dual of the operator complexity and extend the relevant holographic dictionary to describe the details of the matter insertions. Additionally, we define a class of two-sided perturbations of the Lanczos algorithm that allows to analyze the switchback effect. In the appropriate semi-classical limit, this perturbed operator complexity is dual to an ERB length in JT gravity with corresponding shockwave insertions. We thus establish that K-complexity exhibits the expected switchback effect and universal late-time linear growth, consistent with previous findings regarding its geometric nature in the holographic bulk-boundary map. 
}

\pgfplotsset{
  every axis plot/.append style={line width=1.2pt},
  every axis plot post/.append style={
    every mark/.append style={line width=2.0pt,draw=green,fill=red}
  }
}

\begin{document}

\maketitle

\section{Introduction and overview}
Connections between quantum information and the building blocks of spacetime have added a powerful ingredient to ongoing research in both quantum gravity and Quantum Field Theory. One such notion that was imported from quantum information theory to express basic geometrical concepts is complexity. The dialogue between geometry and complexity has led to mutual expectations and constraints on their properties. A certain notion, termed Krylov complexity (``K–complexity"), emerged, that has been shown to fulfill a significant part of those expectations for both finite and infinite systems. Recent work \cite{Rabinovici:2023yex,Ambrosini:2024sre} has established the fundamental role played by Krylov complexity, \cite{Parker:2018yvk,Barbon:2019wsy,Rabinovici:2020ryf}, in two-dimensional holographic duality, and beyond \cite{Rabinovici:2025otw, Nandy:2024evd}. In particular, as is directly manifested in the `cordial' treatment of double-scaled SYK \cite{Berkooz:2018jqr,Berkooz:2024lgq}, a natural basis of the bulk JT-gravity Hilbert space under the homolographic mapping is given by the Krylov basis of the boundary DSSYK  model \cite{Lin:2022rbf,Rabinovici:2023yex} and Krylov complexity itself is mapped to total chord number under this encoding. This total chord length can then be shown to become the natural continuous variable parameterizing wormhole length in semi-classical JT gravity. A basic requirement of any bulk notion of complexity, is that it manifest the so-called switchback effect, which causes a universal delay in the growth of complexity subject to precursor operator insertions \cite{Stanford:2014jda,Susskind:2018pmk,Baiguera:2025dkc} (see Sec. \ref{sec:switch_grav_intro} for an elementary discussion of this effect). From a gate-complexity point of view, this behavior is due to the cancellation of quantum gates on adjacent forward and backwards parts of a Schwinger-Keldysh type `timefold' contour\footnote{A discussion of such contours in terms of the Schwinger-Keldysh approach is given in \cite{Haehl:2017qfl}.} (see Fig. \ref{fig:timefold_example} below), before the characteristic scrambling dynamics of chaotic systems causes the two evolutions to be sufficiently de-cohered that no such cancellation can occur. However, from the point of view of Krylov complexity, such a switchback behavior at first seems unlikely: Operator Krylov complexity relies on a basis adapted to a particular `seed' operator, ${\cal O}_0$, and the complexity of any other operator in such a basis tends to be near-maximal. In other words, adding (naively) a second operator in the evolution along a Krylov chain adapted to ${\cal O}_0$ causes immediate saturation, rather than the switchback delay followed by linear growth one would like to see. However, as we establish in this paper, this conclusion is not inevitable, and a suitably defined Krylov protocol does in fact show the switchback behavior, in situations where there is a holographically dual description. Given that the bulk dual of \cite{Rabinovici:2023yex,Ambrosini:2024sre} clearly shows the appropriate switchback phenomenology, it follows from the bulk-to-boundary mapping \cite{Lin:2022rbf, Rabinovici:2023yex, Ambrosini:2024sre} that the boundary also exhibits it. The question is then only which aspect of boundary Krylov dynamics exhibits the dual manifestation of the bulk switchback effect. We answer this question in detail in this paper.\\

\subsection{Overview}
In this work we establish two basic properties of K-complexity, and explore their consequences as pertains to its behavior as a holographic measure of complexity. Firstly, we derive the bulk-dual description of operator K-complexity, in terms of gravitational shockwave geometries and geodesics lengths therein, and secondly we demonstrate that boundary K-complexity exhibits the switchback effect, consistently with its geometric bulk dual representation.

We recall that Krylov complexity, whether applied to states (also known as `spread complexity' \cite{Balasubramanian:2022tpr}), or to operators, has a simple intuitive picture \cite{Nandy:2024evd,Rabinovici:2025otw}: given a Hamiltonian $H$, its associated Hilbert space ${\cal H}$, and a seed state $|\psi\rangle$ or operator ${\cal O}_0$, one defines the ordered basis $\{  |n \rangle\}$, respectively $\{ |n ) \}$, as the orthonormalised set of basis vectors, made from the set $\{|\psi\rangle, H |\psi \rangle, \ldots ,  H^n | \psi \rangle,\ldots\}$, or respectively the Liouvillian operating repeatedly on the operator $\left\{{|\cal O}_0), |{\cal L}{\cal O}_0)\,\ldots , |{\cal L}^n{\cal O}_0), \ldots\right\}$. The actual quantum dynamics of the system is then efficiently encoded in an auxiliary quantum dynamical system on the Krylov chain, in terms of which we define Krylov complexity as the expected value of position of the quantum system along the Krylov chain, with respect to the system's wave function $\varphi_n(t)$,
\begin{equation}
    C_K(t) = \left\langle n \right \rangle = \sum_n n \varphi_n^\dagger(t) \varphi_n(t)\,.
\end{equation}
Many of the results concerning Krylov complexity, and in particular the ones established in this article can be understood in terms of the effective dynamics along the Krylov chain (see Figures \ref{fig:ShockwaveOverview} and 
\ref{fig:KrylovOverview}). The bulk of this article analyses this effective Krylov-space dynamics and its bulk-dual representation, but before delving into the details, we first give an overview of the main results, ideas and methods employed. For a much more detailed exposition of Krylov methods and K-complexity we refer the Reader to \cite{Rabinovici:2025otw,Nandy:2024evd,Sanchez-Garrido:2024pcy}.\\

\subsubsection*{Shockwave dual of operator complexity}
\begin{figure}
\centering
\includegraphics[width=0.8\linewidth]{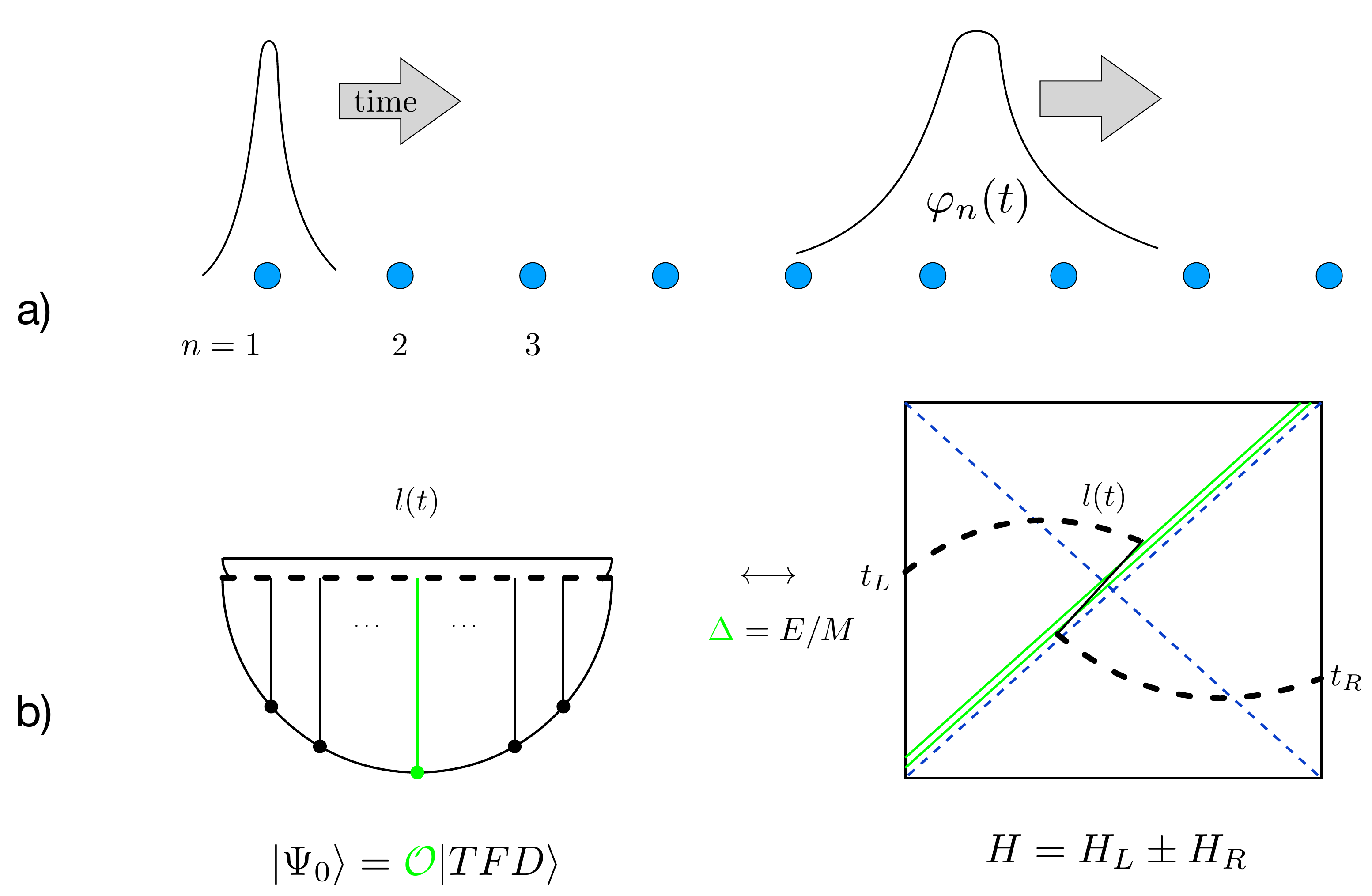}
\caption{Krylov complexity can be understood as the propagation of a wave packet along the Krylov chain, here labeled by the discrete position $n$ and shown in 1a). In \cite{Ambrosini:2024sre} it was shown that, in a suitable continuum limit, $\lambda\rightarrow 0$, this propagation becomes ballistic, corresponding to the position of a fully localised wave packet evolved by a Liouville-like Hamiltonian $H_L \pm H_R$ evolving a general class of states, obtained by perturbing the thermofield double state, (see 1b). The choice of sign corresponds to different bulk dual prescriptions, as we show in Section \ref{sec:bulk_dual}. In all cases, the bulk geometry is that of a shockwave, with energy $E$ inserted above the black hole mass $M$, where the boundary operator dimension is given by $\Delta = E/M$.  \label{fig:ShockwaveOverview} }
\end{figure}
 K-complexity of a wide class of states, as well as operator K-complexity \cite{Parker:2018yvk,Barbon:2019wsy,Rabinovici:2020ryf} are given by total chord length, \cite{Rabinovici:2023yex,Ambrosini:2024sre, Lin:2022rbf}, which implies that under the holographic encoding these map to a geometric length of the dual gravity theory -- here the two-dimensional JT theory of gravity. In this article we make the bulk-boundary mapping explicit, by demonstrating the equivalence of the propagation of K-complexity along the Krylov chain that can be described by a Liouville-like Hamiltonian and the  geodesic lengths of certain bulk operators in shockwave geometries, see Figure \ref{fig:ShockwaveOverview}. Our construction of this correspondence proceeds as follows. By evaluating the chord diagrams generated by the Lanczos recursion, we obtain the following Lanczos coefficients, \cite{Ambrosini:2024sre},
 
 \begin{equation}
b_n = \frac{2J}{\sqrt{\lambda}}\sqrt{\left[\frac{n}{2}\right]_q \left(1 \pm \tilde q q^{n/2} \right) }\,.
 \end{equation}
 
 The expression $[\cdot]_q$ denotes the $q-$number, $[m]_q =\frac{1-q^m}{1-q}$, which becomes simply the ordinary number $m$ in the limit $q\rightarrow 1$. For a definition of the various other parameters and couplings in the expression, please consult the glossary in appendix \ref{app:glossary}. The choice of sign corresponds to two different possible time evolution prescriptions, which in the holographic bulk picture result in evolution by $H_L \pm H_R$, respectively. The negative sign gives evolution along Killing directions, when applied to the thermofield double state, that is the dual eternal black hole. In this case it would act as a symmetry generator, but in the presence of the additional operator insertion (green in Figure \ref{fig:ShockwaveOverview}) the dynamics is non-trivial for both choices of the sign. In order to map to a bulk evolution describing geometric quantities in JT gravity, we have to take a number of limits, which we now summarize. Firstly, we need to take the double-scaling limit, where the number of degrees of freedom of the boundary quantum mechanics $N\rightarrow \infty$, and simultaneously, where the range of the $p-$local interaction (see Section \ref{sec:background_recap} for a precise definition) also tends to infinity, with $2p^2/N =\lambda$ held fixed. Finally we need to zoom in on the spectral edge, in order to isolate pure JT gravity dynamics in the bulk, which is achieved by sending $\lambda \rightarrow 0$, also referred to as the triple-scaling limit \cite{Lin:2022rbf}. In this limit we obtain the Liouville Hamiltonian
 \begin{equation}
     H \propto \frac{\tilde k^2}{2} + \Delta e^{-\tilde l/2} + 2 e^{-\tilde l}\,,
 \end{equation}
with $\tilde l$ and $\tilde k$ canonically conjugate length and momentum operators. In a final step to establish the precise complexity-length duality, we show that the semi-classical length evolution of this Hamiltonian agrees precisely with the evolution of boundary anchored geodesics in JT gravity shockwave geometries, which we obtain via dimensional reduction from 3D BTZ-like shockwaves, where we identify the added energy along the shock, $E$, over and above the mass of the black hole $M$, with the operator dimension, as
\begin{equation}\label{eq:DeltaEnergyMass}
    \Delta = E/M\,.
\end{equation}
In terms of this operator dimension, we show that the operator scrambling time in DSSYK takes the form
\begin{equation}
    t_{scr} = \frac{1}{2J\lambda}\log\left(1/\Delta  \right)\,.
\end{equation}
 Equation \eqref{eq:DeltaEnergyMass} summarizes the bulk-to-boundary map that was proposed in previous work, \cite{Ambrosini:2024sre}, connecting operator complexity with its bulk dual description. 
 Subsequently, this has also been extended to the finite temperature case in \cite{Aguilar-Gutierrez:2025mxf}, and outside the strict shockwave limit in \cite{aguilargutierrez2025buildingholographicdictionarydssyk}.\\
\subsubsection*{K-complexity switchback effect}
 We now summarize the second main result obtained in this article, namely the behavior of K-complexity in the presence of precursor operators. K-complexity is by now extensively used as a sensitive tool describing and distinguishing chaotic and integrable quantum dynamics. Specialized to the double-scaled SYK model, its identification as a total chord number means that its dual description falls into the class of geometric quantities which in the semi-classical limit exhibit the switchback effect \cite{Belin:2021bga, Belin_2023}, which was originally motivated by the analogous behavior of gate complexity \cite{Susskind:2014rva,Stanford:2014jda}.\\ Some imprints of the switchback effect on K-complexity were discussed in the recent literature, \cite{Heller:2024ldz,Aguilar-Gutierrez:2025mxf}. In particular, in \cite{Aguilar-Gutierrez:2025mxf}, the authors discuss how the factorization of higher-point correlators in DSSYK can have a possible complexity-theoretic connection to the switchback effect. In this paper, we adopt a different approach which directly shows that Krylov complexity in DSSYK satisfies the switchback effect, and is thus a good measure of holographic complexity, as per \cite{Belin:2021bga}.\\
 
 We now give a heuristic picture of our findings.
 
\begin{figure}
\centering
\includegraphics[width=1\linewidth]{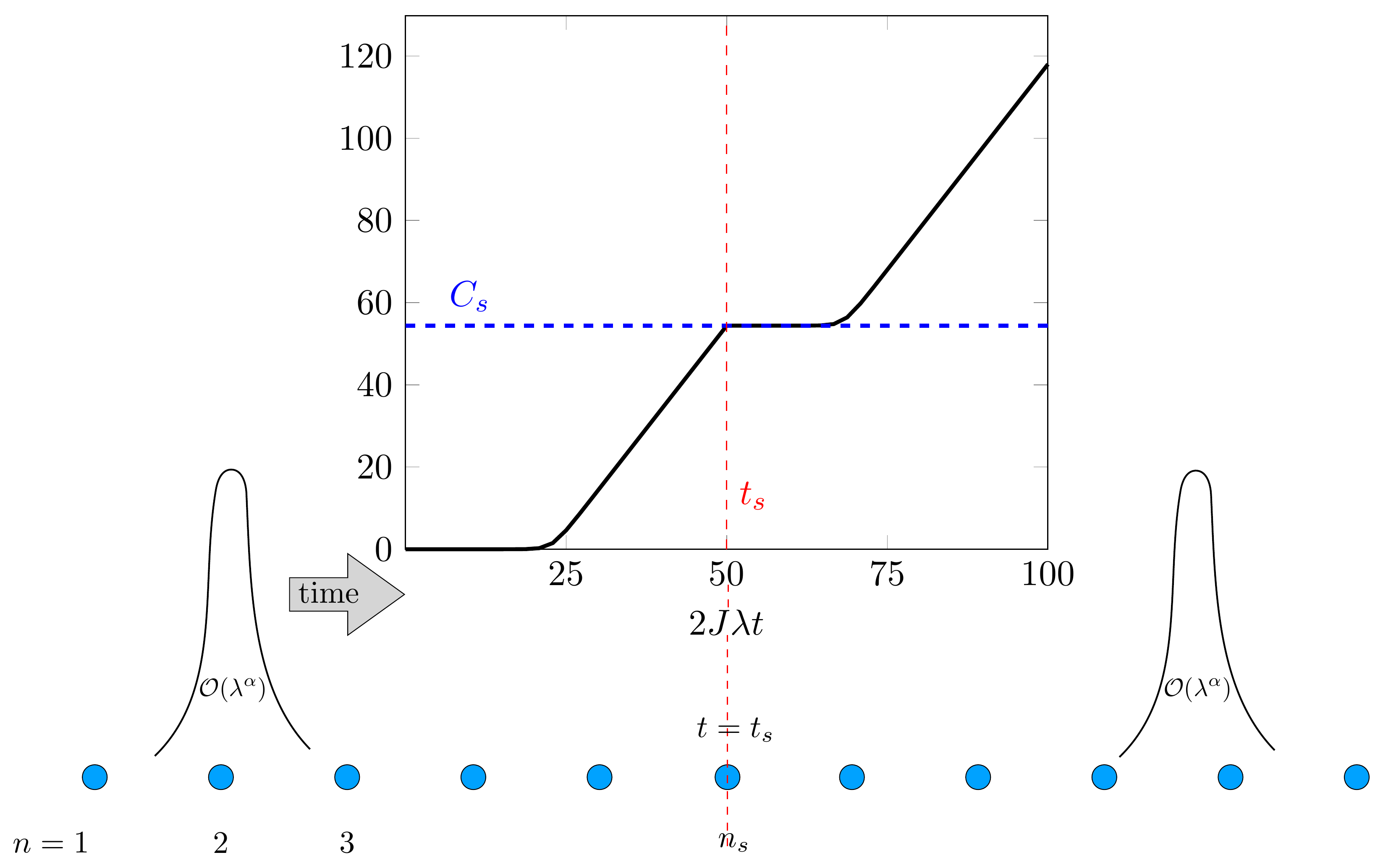}
\caption{Krylov complexity can be understood as the propagation of a wave packet along the Krylov chain, here labeled by the discrete position $n$. In \cite{Ambrosini:2024sre} it was shown that, in a suitable continuum limit, $\lambda\rightarrow 0$, this propagation becomes ballistic, corresponding to the position of a fully localized wave packet whose width goes to zero with $\lambda$, and can be understood as the growth in time of Krylov complexity in the presence of precursor operator insertions. In the so-called triple-scaled limit of DSSYK, which maps to the semiclassical description of JT gravity in an AdS$_2$ bulk, this complexity shows the switchback effect, triggered by the wave packet hitting the critical position $n_s = n(t_s)$ along the, now continuous, Krylov chain which induces the characteristic delay of complexity growth of order of the scrambling time. Indeed, as shown in the complexity profile, $C_K$ lingers around the critical value $C_s$ for a scrambling time. Here we show the case of a single precursor operator, but the picture generalizes to several such insertions.\label{fig:KrylovOverview}}
\end{figure}

 As shown in \cite{Rabinovici:2023yex,Ambrosini:2024sre}, the time evolution of K-complexity in the relevant triple-scaled limit dual to JT gravity, is given by the ballistic propagation of a wave packet on the Krylov chain, as shown in figure \ref{fig:KrylovOverview}. Furthermore, the width of this wave packet tends to zero as the semiclassical limit is taken, so that the ballistic propagation becomes effectively that of a particle moving along the chain. The insertion of a (precursor) operator at a particular instant of time, $t_s$, can thus be understood as a modification of the Lanczos algorithm at a well-defined step $n_s = n(t_s)$, which indeed is the position of the wave packet on the Krylov chain at time $t_s$. As we show in section \ref{sec:switchback_compl}, the ballistically propagating wave packet hits the Krylov position  $n=n_s$ at $t=t_s$, and is effectively slowed down before beginning to propagate further along the chain ballistically exploring the part of the chain with $n>n_s$, after a scrambling time $t_{scr}$ has elapsed.  More concretely, we consider the insertion of perturbation operators on the background considered in \cite{Ambrosini:2024sre}, accompanied by modifying the evolution operator in the Lanczos algorithm. In order to get an analytically tractable recursion describing the heuristic situation above, we consider a class of two-sided insertions, for which we can explicitly solve the Lanczos algorithm in the triple-scaled limit. 
 We prove that at the moment of the perturbation, the chord number between the perturbation matter chords freezes, and we create new dynamical boundary-anchored lengths. The operator complexity can be computed, as in the single-operator case, by summing the expectation values of these lengths, and, in particular, will present the same scrambling time delay. If we translate the two-sided perturbation we described in the timefold picture of \cite{Stanford:2014jda}, we can understand that it adds a switchback to the evolution. Then, the freezing described above, together with the creation of a new dynamical region in the DSSYK disc, is the mechanism manifesting the expected switchback effect behavior of operator complexity on the boundary side of the holographic duality.

 By solving this modified Lanczos algorithm, we obtain the associated set of Lanczos coefficients, $b_n$, from which we can directly deduce the time dependence of K-complexity. The main technical result is the analytical expression of the $b_n$ coefficients solving the Lanczos algorithm perturbed by a precursor operator,
 \begin{equation}
b_n=2\frac{J}{\sqrt{\lambda}}\sqrt{\biggr[\frac{n-n_s}{2}\biggr]_q\biggr(1\pm\Tilde{q}'^2\Tilde{q}q^{\frac{n_s+n}{2}}\biggr)+\biggr[\frac{n_s}{2}\biggr]_q\Tilde{q}'\biggr(\Tilde{q}q^{n/2}\pm q^{\frac{n-n_s}{2}}\biggr)} \,,
 \end{equation}
 where again the choice of sign corresponds to the two possible time evolution prescriptions of $H_L \pm H_R$.
As before, the expression $[\cdot]_q$ denotes the $q-$integer, while the remaining symbols to be introduced in the bulk analysis of this paper, are also defined in the glossary in \ref{app:glossary}. We can then take a continuum limit and derive the corresponding Hamiltonian describing the continuum Krylov chord dynamics. This turns out to be of a generalized Liouville form, resulting in particle moving in a Morse potential. By solving the corresponding equations of motion we find perfect agreement with the dynamics of bulk geodesic length in shockwave geometries corresponding to the backreacted influence of the precursor operators. We illustrate this phenomenon heuristically in Figure \ref{fig:KrylovOverview}, with an actual K-complexity profile superimposed above the dynamics on the Krylov chain.

Together with the shockwave bulk picture of single precursor operators, this analysis further cements the finding that Krylov complexity, in addition to its wide applicability and usefulness in quantum chaotic dynamics, reduces to a full-fledged holographic complexity, at least in the concrete arena of the DSSYK-JT duality, where the requisite calculations can be performed. Of course, the direct usefulness of the Krylov basis and associated complexity in this case gives hope and impetus to establishing an equally central role of Krylov methods in higher-dimensional holographic duality.\\

 This paper is organized as follows: we begin by reviewing, in \cref{sec:background_recap}, the DSSYK model together with its holographic dual of JT gravity, and summarizing the results obtained in \cite{Ambrosini:2024sre} for the operator complexity in this framework. The Reader already familiar with these topics can move to \cref{sec:bulk_dual}, where we describe how to compute the triple-scaled operator complexity in DSSYK from its \textit{total chord length} property, and identify its bulk dual geodesic in JT gravity with a shockwave insertion. The duality presented in the \textit{shockwave approximation}, allows us to update the DSSYK-JT holographic dictionary by including the details of matter. Then, in \cref{sec:switchback_compl}, we describe the modified Lanczos algorithm that adds a two-sided operator perturbation in the evolution. We repeat the analysis of \cref{sec:bulk_dual} in this multi-operator case, yielding a \textit{perturbed} operator K-complexity that shows the expected switchback effect. We continue by describing how to generalize this procedure for an arbitrary number of insertions of the discussed kind. We end the paper with a discussion on possible venues of future study, in \cref{sec:final_discussion}, followed by a number of appendices containing more details on the analytical computations of sections \ref{sec:bulk_dual} and \ref{sec:switchback_compl}. \\

\section{Background material}\label{sec:background_recap}
The model in which we stage our computation, double-scaled SYK (DSSYK), is defined as a particular limit of the SYK model itself \cite{kitaev_talks, Maldacena:2016hyu}.
This is an ensemble averaged many-body system of $N$ Majorana fermions interacting all-to-all with range $p$ and Hamiltonian:
\begin{eqnarray} \label{H_SYK}
    H_{SYK} = i^{p/2} \sum_{1\leq i_1 < \dots < i_p \leq N} J_{i_1 \dots i_p}\,\psi_{i_1}\dots \psi_{i_p},
\end{eqnarray}
where $\left\{\psi_i,\psi_j\right\}=2\delta_{ij}$, the coefficients $J_{i_1 \dots i_p}$ are random and sampled from a distribution of zero mean and variance given by $\langle J_{i_1\dots i_{\tilde{p}}} J_{j_1\dots j_{\tilde{p}}}\rangle = \frac{J^2}{\lambda}\binom{N}{p}^{-1} \delta_{i_1 j_1}\delta_{i_2 j_2} \dots \delta_{i_p j_{\tilde{p}}}$\footnote{for simplicity in the computation of the Lanczos coefficients, as in \cite{Berkooz:2018jqr}, we will instead consider the variance $\langle J_{i_1\dots i_{\tilde{p}}} J_{j_1\dots j_{\tilde{p}}}\rangle = \binom{N}{p}^{-1}\delta_{i_1 j_1}\delta_{i_2 j_2} \dots \delta_{i_p j_{\tilde{p}}}$, where we set $J=1$. We will go back to the original normalization before taking the triple-scaling limit and discussing the holographic relation with gravity. In order to restore the desired normalization, one multiplies the Lanczos coefficients by $J/\sqrt{\lambda}$.}, and $\lambda$ is defined below.\\
In order to obtain DSSYK one performs the following double-scaling limit \cite{Berkooz_chords, Berkooz:2018jqr}:
\begin{equation} N\to\infty,\;p\to\infty\quad\mathrm{while}\quad \lambda \equiv 2p^2/N\;\mathrm{fixed}
\end{equation}
DSSYK is a particularly useful model because in this limit the effective coupling-averaged theory can be described by only the combinatorics of objects called chord diagrams \cite{Berkooz:2018jqr, Berkooz:2022mfk, Berkooz:2024ofm}. The diagrammatic rule is that each Hamiltonian insertion either creates or annihilates a chord on a circle, representing a trace, and observables are computed by summing over all possible diagram configurations weighted by $q^{\#\mathrm{chord\;intersections}}$, where $q=e^{-\lambda}$ (see \cite{Berkooz:2024lgq} for a recent review).\\

In this section, we briefly introduce some background material regarding chord-diagram techniques in DSSYK and its holographic duality with JT gravity. We will also summarize the results for operator Krylov complexity obtained in \cite{Ambrosini:2024sre}, where one can find a more extended and pedagogical version of many of the topics presented here in \cref{sec:background_recap}. \\

\subsection{Matter insertions in DSSYK}\label{sec:dssyk_matter_recap}
The insertion of matter in DSSYK can be considered via the insertion of operators of the form \cite{Berkooz:2018jqr}:
\begin{equation}\label{eq:random_operator_def}
    \mathcal{O} = i^{\tilde{p}/2}\sum_{1\leq i_1< \dots < i_{\tilde{p}} \leq N} O_{i_1\dots i_{\tilde{p}}} \psi_{i_1} \psi_{i_2}\dots \psi_{i_{\tilde{p}}},
\end{equation}
where $O_{i_1\dots i_{\tilde{p}}}$ are random, taken from a distribution with zero mean and variance given by
\begin{equation}
    \langle O_{i_1\dots i_{\tilde{p}}} O_{j_1\dots j_{\tilde{p}}}\rangle = \binom{N}{\tilde{p}}^{-1} \delta_{i_1 j_1}\delta_{i_2 j_2} \dots \delta_{i_p j_{\tilde{p}}}.
\end{equation}
These random couplings are independent of the random couplings contained in the Hamiltonian, and thus have to be averaged over separately. This procedure adds a single $\mathcal{O}-\mathcal{O}$ matter chord to the chord diagrams, and the prescription is that intersections between the operator and Hamiltonian chords are weighted by $\Tilde{q}=q^{\Delta}$, where $\Delta=\Tilde{p}/p$. \\

In \cite{Lin:2022rbf}, it was proposed that the one-particle sector of the DSSYK Hilbert space, created upon the action of operators \eqref{eq:random_operator_def} on the TFD state, is described by states characterized by $n_L$ and $n_R$, the number of open chords, respectively on the left/right of the operator chord. These states are denoted by $|n_L,n_R\rangle$ and have the following chord diagrammatic representation:\\
\begin{center}
  \begin{tikzpicture}[scale=0.85]
    \node[left] at (-2.0, -0.5) {$|n_L,\,n_R\rangle = $};
    
    \draw[thick] (-2,0) arc[start angle=180, end angle=360, radius=2];
    
    \foreach \x in {-1.5,-1.0,1.0, 1.5} {
        \fill[black] (\x,-{sqrt(4-(\x)^2)}) circle (2pt); 
        \draw[thick] (\x,-{sqrt(4-(\x)^2)}) -- (\x,0); 
    }
    \fill[green] (0,-2) circle (2pt); 
    \draw[thick, green] (0,-2) -- (0,0); 
    
    \node at (-0.5,-0.5) {\tiny{$\cdots$}};
    \node at (0.5,-0.5) {\tiny{$\cdots$}};
    
    \draw[thick] (-2.0,0.2) -- (-0.1,0.2);
    \draw[thick] (-2.0,0.2) arc[start angle=180, end angle=230, radius=0.3];
    \draw[thick] (-0.1,0.2) arc[start angle=0, end angle=-50, radius=0.3];
    \node[above] at (-1,0.5) {$n_L$};

    \draw[thick] (0.1,0.2) -- (2.0,0.2);
    \draw[thick] (0.1,0.2) arc[start angle=180, end angle=230, radius=0.3];
    \draw[thick] (2.0,0.2) arc[start angle=0, end angle=-50, radius=0.3];
    \node[above] at (1,0.5) {$n_R$};
    \node[right] at (2,-0.5){.};
\end{tikzpicture}\\
\end{center}

It is possible to compute recursively the overlap $\braket{n_L',n_R'|n_L,n_R}$ of such states, and obtain, as shown in \cite{Lin:2023trc}, the inner product:
\begin{equation}
\label{eq:inner_product_recursion_solution}
    \langle n_L', n_R' | n_L,n_R\rangle = \sum_{0\leq k \leq n_R} q^{k^2+k(2\Delta + y-y')+\Delta(y-y')} \frac{[n_L]![n_R]![n_L']![n_R']!}{[k]![y-y'+k]![n_L'-k]![n_R-k]!}~,
\end{equation}
where $n_L-n_R = 2y,\; n_L'-n_R' = 2y', \;n_L'+n_R'=n=n_R+n_L$ and $[n]_q=(1-q^n)/(1-q)$ is the $q$-integer. Notice that an important property of the inner product \eqref{eq:inner_product_recursion_solution} is that states $\ket{n_L,n_R}$ are orthogonal whenever they belong to different total chord number $n_L+n_R$ sectors:
\begin{equation}
    \label{orthog_chord_sectors}
    \langle n_L^\prime,n_R^\prime|n_L,n_R\rangle = 0 \qquad \text{if}\quad n_L^\prime + n_R^\prime \neq n_L+n_R~.
\end{equation}
We can use such states in the one-particle sector as a useful basis to describe the time-evolved state
\begin{equation} \label{operator_to_HR_HL}
    e^{i t H_\textsc{syk}} \mathcal{O} e^{-itH_\textsc{syk}} |\mathrm{TFD}\rangle \to e^{it H_L} e^{-it H_R }|n_L=0,n_R=0\rangle ,
\end{equation}
where $H_\textsc{syk}$ is given by \eqref{H_SYK} and $H_L$ and $H_R$ provide the effective description in the ensemble-averaged theory for the operator time evolution.\\
We can write the left and right Hamiltonians as:
\begin{equation}\label{eq:HlHr_def}
    H_L =\frac{J}{\sqrt{\lambda}} \left( a_L^\dagger +a_L \right)\,,\quad\quad
    H_R = \frac{J}{\sqrt{\lambda}} \left( a_R^\dagger +a_R \right)\,.
\end{equation}
 In \eqref{eq:HlHr_def} $a_{L/R}^\dagger$ are operators creating a Hamiltonian chord to the left/right of all the existing chords
\begin{equation}\label{eq:adag_leftright_def}
    a_L^\dagger |n_L, n_R\rangle = |n_L+1, n_R\rangle \quad\quad
     a_R^\dagger |n_L, n_R\rangle = |n_L, n_R+1\rangle,
\end{equation}
   while, if we introduce the left and right annihilation operators
    \begin{equation}\label{eq:alpha_lr_def}
          \alpha_L |n_L, n_R\rangle= |n_L-1, n_R\rangle\quad\quad
    \alpha_R |n_L, n_R\rangle = |n_L, n_R-1\rangle ~,
    \end{equation}
    one has that the $a_{L,R}$, defined as the Hermitian conjugates of the creation operators \eqref{eq:adag_leftright_def}, appearing in \eqref{eq:HlHr_def} can be written as:
    \begin{align}
         a_L &= \alpha_L \,\frac{1-q^{n_L}}{1-q} +\alpha_R \, \tilde{q} \, q^{n_L} \frac{1-q^{n_R}}{1-q} \label{aL_long} \\
    a_R &= \alpha_R \,\frac{1-q^{n_R}}{1-q} +\alpha_L \, \tilde{q} \, q^{n_R} \frac{1-q^{n_L}}{1-q}\nonumber~.
    \end{align}
    To obtain \eqref{aL_long}, consider that the action of $a_{L}$, for example, has the diagrammatic interpretation of closing an arbitrary chord by taking it all the way to the left:
\begin{eqnarray*}
\begin{aligned}
    & a_L \ket{n_L,n_R}\equiv\\&\equiv a_L \raisebox{-8pt}{\includegraphics[scale=0.3]{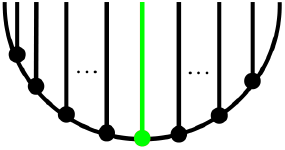} } =\raisebox{-8pt}{\includegraphics[scale=0.3]{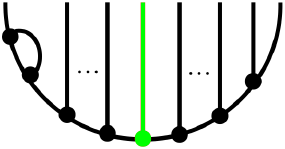} }+\raisebox{-8pt}{\includegraphics[scale=0.3]{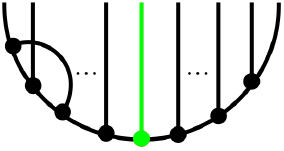} } +\dots +\raisebox{-8pt}{\includegraphics[scale=0.3]{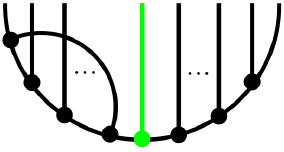} }+\raisebox{-8pt}{\includegraphics[scale=0.3]{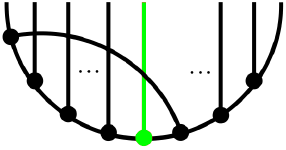} }+\dots+ \raisebox{-8pt}{\includegraphics[scale=0.3]{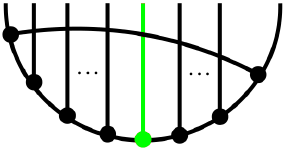} }\equiv\\&
    = (1+q+\dots+q^{n_L-1})\ket{n_L-1,n_R}+(\tilde{q}q^{n_L}+\dots+ \tilde{q}q^{n_L}q^{n_R-1})\ket{n_L,n_R-1},
\end{aligned} 
\end{eqnarray*}
which, re-summed, indeed gives \eqref{aL_long}.\\

So, the left and right Hamiltonians, $H_L$ and $H_R$, can be written as \cite{Lin:2022rbf}:
\begin{align}
     H_L & = \frac{J}{\sqrt{\lambda}} \left( a_L^\dagger +\alpha_L \,\frac{1-q^{n_L}}{1-q} +\alpha_R \, \tilde{q} \, q^{n_L} \frac{1-q^{n_R}}{1-q} \right)\label{HL}\\
     H_R & = \frac{J}{\sqrt{\lambda}} \left( a_R^\dagger +\alpha_R \,\frac{1-q^{n_R}}{1-q} +\alpha_L \, \tilde{q} \, q^{n_R} \frac{1-q^{n_L}}{1-q} \right) \label{HR}
\end{align}

Since $[H_L,H_R]=0$, the time-evolution of an operator inserted on the TFD state of DSSYK, as in \eqref{operator_to_HR_HL}, is described by the time evolution of $|n_L=0, n_R=0\rangle$ under $H_R-H_L$:
\begin{equation} \label{operator_evolution}
    e^{i t H_{\textsc{syk}}} \mathcal{O} e^{-itH_{\textsc{syk}}} |\mathrm{TFD}\rangle \to e^{it H_L} e^{-it H_R }|n_L=0,n_R=0\rangle = e^{-it(H_R-H_L)}|0,0\rangle ~.
\end{equation}
Then, solving the Lanczos algorithm with seed $\ket{0,0}\equiv \mathcal{O}\ket{TFD}$ and evolution operator $H_R-H_L$ will give the operator Krylov complexity of $\mathcal{O}$. If we consider instead the evolution operator $H_R+H_L$ we will obtain the state K-complexity, or ``spread complexity" \cite{Balasubramanian:2022tpr}, of $\mathcal{O}\ket{TFD}$. These computations, whose results we summarize in the next section, were performed in \cite{Ambrosini:2024sre}, to which we refer for a more detailed discussion.\\

\subsection{$\mathcal{O}$TFD and Operator K-complexity}\label{sec:op_compl_recap}
In this section, we briefly summarize how to solve the Lanczos algorithm for evolutions $H_R\pm H_L$ and seed $\ket{0,0}$ \cite{Ambrosini:2024sre}. The orthonormal Krylov basis elements $\ket{\psi^\pm_n}$ are built recursively via the following Lanczos recursion:
\begin{equation}
    \label{eq:lanczos_recursion_1p}
    b_n^\pm|\psi^\pm_n\rangle = (H_R\pm H_L)|\psi^\pm_{n-1}\rangle - b^\pm_{n-1}|\psi^\pm_{n-2}\rangle~,
\end{equation}
with boundary conditions $b_{0}^\pm=1$ and $|\psi^\pm_0\rangle = |0,0\rangle$, $|\psi^{\pm}_{-1}\rangle = \mathbf{0}$. Here $b_n^+$, $b_n^-$ and $\ket{\psi^+}$, $\ket{\psi^-}$ are the Lanczos coefficients and Krylov basis elements associated to the evolution with $H_R+H_L$ and $H_R-H_L$ respectively.\\
Let us consider the following ansatz for the un-normalized Krylov basis $\ket{\chi^\pm_n}$:

\begin{equation}
    \label{eq:binomial_ansatz}
    |\psi^\pm_n\rangle = \frac{1}{\prod_{k=0}^n b^\pm_k}|\chi^\pm_n\rangle~,\qquad |\chi^\pm_n\rangle := \sum_{k=0}^n (\pm)^k \binom{n}{k}|k,n-k\rangle~.
\end{equation}
Inserting this ansatz and the expression for the Hamiltonians \eqref{eq:HlHr_def} in the recursion \eqref{eq:lanczos_recursion_1p}, we notice
\begin{equation}
    \label{eq:adagger_gives_binomial}
    (a_R^\dagger \pm a_L^\dagger)|\chi^\pm_n\rangle = |\chi^\pm_{n+1}\rangle~,
\end{equation}

so that we can recast the condition that the basis $\ket{\chi^\pm_n}$ solves the Lanczos recursion as the cancellation condition:
\begin{equation}
\label{eq:binom_recursion_cancellation_cond}
     (a_R\pm a_L)\ket{\chi^\pm_n} \overset{!}{=} (b^\pm_n)^2 \ket{\chi^\pm_{n-1}}~.
\end{equation}

Now we proceed to give some details on how to prove the above identity in a particular limit. Let us start by rewriting \eqref{eq:binom_recursion_cancellation_cond} in the form:
\begin{equation}
    \label{eq:ck_def_1p}
    (a_R\pm a_L)|\chi^\pm_n\rangle = \sum_{k=0}^{n-1} c^\pm_k(n)~ (\pm)^k\binom{n-1}{k}|k,n-1-k\rangle~,
\end{equation}
where after some manipulation we find the following expression for $c_k(n)$:
    \begin{eqnarray}
     \label{eq:ck_expression_1p}
     c^\pm_k(n) = \frac{n}{n-k}\frac{1- q^{n-k}}{1-q}(1\pm\tilde{q}q^k) + \frac{n}{k+1}\frac{1-q^{k+1}}{1-q}(1\pm\tilde{q}q^{n-1-k})~.
 \end{eqnarray}
 Using the above expression for $c_k(n)$, we can compute the Lanczos coefficients as
 \begin{align}
\label{eq:lower_diag_L}
\begin{aligned}
     (b^\pm_{n+1})^2&= \frac{\langle\chi^\pm_{n+1}|\chi^\pm_{n+1}\rangle}{(b^\pm_1)^2\dots (b^\pm_n)^2} =\\&= \frac{1}{(b^\pm_1)^2\dots (b^\pm_n)^2}
    \sum_{k^\prime,k=0}^{n} (\pm)^{k^\prime+k} c^\pm_{ \scriptscriptstyle k}{ (n+1)}\,\binom{n}{k^\prime}\binom{n}{k}\braket{k^\prime,n-k^\prime|k,n-k}~.
\end{aligned}
\end{align}
However, we note that the dependence of $c_{k}(n)$ on the summation index $k$ is an indication that, in general, the condition \eqref{eq:binom_recursion_cancellation_cond} will not be satisfied. In particular, the correct Krylov basis that solves \eqref{eq:lanczos_recursion_1p} will be a linear combination of states with different total chord numbers \cite{Ambrosini:2024sre}.\\
We can gain analytical control on the binomial ansatz if we consider the following semiclassical limit:
 \begin{equation}
    \label{eq:semiclassical_limit}
    \lambda \to 0~,\quad n_L,\;n_R\to\infty~,\quad \lambda n_{L,\,R}\equiv l_{L,R}~ \text{fixed}~,
\end{equation}
where $l=\lambda n=l_L+l_R$ is the total length.
At this point, the crucial idea is that, in the semiclassical limit, we have the following asymptotic expansion of the binomial $\binom{n}{k}$ near $\lambda\sim 0$:
\begin{align}
    &\qquad \qquad \qquad \qquad \qquad \qquad \qquad\qquad\binom{n}{k}\overset{\text{semicl.}}{=} \binom{l/\lambda}{l_L/\lambda} \nonumber \\
     &\underset{\lambda\sim 0}{\sim} \sqrt{\frac{2l}{\pi\lambda(l^2-4x^2)}} ~ \text{exp}\left\{ \frac{1}{\lambda} \Big( l\log(2l)-(l-2x)\log(l-2x)-(l+2x)\log(l+2x) \Big) \right\}~, \label{eq:binom_asymptotics_1p}
\end{align}
where $x=l_L-l/2$ and $l_L=\lambda n_L$. We notice that a large factor $1/\lambda$ appears in the exponent of \eqref{eq:binom_asymptotics_1p}. This means that we can interpret the $\lambda\to0$ limit as a saddle point approximation that localizes the contributions in the integrals, obtained from considering the limit \eqref{eq:semiclassical_limit} of \eqref{eq:lower_diag_L}, to $l_L=l/2$. More intuitively, in the semiclassical limit, the binomial coefficients get squeezed, and effectively become Dirac delta functions around the middle of the domain of $l_L=\lambda k$.\\

At this point, we raise an important technical difference between the evolution with $H_R+H_L$ and $H_R-H_L$: for $H_R-H_L$ in order for the binomial coefficients to act as Dirac delta functions, in the limit \eqref{eq:semiclassical_limit}, it is necessary that the inner product is asymmetric around $l_L=l/2$\footnote{this is needed in order to spoil the complete cancellation originated, for example in \eqref{eq:lower_diag_L}, because of the alternating sign present in the summation when considering $H_R-H_L$ evolution.}. Fortunately this property is indeed respected by the semiclassical limit of \eqref{eq:inner_product_recursion_solution}:
\begin{equation}\label{eq:scal_prod_semiclass_1p}
   \braket{x'|x}= \left\langle n_L^{\prime}, n_R^{\prime} \mid n_L, n_R\right\rangle=[n]_q !\left(\frac{\left(1-c^2\right) / 2}{\cosh \frac{x-x^{\prime}}{2}-c \cosh \frac{x+x^{\prime}}{2}}\right)^{2 \Delta},
\end{equation}
where $c^2=q^n$ and $x=\lambda\frac{n_L-n_R}{2}$. For further details on this asymptotic analysis please refer to appendix D in \cite{Ambrosini:2024sre}, or to \cref{app:recap_asympt} here.\\
 By virtue of this saddle point approximation, only terms with $k\sim n/2$ contribute to \eqref{eq:lower_diag_L} and we obtain the following Lanczos coefficients:
 
\begin{equation}
\label{eq:b_nplus1_induction_proof}
\begin{aligned}
    (b^\pm_{n})^2\underset{\lambda\sim 0}{\sim} &\frac{c^\pm_{\scriptscriptstyle\frac{n}{2}}{\scriptstyle (n)}~\langle \chi^\pm_{n-1}|\chi^\pm_{n-1}\rangle}{(b^\pm_1)^2\dots (b^\pm_{n-1})^2}  \underset{\lambda\sim 0}{\sim} c^\pm_{\frac{n}{2}}(n)\\&\implies b^\pm_n=2\sqrt{\frac{1-q^{n/2}}{1-q}\bigr(1\pm\Tilde{q}q^{n/2}\bigr)},
\end{aligned}
\end{equation}
Likewise, the norm of the lower $n$ total chord number tail is expressed as a sum, localized around $k\sim n/2$ because of the same saddle point approximation, containing a multiplicative coefficient $\propto(c^\pm_k(n)-(b^\pm_n)^2)$ \cite{Ambrosini:2024sre}. This means that in the limit \eqref{eq:semiclassical_limit}, the lower total chord number tail is suppressed and the Krylov basis is indeed given by the binomial ansatz \eqref{eq:binomial_ansatz}. We argue that the fact that the Krylov basis is built out of states in the same total chord number sector, is the fundamental indication of the complexity gaining a geometric interpretation when $\lambda\to 0$, as indeed in \cref{sec:bulk_dual} we will use it to find the appropriate dual bulk length.\\

We summarize in \cref{tab:op_OTFD_compl} below the results obtained in \cite{Ambrosini:2024sre} in the semiclassical limit \eqref{eq:semiclassical_limit}.
\begin{table}
\centering
\begin{tabular}{|p{7cm}|p{7cm}|}
\hline
    \textbf{Operator K-complexity} & ${\cal O}|$TFD$\rangle$ \textbf{K-complexity}  \\ \hline
    $\mathcal{O}\ket{0}=\ket{0,0}$ seed & $\mathcal{O}\ket{0}=\ket{0,0}$ seed  \\
    $H = H_R - H_L$ evolution & $H = H_R + H_L$ evolution \\ \hline
    Lanczos coefficients: & Lanczos coefficients:\\
    $b_n = \frac{2J}{\sqrt{\lambda}} \sqrt{\frac{1-q^{n/2}}{1-q} \left( 1-\tilde q q^{n/2}  \right)}$ & 
    $b_n^+ = \frac{2J}{\sqrt{\lambda}} \sqrt{\frac{1-q^{n/2}}{1-q} \left( 1 + \tilde q q^{n/2}  \right)}$ \\ \hline
    Krylov basis states are eigenstates of total chord number: &    Krylov basis states are eigenstates of total chord number:\\
    $\ket{\psi_n}=\frac{J^n}{b_1\dots b_n}\sum_{k=0}^n (-1)^k\binom{n}{k}\ket{k,n-k}$& $\ket{\psi_n^+}=\frac{J^n}{b_1^+\dots b_n^+}\sum_{k=0}^n \binom{n}{k}\ket{k,n-k}$
    \\ \hline
    Operator complexity in the semiclassical limit: &  ${\cal O}|$TFD$\rangle$-complexity in the semiclassical limit: \\ \parbox{5cm}{\begin{align*}
    \lambda C_K(t) =&  2 \log \Big[ 1 + (1-\tilde q) \sinh^2 Jt\Big],
    \end{align*}} & \parbox{5cm}{\begin{align*}
    \lambda C_K^+(t) =&  2 \log \Big[ 1 + (1+\tilde q) \sinh^2 \left(Jt\right)\Big],
    \end{align*}}\\
    when $\Tilde{q}\to 1$ has long exponential time behavior & No exponential time behavior\\
\hline
\end{tabular}
\caption{Summary of the results of \cite{Ambrosini:2024sre} for the operator and $\mathcal{O}\ket{TFD}$ K-complexities.}
\label{tab:op_OTFD_compl}
\end{table}

\vspace{6mm}
The last result from \cite{Ambrosini:2024sre} we will need is the expression of the Hamiltonian $H_L+H_R$, in the triple-scaled limit, defined as:
\begin{equation}\label{eq:triple_scaling_nomatt}
    \lambda \to 0,\;l \to \infty\quad\mathrm{while}\quad e^{-l}/(2\lambda)^2\equiv e^{-\tilde{l}}\;\mathrm{fixed},
\end{equation}
where we can put our DSSYK discussion in contact with gravity. We can write the triple-scaled Hamiltonian $H=H_L+H_R$ we are searching for, by using $b^{+, TS}(\Tilde{l})$, the triple-scaled limit of Lanczos coefficients $b_n^+$, obtained by such an evolution operator:
\begin{align}
    b^+_n\longmapsto_{\eqref{eq:triple_scaling_nomatt}} b^{+,TS}(\widetilde{l}) &=  \frac{2J}{\sqrt{\lambda (1-q)}}\sqrt{(1-2 \lambda e^{-\widetilde{l}/2})(1+2 \lambda e^{-\lambda \Delta - \widetilde{l}/2})}= \nonumber\\
    &=b_0(\lambda) - 2 \lambda J \Big( \Delta e^{-\widetilde{l}/2} + 2 e^{-\widetilde{l}} \Big) + \mathit{O}(\lambda^2)~, \label{eq:triple_scaled_b+}
\end{align}
where we defined $b_0(\lambda)=2J/\lambda+O(\lambda^0)$. The triple-scaled Hamiltonian, denoted by $H^{(+)}$, can be derived out of these triple-scaled Lanczos coefficients as:
\begin{equation}
    \label{eq:triple_scaled_OTFD_Ham_def}
    H^{(+)}\equiv H_R+H_L = e^{i\lambda \widetilde{k}} b^{+,TS}(\widetilde{l}) + b^{+,TS}(\widetilde{l}) e^{-i\lambda \widetilde{k}}~.
\end{equation}
As usual, similarly to \cite{Rabinovici:2023yex,Ambrosini:2024sre}, we perform a redefinition of $H^{(+)}$ by subtracting the divergent ground state energy $2b_0(\lambda)$, and changing sign in \eqref{eq:triple_scaled_OTFD_Ham_def}, in order to achieve a low-energy Hamiltonian bounded from below. If we substitute  \eqref{eq:triple_scaled_b+} in \eqref{eq:triple_scaled_OTFD_Ham_def} after this redefinition, we obtain the expression
\begin{equation}
\label{eq:2triple_scaled_Htot}
H^{(+)}=H_L+H_R= 4\lambda J \left( \frac{\widetilde{k}^2}{2} + \Delta e^{-\widetilde{l}/2} + 2 e^{-\widetilde{l}} \right) + {\cal O}\left( \lambda^2  \right)~.
\end{equation}
As a final step, both in the rest of this section and in \cref{sec:bulk_dual}, we will be interested in the solutions to the equations of motion of the Hamiltonian $H$ defined as
\begin{equation}
\label{eq:triple_scaled_Htot}
H\equiv\frac{H_L+H_R}{2}= 2\lambda J \left( \frac{\widetilde{k}^2}{2} + \Delta e^{-\widetilde{l}/2} + 2 e^{-\widetilde{l}} \right) + {\cal O}\left( \lambda^2  \right)~.
\end{equation}
Notice that, by virtue of this division by 2 performed above with respect to \eqref{eq:2triple_scaled_Htot}, in the limit $\Delta\to 0$, \eqref{eq:triple_scaled_Htot} correctly reproduces the matterless triple-scaled Liouville Hamiltonian, that in \cite{Lin:2022rbf}\cite{Rabinovici:2023yex} was matched to the one of JT gravity.\\

Now we want to search for solutions to the equations of motion of \eqref{eq:triple_scaled_Htot} having the same energy as the matterless TFD complexity \cite{Rabinovici:2023yex}, which is itself computed in the same way from the matterless Liouville Hamiltonian. The idea when we request this condition is that we are considering a low-energy approximation where we can neglect the additional (small) energy coming from the operator insertion on the boundary, while searching for this solution. As we will motivate in \cref{sec:bulk_dual}, this is an interesting approximation from the bulk perspective, because it is analogous to the approach used in \cite{Shenker_2014} for energy insertions in JT gravity creating the shockwave setup.\\ 
We recognize that \eqref{eq:triple_scaled_Htot} is a Morse potential, whose general solutions are discussed, for example, in \cite{Gao_2022}\cite{Ambrosini:2024sre}. We can verify that the following is a solution of the equations of motion (\cref{app:morse_sol}):
\begin{equation}\label{eq:opstate_compl_eom}
    \Tilde{l}_+(t)=2\log\left(\sqrt{1+\left(\frac{\Delta}{4}\right)^2}\cosh{2J\lambda t}+\frac{\Delta}{4}\right),
\end{equation}
and thus is the $\mathcal{O}$TFD complexity in the triple-scaled limit. By substituting \eqref{eq:opstate_compl_eom} back into \eqref{eq:triple_scaled_Htot}, we can verify that this solution has boundary energy:
\begin{equation}\label{eq:matterless_energy}
    E_b=4J\lambda,
\end{equation}
which is equal to the one obtained for the triple-scaled infinite TFD state complexity that in \cite{Rabinovici:2023yex} was matched to the length of the matterless wormhole in JT gravity.
Notice that as expected, when $\Delta\to 0$ the length \eqref{eq:opstate_compl_eom} correctly reproduces the aforementioned known matterless TFD state complexity\footnote{Notice that we have that the timescales appearing in the hyperbolic cosine are the same in the two cases, and in particular they do not differ by a factor of $2$, by virtue of the extra factor of $2$ redefinition introduced in \eqref{eq:triple_scaled_Htot}.}. 
\subsection{The holographic duality between DSSYK and JT gravity}\label{sec:jt_dual_recap}

In \cite{Lin:2022rbf} it was shown that in the triple-scaled limit \eqref{eq:triple_scaling_nomatt} the Hilbert space of DSSYK without matter is isomorphic to the Hilbert space of JT gravity \cite{Jackiw:1984je, Teitelboim:1983ux, maldacena2016conformalsymmetrybreakingdimensional}. This model has attracted considerable attention in recent years as a useful solvable toy model for quantum gravity in two dimensions, \cite{saad2019jtgravitymatrixintegral, Jafferis:2022wez,Jafferis:2022uhu, Harlow:2018tqv, Iliesiu:2024cnh, Mertens:2022irh}, exploring the connections to chaotic dynamics \cite{saad2019latetimecorrelationfunctions, Altland:2022xqx, Jafferis:2022uhu}. Here we limit ourselves to briefly reviewing the facts that will be relevant to our discussion, for a much more extensive introduction to the topic see, for example \cite{Mertens:2022irh}. JT gravity is characterized by an action, written in terms of the 2-dimensional metric $g$ and the dilaton field $\Phi$, in the following form:
\begin{equation}\label{eq:JT_action}
    S_{JT} =   \int_\mathcal{M} d^2 x \sqrt{-g}\Big[\Phi_0R +  \Phi (R+2/l_{AdS}^2)\Big] +  2 \int_{\partial\mathcal{M}} dx \sqrt{\gamma} \Big[  \Phi_0 K + \Phi_b (K-1/l_{AdS}) \Big]~,
\end{equation}
where $\gamma$ is the induced metric on the boundary, $K$ is its extrinsic curvature, and $\Phi_b$ is the value assumed by the dilaton on it. The terms proportional to $\Phi_0$ recombine into the topological part of the Einstein-Hilbert action. We restrict to the case where $\Phi_0$ is a large constant, so we can neglect higher genus contributions and limit our analysis to the disk topology. At this point, the dynamical dilaton field $\Phi$ acts as a Lagrange multiplier and fixes $R=-2/l_{AdS}^2$, describing an hyperbolic disk, on whose boundary both dilaton profile and metric will blow up. Thus, we need to come up with a cut-off procedure defining a regularized boundary characterized by a small parameter $\epsilon$ and boundary conditions:
\begin{equation}\label{eq:reg_bdry_condition}
    \Phi|_{\mathrm{boundary}}=\Phi_b/\epsilon,\quad ds^2|_{\mathrm{boundary}}=-\frac{dt_b^2}{\epsilon^2},
\end{equation}
where $t_b$ is the boundary time. These regularized boundary conditions together with $R=-2/l_{AdS}^2$, determine the metric in the interior to be the one of empty AdS$_2$ \cite{Mertens:2022irh}. In the black-hole patch, if we define $r_s$ to be the Schwarzschild radius, we can write the metric using the following Schwarzschild coordinates: 
\begin{equation}\label{eq:metric_schw}
    ds^2=-\frac{r^2-r_s^2}{l_{AdS}^2}dt^2+\frac{l_{AdS}^2}{r^2-r_s^2}dr^2,
\end{equation}
and in this coordinates the solution to the dilaton equation of motion is $\phi(r,t)=\Phi_b r/l_{AdS}$.\\
Now we can perform the following change of coordinates $(r,t)\to (\rho,t)$, where $\rho$ is implicitly defined by $\sinh{\rho}=(\sinh(r_s z/(2l_{AdS}^2))^{-1}$ and $r=r_s \coth{r_s z/l_{AdS}^2}$, in order to obtain the metric:
\begin{equation}\label{eq:metric_rindler}
    ds^2=-\frac{r_s^2}{l_{AdS}^2}\sinh^2\rho dt^2+l_{AdS}^2d\rho^2
\end{equation}
This metric, in the near-horizon limit $\rho\ll 1$, is in the form of a Rindler metric.  \\
At this point, for example in the metric \eqref{eq:metric_schw}, we can compute the length of geodesics anchored at times $t_L=t_R=t$ on the regularized boundary\footnote{to obtain this result we need to use that the regularized boundary position is $r=l_{AdS}/\epsilon$ and, in the $\epsilon\to0$ limit, the Schwarzschild time coordinate near the regularized boundary coincides with the boundary time $t_b$ \eqref{eq:reg_bdry_condition}.}:
\begin{equation}\label{eq:jt_geodesic}
    l(t)/l_{AdS}=2\log\frac{2l_{AdS}}{\epsilon r_s}+2\log\cosh{r_st/l_{AdS}^2}\implies \Tilde{l}(t)/l_{AdS}=2\log\cosh{r_st/l_{AdS}^2},
\end{equation}
where we defined the re-normalized length $\Tilde{l}(t)$ by subtracting the divergent constant when $\epsilon\to 0$. Using this length as a canonical variable, together with its associated momentum $P$, we can write the classical Hamiltonian for JT gravity as \cite{Harlow:2018tqv}:
\begin{equation}\label{eq:jt_ham}
    H=\frac{1}{l_{AdS}\phi_b} \left( \frac{l_{AdS}^2P^2}{2} + 2 e^{-\tilde{l}/l_{AdS}} \right)
\end{equation}
The geodesic length \eqref{eq:jt_geodesic} is a classical solution of the above Hamiltonian. If we define $\Phi_h=\Phi_b r_s/l_{AdS}$ as the value of the dilaton at the horizon, by substituting \eqref{eq:jt_geodesic} back in \eqref{eq:jt_ham}, we find that the energy associated with the solution \eqref{eq:jt_geodesic} is $2E_b$, where:
\begin{equation}\label{eq:bdry_energy}
    E_b=\frac{\Phi_h^2}{l_{AdS}\Phi_b}=\frac{\Phi_b}{l_{AdS}^3}r_s^2,
\end{equation}
This energy, consistently, is the same we obtain by evaluating the boundary stress-energy tensor on each boundary \cite{Harlow:2018tqv}.\\
Via canonical quantization of \eqref{eq:jt_ham}, we can build an Hilbert space for the quantized theory which consists of eigenstates $\{\ket{\Tilde{l}}\}$ of the geodesic length operator.\\

In \cref{tab:bulk_to_bdry}, we summarize how the bulk picture summarized above can be equivalently portrayed in the triple-scaled limit of DSSYK. In particular, in  \cite{Rabinovici:2023yex}\cite{Lin:2022rbf}, it was understood that the geodesic length \eqref{eq:jt_geodesic} is dual to the K-complexity of the infinite temperature TFD state $\ket{0}$ in DSSYK. In this framework, the bulk-to-boundary map that reorganizes the open chord number basis, in the triple-scaled limit, into bulk states is precisely the orthonormalization procedure of the Lanczos algorithm.\\
With respect to the parameters introduced above, in \cref{tab:bulk_to_bdry}, as well as in the rest of the paper, we have set $\phi_h=1$, which corresponds to choosing null initial values on the complexity side of the duality. The bulk-to-boundary duality between JT gravity and triple-scaled DSSYK requires the following holographic dictionary:
\begin{equation}\label{eq:hol_dict_nomatt}
    2\lambda J =\frac{r_s}{l_{AdS}^2}\quad \text{and}\quad l_f=l_{AdS} ~.
\end{equation}

\begin{figure}
\begin{minipage}[t]{0.49\textwidth}
\centering
     \begin{tikzpicture}[scale=1.2]
    \draw[thick] (-2,0) arc[start angle=180, end angle=360, radius=2];
    \foreach \x in {-1.5,-1.0,1.0, 1.5} {
        \fill[black] (\x,-{sqrt(4-(\x)^2)}) circle (2pt); 
        \draw[thick] (\x,-{sqrt(4-(\x)^2)}) -- (\x,0); 
    }
    \fill[black] (0,-2) circle (2pt); 
    \draw[thick] (0,-2) -- (0,0); 
    
    \node at (-0.5,-0.5) {\tiny{$\cdots$}};
    \node at (0.5,-0.5) {\tiny{$\cdots$}};
    \draw [line width=2pt, black, line cap=round, dash pattern=on 4pt off 4\pgflinewidth] (-2.0,0) -- (2.0,0);
    \draw[thick] (-2.0,0.2) -- (2.0,0.2);
    \draw[thick] (-2.0,0.2) arc[start angle=180, end angle=230, radius=0.3];
    \draw[thick] (2.0,0.2) arc[start angle=0, end angle=-50, radius=0.3];
    \node[above] at (0,0.5) {$\Tilde{l}(t)$};
    \node[right] at (3.3, 0) {$\longleftrightarrow$};
\end{tikzpicture}
\end{minipage}\hfill
\begin{minipage}[t]{0.49\textwidth}
\centering
\begin{tikzpicture}[scale=0.5]
\draw [ color={rgb,255:red,26; green,95; blue,180} , line width=0.9pt ] (6.25,15.25) rectangle (6.25,15.25);
\draw [ line width=0.9pt ] (6.25,15.5) rectangle (16.5,6.25);
\draw [ color={rgb,255:red,11; green,64; blue,201}, line width=0.9pt, dashed] (6.25,15.5) -- (16.5,6.25);
\draw [ color={rgb,255:red,11; green,64; blue,201}, line width=0.9pt, dashed] (16.5,15.5) -- (6.25,6.25);

    
\node (f) [above] at (12.5,12.5){$\Tilde{l}(t)$};
\node (l1) [left] at (6.25,12){$t$};
  \node (l4) [right] at (16.5,12){$t$};
  \draw [line width=2pt, black, line cap=round, dash pattern=on 4pt off 4\pgflinewidth] (l1) -- (l4);
\end{tikzpicture}
\end{minipage}\hfill
\caption{Intuitive representation of the DSSYK-JT gravity duality, where the blue dashed lines represent the event horizon of the 2D black hole. The re-normalized length obtained by considering the triple-scaled limit of K-complexity, that is the number of open chords intersected by the black dashed line, is dual to the re-normalized geodesic length anchored at times $t_L=t_R=t$ on the regularized boundary of AdS$_2$. Upon canonical quantization, by the matching of the Hamiltonians with the holographic dictionary \eqref{eq:hol_dict_nomatt}, this duality is uplifted to an isomorphism between the Hilbert spaces of the quantized theories \cite{Rabinovici:2023yex}.}
\label{fig:jt_duality_matterless}
\end{figure}
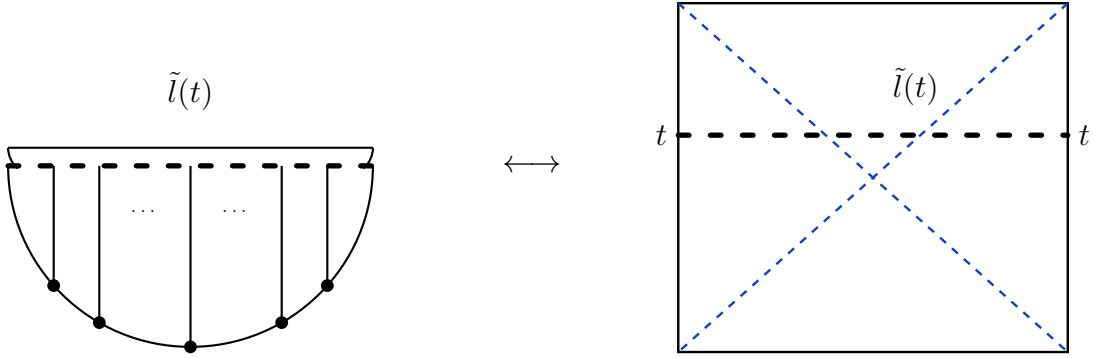

\begin{table}[h!]
  \centering
  \begin{tabular}{|p{7cm}|p{7cm}|}
\hline
    \textbf{Boundary} & \textbf{Bulk}  \\ \hline
    triple-scaled DSSYK & JT gravity  \\ \hline
    $H = - \frac{2J}{\lambda} +2 \lambda J \left(\frac{l_f^2 k^2}{2}+2 e^{-\tilde{l}/l_f} \right)$ & 
    $H = \frac{1}{l_{AdS}\phi_b} \left( \frac{l_{AdS}^2P^2}{2} + 2 e^{-\tilde{l}/l_{AdS}} \right)$ \\ \hline
    Krylov basis are $|\tilde{l}\rangle$ states & Hilbert space consists of states with well defined wormhole length $|\tilde{l}\rangle$ \\ \hline
    K-complexity of $\ket{0}$ in triple-scaled limit & Normalized wormhole length in JT gravity \\ \parbox{5cm}{\begin{align*}
    \lambda \widetilde{C}_K(t) /l_f=&  2 \log \left[ \cosh \left( 2\lambda J t \right)\right]
    \end{align*}} & \parbox{5cm}{\begin{align*}
   \frac{\tilde{l}(t)}{l_{AdS}} =& 2 \log\left[\cosh\left(\frac{r_s}{l_{AdS}^2} t \right) \right]
    \end{align*}}\\
\hline
  \end{tabular}
  \caption{Summary of the bulk-to-boundary dictionary between DSSYK and JT gravity.}
  \label{tab:bulk_to_bdry}
\end{table}

\subsubsection{A note on dimensional reduction of the BTZ black hole}\label{sec:jt_to_btz_dimred}
As reviewed in the previous section, we understand from \eqref{eq:metric_rindler} that what is referred to as a `black hole' in two-dimensional gravity is a coordinate patch of AdS$_2$ associated to an accelerated observer. In particular, there is no singularity or mass in the origin, and strictly speaking the event horizon would be a Rindler horizon. However, it is understood that we can still interpret these solutions as black holes, because we can obtain them via a dimensional reduction of near-horizon limits of higher dimensional black holes, inequivalent to empty AdS. In this section, we will first review how JT gravity can be obtained via a dimensional reduction of the three-dimensional BTZ black hole (see \cite{Verheijden:2021yrb} for a more extended discussion). Then, we will define, purely from two-dimensional gravity grounds, the mass and temperature of the would-be black hole, and argue that they coincide with the prescription obtained from dimensional reduction.\\

We want to show that it is possible to recognize JT gravity as the AdS$_2$ sector of the BTZ black hole. Let us recast the metric \eqref{eq:metric_schw} in the right patch using the Kruskal coordinates, that, similarly to their more common 4D Schwarzschild black hole counterparts, allow a maximal extension to the whole two-sided spacetime:
\begin{equation}\label{eq:kruskal_coord_def}
    U=-e^{-\frac{r_s}{l_{AdS}^2}(t-r^*)},\quad V=e^{\frac{r_s}{l_{AdS}^2}(t+r^*)},
\end{equation}
where $r^*$ is a tortoise coordinate such that, when $r\to r_s$, $r^*\to-\infty$:
\begin{equation}
    r^*(r)=\int \frac{l_{AdS}^2}{r^2-r_s^2}dr=\frac{l_{AdS}^2}{2r_s}\log\biggr|\frac{r-r_s}{r+r_s}\biggr|
\end{equation}
 The metric in the coordinates \eqref{eq:kruskal_coord_def} becomes:

\begin{equation}\label{eq:metric_kruskal}
    ds^2=\frac{-4l_{AdS}^2}{(1+UV)^2}dUdV,
\end{equation}
where the boundaries are at $UV=-1$, the singularities at $UV=1$, and the horizons at $U=0$ and $V=0$. \\

Considering $U,\,V$ defined as in \eqref{eq:kruskal_coord_def}, but with $r$ being the three-dimensional radial coordinate, together with an angular coordinate $\phi$ with periodicity $\phi\sim\phi+2\pi$, we can write the BTZ black hole metric as \cite{Shenker_2014}:
 \begin{equation}\label{eq:BTZ_bh_metric}
     ds^2=\frac{-4l_{AdS}^2}{(1+UV)^2}dUdV+\frac{r_s^2(1-UV)^2 d\phi^2}{(1+UV)^2},
\end{equation}
The metric \eqref{eq:BTZ_bh_metric} is associated to a black hole with mass $M$ given by $r_s^2=8G^{(3)}Ml_{AdS}^2$, where $G^{(3)}$ is the $3D$ Newton constant, that can be linked to the $2D$ one via $G^{(3)}\sim 2\pi l_{AdS} G^{(2)}$ \cite{Mertens:2022irh}, where we compactified a dimension of characteristic scale $l_{AdS}$. At this point, we recognize that dimensional reduction on the $\phi$ coordinate in \eqref{eq:BTZ_bh_metric} yields \eqref{eq:metric_kruskal}, so that JT gravity can be seen as the AdS$_2$ sector of the BTZ black hole. Notice that, by what we discussed above, if we obtain a two-dimensional theory of gravity with this prescription, the radius of the event horizon $r_s$ appearing for example in \eqref{eq:metric_schw}, is linked to the mass of the BTZ black hole.\\

Now before closing the section, we want to link the $2D$ definition of black hole mass and temperature to the ones we gave above coming from the $3D$ BTZ black hole mass, and show their consistency. From purely, $2D$ grounds, we can define the Hawking temperature of the black hole to be the one associated to the Rindler observer in \eqref{eq:metric_rindler}:
\begin{equation}\label{eq:2d_hawk_T}
    T = \frac{r_s}{2\pi l_{AdS}^2}=\frac{ \phi_h}{2\pi l_{AdS}~ \phi_b}~.
\end{equation}
Then, can define the mass of the $2D$ black hole, by evaluating the stress energy tensor on the boundary \eqref{eq:bdry_energy}:
\begin{equation}\label{eq:2d_bh_mass}
    M\equiv E_b=\frac{\Phi_b}{l_{AdS}^3}r_s^2
\end{equation}
Notice that, upon re-normalizing\footnote{this is needed because the Einstein-Hilbert 3D gravity action used to obtain \eqref{eq:BTZ_bh_metric} is defined with a prefactor of $(16\pi G^{(3)})^{-1}$\cite{Verheijden:2021yrb}\cite{Mertens:2022irh}. Then via dimensional reduction it can be associated to a JT gravity action with an extra multiplicative coefficient $(16\pi G^{(2)})^{-1}$, with respect of the convention \eqref{eq:JT_action} we derived our 2D results in.} $\Phi\to\Phi/(16 \pi G^{(2)})$, \eqref{eq:2d_bh_mass} and \eqref{eq:2d_hawk_T} coincide with the corresponding ones obtained in \cite{Verheijden:2021yrb} from dimensional reduction of the BTZ metric. This means that the mass of the $2D$ black hole, via matching of the radius of the event horizons featuring in the AdS$_2$ Rindler wedge \eqref{eq:metric_rindler} and the BTZ metric \eqref{eq:BTZ_bh_metric}, is directly inherited from the mass of the actual black hole, whose dimensional reduction gives JT gravity.

\subsubsection{The switchback effect of ERB lengths}\label{sec:switch_grav_intro}
The switchback effect is one of the peculiar properties of ERB lengths in gravitational models, observed under specific matter perturbations \cite{Stanford:2014jda,susskind2014switchbacksbridge}. In particular, we review the configuration of \cite{Stanford:2014jda}, where spherically-symmetric low-energy $E$ quanta are inserted, say on the left gravitational boundary, by acting with an operator $W_L(t_s)$ at a certain time $t_s$. The dynamics of these particles is controlled by the scrambling timescale $t_{scr}\propto \log(M/E)$: these quanta will fall into the black hole and get blue-shifted, becoming more energetic, and if we wait a long time $t\gg t_{scr}$, they will start to backreact in the form of a high-energy shockwave insertion (as we review in \cref{sec:shock_model_def}). Next we consider many insertions of this kind on the TFD state:
\begin{equation}\label{eq:multi_shock_state}
    W_L(t_n)\dots W_L(t_1)\ket{TFD},
\end{equation}
 where $t_1,\dots t_n$ are separated by more than the scrambling time and are not necessarily time-ordered. As noticed in \cite{Stanford:2014jda}, the set of states \eqref{eq:multi_shock_state}, obtained by acting only on the left boundary, also describes the insertion of small energy quanta on the right. This happens because states where we insert low-energy shockwaves, separated by more than the scrambling time, have approximately maximal entanglement \cite{Stanford:2014jda}. For such states $\ket{\psi}$, one has the following `\textit{reflection} identity':
 \begin{equation}\label{eq:reflection_princ}
     W_L(t_1)\ket{\psi}=W_R(-t_1)\ket{\psi}
 \end{equation}
 Using this `\textit{reflection} identity', one can represent, in the aforementioned approximations, all multi-shock states in the form \eqref{eq:multi_shock_state}.\\
 
\begin{figure}
\centering
        \includegraphics[width= 0.3\textwidth]{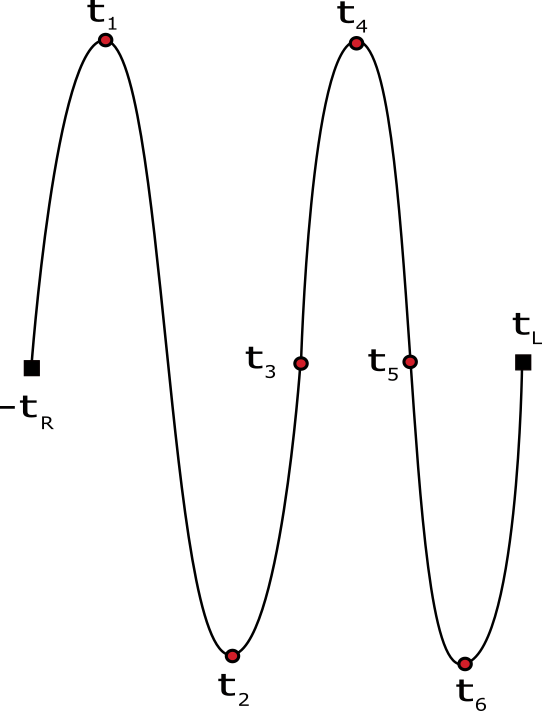} 
    \caption{A timefold with six operator insertions, denoted by the red dots, between times $t_L$ and $-t_R$: at $t_1$, $t_2$, $t_4$ and $t_6$ we have switchback insertions, while $t_3$ and $t_5$ are through-going.} 
    \label{fig:timefold_example}
\end{figure}

 Now, as in \cite{Stanford:2014jda}, we allow the state above to be evolved for times $t_L$ and $t_R$ using respectively the evolution operators $U_{L,R}(t_{L,R})=e^{-iH_{L,R}t_{L,R}}$ on the left/right boundaries:
 \begin{equation}\label{eq:evolution_example}
     U_R(t_R)U_L(t_L)W_L(t_n)\dots W_L(t_1)\ket{TFD}=U_L(t_L)W_L(t_n)\dots W_L(t_1)U_L^\dag(-t_R)\ket{TFD},
 \end{equation}
 where in order to obtain the second expression we used that $H_L-H_R$ annihilates the TFD.
 We can define from the evolution prescription \eqref{eq:evolution_example}, a timefold contour. The contour is outlined by the ordered sequence of times $t_0,t_1\dots t_{n+1}$, where $t_{n+1}\equiv t_L$ and $t_0=-t_R$ and the other $t_i$'s are the times at which the perturbations $W_L$ are inserted. We can use the values of these times and their position in the sequence to define two kinds of operator insertions: switchback and through-going. In particular, we say that the $j$-th insertion is through-going if $t_{j-1}<t_j<t_{j+1}$, while it is a switchback if we have $t_{j-1},t_{j+1}<t_{j}$ (\cref{fig:timefold_example}). 
 In \cite{Stanford:2014jda}, it has been observed that the length of boundary-anchored geodesics presents a specific delay of a scrambling time anytime you encounter a switchback insertion. So, the switchback effect, at the level of an ERB length $C_W(t_0,t_1,\dots t_{n+1})$, characterized by insertions of $W$ at $t_1,\dots t_n$, the switchback effect is defined as showcasing the following asymptotic behavior:
 \begin{equation}\label{eq:switchback_erb_def}
    C_W(t_0,\dots,t_{n+1})\propto t_f-2n_{sb}t_{scr},
\end{equation}
where $n_{sb}$ is the number of switchbacks inserted and $t_f$ is the total length of the timefold $t_f=|t_1-t_0|+|t_2-t_1|+\dots|t_{n+1}-t_n|$. See \cref{app:switch_erb} for a review on how to compute this result from \cite{Stanford:2014jda}, in particular for the cases of interest in this paper. However, let us stress here that, in order to obtain \eqref{eq:switchback_erb_def}, we require that all the perturbations inserted define parametrically equivalent scrambling times.\\

To summarize, the effect of a switchback insertion is causing a delay of a scrambling time in the asymptotic late-time linear behavior of ERB lengths. This property has obtained considerable attention because it is also showcased by the gate complexity of the precursor operator $W(t)=U^\dag(t)WU(t)$, where $U(t)$ is a unitary time evolution and $W$ is a perturbation acting on a small number of qubits. This fact has been one of the motivations to study this notion of complexity in holographic systems \cite{Susskind:2014rva, Stanford:2014jda, susskind2014switchbacksbridge, Haferkamp_2022, Jian:2020qpp, Brown:2015bva, Kar_2022, PhysRevD.99.046016,Brown_2016,Chapman_2022,Carmi_2017,Harlow_2013}.\\The gate complexity of $W(t)$ is defined as the minimum number of unitary gates necessary to reproduce it. The crucial idea now is noticing that if $W=\mathbf{1}$ then the unitary gates reproducing $U(t)$ cancel against the ones for $U^\dag(t)$, and the complexity does not grow. However, if $W$ perturbs even a small number of qubits, because of the chaotic nature of the evolution, it spoils this cancellation \cite{Susskind:2014rva}. If we imagine that $W$ infects some small number of qubits, we can compute the complexity via an epidemic model where qubits become infected if they are acted upon by a unitary gate that simultaneously acts on an infected qubit \cite{susskind2014switchbacksbridge}. The number of unitary gates needed to reproduce $W(t)$, that is the complexity of the precursor, will be the number of infected qubits. We imagine to model the chaotic dynamics with $k$-local random gates, that is a unitary gate that acts on $k$ qubits chosen at random, then at early times the complexity has an exponential growth behavior, because generally an infected qubit will be coupled with non-infected ones. Then, when the epidemic has spread to a big enough fraction of the total qubits, condition which defines the scrambling time $t_{scr}$, the exponential behavior will stop. However, complexity will still rise, namely when $t\gg t_{scr}$, the cancellation between unitary gates is totally destroyed, and the complexity of $W(t)$ will just be twice the number of unitary gates needed to approximate $U(t)$, which is linear in $t$. So the complexity of the precursor will be \cite{Stanford:2014jda}:
\begin{equation}\label{eq:precursor_compl}
    C_W(t)\propto 2(t-t_{scr}),
\end{equation}
where the $t_{scr}$ delay is accounting for the early times partial cancellation during the exponential growth regime, before the butterfly effect sets in. Notice that the evolution prescription for the precursor is associated to a timefold of length $t_f=2t$ and one switchback insertion $n_{sb}=1$, so that \eqref{eq:precursor_compl} is precisely the behavior predicted by the switchback effect \eqref{eq:switchback_erb_def}.\\

The crucial observation that operator complexity showcases the switchback effect, similarly to ERB lengths, was what originated the idea of this quantity having a geometric dual, in the context of an holographic correspondence. In particular, any good notion of complexity, which one wants to relate with a bulk dual consisting of an ERB length, is expected to showcase the switchback effect. One of the purposes of this work is proving that this is the case for Krylov complexity, whose definition does not depend on arbitrary control parameters.\\ 
As discussed in \cite{Ambrosini:2024sre}, there is a possible correspondence between this epidemic model and Krylov operator complexity in the semiclassical limit. Armed with this intuition, we can heuristically expect that Krylov complexity shows the same behavior upon perturbation insertion. The purpose of \cref{sec:switchback_compl} will be to rigorously prove that, even for multi-operator insertions, operator Krylov complexity showcases the switchback effect in the triple-scaled low-energy limit where we state the duality with JT gravity. This results shows that the geometric nature of K-complexity, expressed by an ERB length in the bulk, persists under operator perturbations, and robustly outlines the holographic dictionary defined in \cref{sec:bulk_dual}.\\

\section{The bulk dual of operator K-complexity}\label{sec:bulk_dual}
In this section we want to describe a bulk dual of a single operator insertion in DSSYK. As reviewed in \cref{sec:jt_dual_recap}, in the triple-scaling limit, matterless DSSYK is dual to JT gravity, and we want to leverage the knowledge of this duality to describe the matching in the presence of matter. We notice however, that the heavy operator insertion $\Delta\to\infty$, which is better controlled analytically \cite{Ambrosini:2024sre}, is describing a large energy insertion that can drastically modify the matterless background. So, for the purpose of describing the duality of an operator insertion on top of the JT gravity background, we need to analyze the triple-scaled limit of light operator insertions with $\Delta\to 0$ in DSSYK. We imagine that an insertion of a light operator does not completely destroy the matterless background, so that a natural expectation for the dual of a light-operator in DSSYK will be a low-energy matter perturbation inserted on a JT gravity background.\\

In \cref{sec:op_compl_recap}, we reviewed how one can use a saddle point approximation to find the Lanczos coefficients, in the $\lambda\to0$ limit, relative to operator complexity. This argument (summarized in \cref{app:recap_asympt}) holds in the $\Delta\to0$ limit as well, so that light-operator Krylov complexity is a total chord number with Lanczos coefficients \eqref{eq:b_nplus1_induction_proof}. This means that, in the semiclassical limit, operator complexity is what we call a \textit{total length}, that is a sum of a left and a right length. From this property, we expect that its bulk dual is a geodesic length separated in two sections by an interface. A possible guess for this interface, obtained, as expected, via a low-energy perturbation of JT gravity, is then a shockwave-like discontinuity. Indeed, this shockwave can be obtained by inserting some low energy quanta on the JT gravity regularized boundary, which will then blue-shift and backreact on the background geometry \cite{dray_shockwave}.\\

 In the following sections, we will discuss how we can access the triple-scaled operator complexity in DSSYK, in the light operator regime $\Delta\to0$. In particular, for low-energy operators with $\Tilde{q}\sim 1$, the leading contribution to the Lanczos coefficients \eqref{eq:b_nplus1_induction_proof}, picked by the aforementioned saddle point approximation, returns a null operator complexity. This means that the value of the complexity when $\Delta\to 0$ is controlled by some subleading contribution in \eqref{eq:lower_diag_L}, which is rather cumbersome to access directly.  We argue that we can capture these contributions, without explicitly performing the sum \eqref{eq:lower_diag_L}, by leveraging the fact that operator K-complexity is a \textit{total chord length}, that is the sum of the solution of the equations of motion for the left/right Hamiltonians. The key points of the procedure, carried out in the following sections, are the following:
 \begin{itemize}
     \item use the $\mathcal{O}$TFD Lanczos coefficients to build the triple scaled effective total Hamiltonian $H=(H_L+H_R)/2$, as summarized in \cref{sec:op_compl_recap}.
     \item Recognize that, to first order in the $\Delta\to0$ approximation, we can consider the symmetric case where $H_L=H_R=H$.
     \item Recognize that the operator complexity is a total chord number, built as a sum of a left/right chord number, respectively, solutions of the $H_{L/R}$ Hamiltonians, which we know from the previous point. In the general case, the equations of motion for $l_{L,R}$ obtained from Hamiltonians $H_{L,R}$ are complicated coupled differential equations \cite{Lin:2022rbf}. However, in the symmetric limit of the point above, these equations decouple, and in particular they are identified with a single equation for $l/2$, whose solutions are the possible choices of $l_{L}$ and $l_R$.
     \item Solve, in a simplifying late-time limit, separately the equations of motion of $H_{L/R}$ to find $l_{L,R}$ respectively, and sum the solutions to build the triple-scaled operator complexity. 
 \end{itemize}
 From this procedure we find the triple scaled length associated to the K-complexity of light operators, of which we proposed to find the holographic dual. Notice that, in the tentative procedure described above, we are proposing to perform at the same time a low-energy and a late-time limit, which we will refer to as the `\textit{shockwave approximation}'. The reason behind this nomenclature is that, as we will review in the next sections, in the putative bulk dual model, an analogous set of assumptions is needed for the shockwave description to hold \cite{Shenker_2014}.\\ 
 
 
 We start the next section by giving a construction of the conjectured bulk dual theory, explaining the assumptions that underlie the proposed description with the \textit{shockwave approximation}. Then we show how to compute geodesics in this background \cite{Stanford:2014jda}\cite{Shenker_2014}, and in particular, we will find a length dual to the DSSYK operator complexity, which confirms its geometric interpretation and provides a new entry in the JT-DSSYK holographic dictionary.\\

\subsection{The shockwave setup}\label{sec:shock_model_def}

In this section, we introduce the bulk gravity model that we claim to be dual to DSSYK with low-energy operator insertions, in the triple-scaled limit. From the knowledge of the dual of matterless DSSYK, we expect that the theory we are searching for will be describing the insertion of some low-energy quanta in a JT background. Indeed, what we consider is a spherically symmetric configuration of low-energy particles inserted on the boundary of the BTZ black hole background. These particles, by getting blue-shifted when falling in the black hole, will appear to a late-time observer as a high-energy shockwave. This is the configuration considered in \cite{Shenker_2014}, which we will briefly review below. Analogously to what we did in \cref{sec:jt_to_btz_dimred}, we can then perform a dimensional reduction and obtain the description of the shockwave insertion in two-dimensional JT gravity. This will be the model that we claim to be the dual of triple-scaled DSSYK with light-operator insertions. \\

In \cite{dray_shockwave}, it has been shown that the insertion of approximately massless particles, that is with energy dominated by the kinetic part, in the far past of the left boundary can be effectively computed with the insertion of a null shockwave on $U\approx0$. This shockwave can be seen as an interface between a theory characterized by a black hole of mass $M$ and one of mass $M+E$ (because the particles have fallen in the black hole). In order to build this geometry, we then need to find the appropriate continuity conditions across the null shockwave.\\
Let us consider a shockwave of energy $E\ll M$ inserted at a time $t_w$ on the left boundary. In this case the gluing between the BTZ metric \eqref{eq:BTZ_bh_metric} with black hole mass $M$, and the one with $M+E$, happens along the null trajectory $U_w=e^{-r_s t_w/l_{AdS}^2}$. Now we want to relate quantities on the right side of this interface, with the corresponding on the left, which we indicate here with a tilde. We have for example, given $r_s^2\sim M$, from what we discussed in \cref{sec:jt_to_btz_dimred}, that $\Tilde{r_s}=\sqrt{1+E/M}r_s$, and the position of the shockwave is $\Tilde{U}_w=e^{-\Tilde{r_s}t_w/l_{AdS}^2}$, as we require the time coordinate to flow continuously when crossing. Then, as the second condition for smooth gluing, we can impose that the radius of the circle spanned by $\phi$ in \cref{eq:BTZ_bh_metric} is continuous across the shockwave:
 \begin{equation}\label{eq:cont_phi}
     \Tilde{r}_s\frac{1-\Tilde{U}_w \Tilde{V}}{1+\Tilde{U}_w \Tilde{V}}=r_s\frac{1-U_w V}{1+U_wV}
 \end{equation}
 Let us now define the scrambling time as $t_{scr}\sim \frac{l_{AdS}^2}{r_s}\log\frac{M}{E}$. The condition \eqref{eq:cont_phi}, together with the one for $\Tilde{U}_w$, in the limits $E/M\ll 1$ and $t_w\gg t_{scr}$\footnote{here we are considering the configuration where we move the operator insertion in the past, and observables are computed on the initial timeslice $t_L=t_R=0$. Because of the symmetry of the TFD under $H_L-H_R$, this is equivalent to inserting the shockwave at $t_w=0$, and then evolving the observables with $H_L-H_R$. This is the configuration we will consider in the next sections, when we will compute the length of geodesics anchored at $t_L=-t_R=t$, with the operator inserted at $t_w=0$. Notice that, in this case, the large $t_w\gg t_{scr}$ limit, underlying the \textit{shockwave approximation}, translates in the fact that the computation of the observables in this shockwave setup will be trustworthy when $t\gg t_{scr}$.}, admits as a solution the coordinate shift \cite{Shenker_2014}:
 \begin{equation}\label{eq:shift_sol}
     \Tilde{V}=V+\alpha\quad\mathrm{with}\quad \alpha=\frac{E}{4M}e^{r_s t_w/l_{AdS}^2}
 \end{equation}
 So from the solution \eqref{eq:shift_sol}, we get the following BTZ black hole metric modified by the insertion of the spherically symmetric shell of particles \cite{Shenker_2014}:
\begin{equation}\label{eq:schock_metr_noshift}
   d s^2=\frac{-4 l_{AdS}^2 d U d V+r_s^2(1-U(V+\alpha \theta(U)))d\phi^2}{[1+U(V+\alpha \theta(U))]^2} .
\end{equation}
We can use a different set of discontinuous coordinates $V\to V+\alpha\theta(U)$ and same $U$ so that the metric:
\begin{equation}\label{eq:shock_metr_shift}
    d s^2=\frac{-4 l_{AdS}^2 d U d V+4 l_{AdS}^2 \alpha \delta(U) d U^2+r_s^2(1-UV)^2d\phi^2}{(1+U V)^2}\,.
\end{equation}
At this point, similarly to \cref{sec:jt_to_btz_dimred}, we can perform a dimensional reduction over the coordinate $\phi$, and obtain the metric for the JT gravity black hole modified by the shockwave insertion as:
\begin{equation}\label{eq:jt_shock_metr_shift}
    d s^2=\frac{-4 l_{AdS}^2 d U d V+4 l_{AdS}^2 \alpha \delta(U) d U^2}{(1+U V)^2}\,,
\end{equation}
which is the matterless JT gravity metric \eqref{eq:metric_kruskal} we exhibited in \cref{sec:jt_to_btz_dimred}, with the addition of an interface where coordinates get shifted (\cref{fig:JT_shockwave}).\\
 
\subsubsection{Geodesic lengths in JT gravity with a shockwave}\label{sec:shockwave_geodesics}

In this section, we want to discuss how to compute geodesic lengths in JT gravity after the shockwave insertion \eqref{eq:jt_shock_metr_shift}. Let us specify the precise order of the low-energy limit we will consider in our computation. Using \eqref{eq:bdry_energy}, we can compute the boundary energy $E_s$, modified by the shockwave insertion, as:
\begin{equation}
    E_s\approx E_b(1+E/M)
\end{equation}
To first approximation in $E/M\ll1$, analogously to the setup considered in \cite{Shenker_2014}, we neglect this change in the boundary energy, so that we have, for example, $r_s=\Tilde{r_s}$. We refer to this low-energy approximation, together with the late-time limit discussed in \cref{sec:shock_model_def}, as the `\textit{shockwave approximation}'. Then in these limits, the effect of crossing the shockwave is implementing a null coordinate shift along its interface close to the horizon \eqref{eq:shift_sol}.

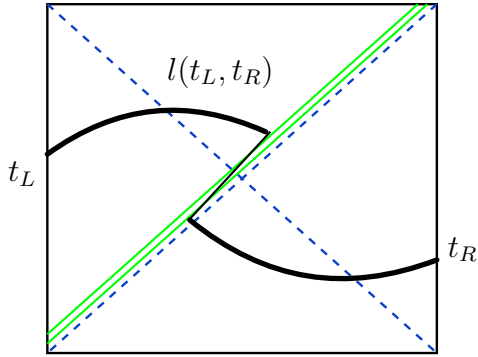
\begin{figure}
\centering
    \begin{tikzpicture}[scale=0.5]
\draw [ color={rgb,255:red,26; green,95; blue,180} , line width=0.9pt ] (6.25,15.25) rectangle (6.25,15.25);
\draw [ line width=0.9pt ] (6.25,15.5) rectangle (16.5,6.25);
\draw [ color={rgb,255:red,11; green,64; blue,201}, line width=0.9pt, dashed] (6.25,15.5) -- (16.5,6.25);
\draw [ color={rgb,255:red,11; green,64; blue,201}, line width=0.9pt, dashed] (16.5,15.5) -- (6.25,6.25);
\draw [ thick,green] (6.25,6.75) -- (16,15.5);
\draw [ thick,green] (6.25,6.5) -- (16.25,15.5);
    
\node (f) [above left] at (12.5,13.0){$l(t_L,t_R)$};
\node (l1) [left] at (6.25,11){$t_L$};
  \node (l2) [right] at (12,12){};
\draw[bend left,line width=2pt,  black, line cap=round]  (l1) to node [auto] {} (l2);
\node (l3) [left] at (10,10){};
  \node (l4) [right] at (16.5,9){$t_R$};
  \draw[bend right,line width=2pt, black, line cap=round]  (l3) to node [auto] {} (l4);
  \draw [thick,black] (12.12,12.12) -- (9.95,9.75);
\end{tikzpicture}
    \caption{A shockwave (green) insertion amounts to a null shift along its interface when computing the length of a geodesic anchored at points $t_L$, $t_R$ on the regularized boundary. } 
    \label{fig:JT_shockwave}
\end{figure}

The length of a geodesic can be computed by summing the lengths from the left/right regularized boundaries up to a point on the shockwave where we impose the shift condition \eqref{eq:shift_sol}(\cref{fig:JT_shockwave}). This crossing point on the shockwave is chosen imposing that the length of the curve is extremal, and this condition guarantees that it is a geodesic. So, via this procedure (summarized in \cref{app:geod_details_single}), it can be obtained that the (renormalized) length of a geodesic anchored on the left/right boundaries respectively at times $t_L$ and $t_R$ is \cite{Shenker_2014}\footnote{for simplicity we have suppressed the $\Tilde{\cdot}$, notation for renormalized length, but, also in this case as in \eqref{eq:jt_geodesic}, one needs to subtract a term $\sim \log(\#/\epsilon)$ to obtain a finite result.}:
\begin{equation}\label{eq:shock_geod_leftright}
    l(t_L,t_R)=2\log\left(\cosh{\frac{r_s}{2l_{AdS}^2}(t_L+t_R)+\frac{E}{8M}e^{\frac{r_s}{2l_{AdS}^2}(t_L-t_R+2t_w)}}\right),
\end{equation}
where $t_w$ is the time of insertion of the particles of low-energy $E$.
In DSSYK we insert the light operator at $t=0$, so, in view of the holographic matching to be performed in \cref{sec:hol_dict_matt}, here we will take $t_w=0$. Then we consider the following two anchoring prescriptions, where the geodesics are anchored respectively at $t_L=t_R=t$, which we call $l_+(t)$, and $t_L=-t_R=t$, which we call $l(t)$:
\begin{equation}\label{eq:shock_lgt_+}
    l_+(t)=2l_{AdS}\log\biggr(\cosh\frac{r_s}{l_{AdS}^2}t+\frac{E}{8M}\biggr)
\end{equation}
\begin{equation}\label{eq:shock_lgt_-}
    l(t)=2l_{AdS}\log\biggr(1+\frac{E}{8M}e^{\frac{r_s}{l_{AdS}^2}t}\biggr)
\end{equation}

Notice that, because of their defining anchoring prescriptions, $l_+(t)$ and $l(t)$, are intuitively associated with the evolutions $H_L+H_R$ and $H_L-H_R$ respectively. As we reviewed in \cref{sec:op_compl_recap}, these are the evolution operators considered in the Lanczos algorithm to obtain respectively the $\mathcal{O}$TFD and operator complexity in DSSYK. We will confirm in \cref{sec:hol_dict_matt} that the appropriate prescription to find the dual of these DSSYK complexities is to consider the bulk anchoring prescriptions mimicking the Lanczos evolution operator.\\

The \textit{shockwave approximation}, underlying the description of the matter backreaction we presented here, is similar to the one we will consider for DSSYK in \cref{sec:3scale_op_compl}. In view of the holographic matching we will perform in the next sections, it is going to be useful to study the regimes of \eqref{eq:shock_lgt_-} and \eqref{eq:shock_lgt_+}, in this late-time and low-energy limit.\\
Notice that, rigorously, the shockwave description we used is valid only for times $t\gg t_{scr}$. However, we do know that the result, obtained from this late-time approximation, needs to match the real solution from the scrambling time onward, where $l(t)$ is still approximately zero. This means that \eqref{eq:jt_op_compl_O0} is able to approximately access also the pre-scrambling regime, when the solution is small, similarly to what we will discuss in \cref{sec:3scale_op_compl} for the triple-scaled operator complexity. In particular, the pre-scrambling regime is reliable at the order $\mathcal{O}((E/M)^0)$ in the low-energy limit. Then, if we keep terms $\sim\frac{E}{M}e^t$, which may become $\sim O(1)$ after the scrambling time, by virtue of the late-time limit, we can observe the asymptotic linear behavior. So, the characteristic regimes of \eqref{eq:shock_lgt_-} in the \textit{shockwave approximation} are:
\begin{equation}\label{eq:jt_op_compl_O0}
      l(t)/l_{AdS}\approx2\log\left(1+\frac{E}{8M}e^{\frac{r_s}{l_{AdS}^2} t}\right)\approx 
    \begin{cases}
      \frac{E}{4M}e^{\frac{r_s}{l_{AdS}^2} t}\sim 0 +O\left(\frac{E}{M}^1\right) & \text{if $t\ll t_{scr}$} \\
      \mathrm{const.}+2\frac{r_s}{l_{AdS}^2}  t & \text{if $t\gg t_{scr}$},
    \end{cases}
\end{equation}
where $t_{scr}\sim\frac{l_{AdS}^2}{r_s}\log\frac{M}{E}$. Note that the prefactor $\frac{l_{AdS}^2}{r_s} = \frac{\beta}{2\pi} = \lambda^{-1}$, where $\lambda_L$ is the maximal Lyapunov exponent of the black hole \cite{Maldacena:2015waa}, and $M/E \sim S$ for a perturbation whose energy $E$ is that of a thermal quantum at the Hawking temperature \cite{Shenker_2014}. This identifies consistently $t_{scr} \sim \lambda_L^{-1} {\rm log} S $. These regimes are analogous to those we will find for operator complexity in \eqref{eq:op_compl_O0}, which will be discriminated by an analogous scrambling timescale. Then, matching the scrambling times defined in the two theories, will ensure that the corresponding characteristic regimes coincide, upon the appropriate parameter identification, encoding the matter insertion details. Notice that, as it will happen for the $\mathcal{O}$TFD complexity in DSSYK, in the \textit{shockwave approximation} described here, $l_+(t)$ reduces to the matterless result \eqref{eq:jt_geodesic}, so it gives no information on new entries in the holographic dictionary.\\

\subsection{Triple-scaled lengths in DSSYK with matter}\label{sec:3scale_op_compl}

In this section, we want to compute the triple-scaled operator complexity in DSSYK.
We restrict our study to the low-energy limit where solutions to the equation of motion have the same boundary energy of the matterless case. This approximation is both a technical and a conceptual requirement, if our objective is to set an holographic duality with the bulk model defined in \cref{sec:shock_model_def}.\\The technical purpose of this low-energy limit is maintaining the symmetry between the left and right sectors of DSSYK, so that we are in the symmetric configuration where $l_L=l_R=l/2$ and $H_L=H_R=\frac{H_L+H_R}{2}$. The procedure we will follow to determine the operator complexity will leverage its property of being a \textit{total length}, meaning that its expectation value can be computed as a sum of those of $l_L$ and $l_R$. The symmetric configuration, originating from the low energy approximation, allows us to compute, from the triple-scaled Hamiltonian \eqref{eq:triple_scaled_Htot}\footnote{and the same expression (apart from normalization dependent factors of $2$) for the symmetric configuration was found in \cite{Lin:2022rbf} with an equivalent construction, without building the triple-scaled Hamiltonian \eqref{eq:triple_scaled_Htot} from the Lanczos coefficients pertaining the computation of the $\mathcal{O}$TFD complexity.}, the expressions of $H_{L,R}$, whose equations of motion are decoupled and thus can be solved separately for $l_{L,R}$.\\ 
From a more fundamental standpoint, on the other hand, the bulk model of \cref{sec:shock_model_def}, contains, in the assumptions underlying the \textit{shockwave approximation}, a similar low-energy limit. So the boundary theory dual to the shockwave insertion needs a low-energy approximation analogous to the one discussed above, in order to mimic the assumptions in the bulk.\\

From \eqref{eq:triple_scaled_Htot}, we write the expression for $H_{L,R}$, respectively, as a function of the canonical variables $l_L$ and $l_R$, and their associated momenta:
\begin{equation}   \label{eq:triple_scaled_Htot_lllr}     
\begin{aligned}
      H_{L,R}&=2J\lambda\left(e^{-l_{L,R}}\Delta+\frac{k_{L,R}^2}{8}+2 e^{-2l_{L,R}}\right)
\end{aligned}
\end{equation}
The prescription is that the sum of two solutions of the above Hamiltonian computes the total lengths we are looking for. The bulk shockwave solutions are accurate only in a late-time limit, thus, here we aim to match them with a boundary computation in their range of validity, that is at late times. In particular, we will consider the late-time approximation in which we can neglect the term $\propto e^{-2l_{L,R}}$ in the Hamiltonian. In \cref{app:details_eom}, we solve the equations of motion of such Hamiltonian in this limit, searching for solutions that, when summed, build a total length with initial value $x_0$ at a certain time $t_0$, and having the same boundary energy as the matterless case, consistently with the low-energy approximation we are considering. We find that, via summing solutions obtained with this prescription, we can build, as expected, two different objects depending on the evolution operator chosen: the $\mathcal{O}$TFD complexity and the operator complexity.\\ 

Using the sum of solutions prescription corresponding to the evolution with $H_R-H_L$, for slightly more general Hamiltonians $H_{L,R}=2J\lambda\left(e^{-l_{L,R}}C+\frac{k_{L,R}^2}{8}+2 e^{-2l_{L,R}}\right)$, in \cref{app:details_eom} we obtain the operator complexity as \eqref{eq:tot_lgt_minus_app}. For the case at hand of \eqref{eq:triple_scaled_Htot_lllr}, we choose $C=\Delta$  and $x_0=t_0=0$ in \eqref{eq:tot_lgt_minus_app}\footnote{a similar expression for the triple-scaled length, modulo renormalization-choice dependent factors of $2$, was obtained in \cite{Xu:2024gfm}, directly from the left/right Hamiltonians of \cite{Lin:2022rbf}.}:
\begin{equation}\label{eq:op_compl_O0}
    l(t)\approx 2\log\left(1+\frac{\Delta}{2}\sinh{(J \lambda t)^2}\right)\approx 2\log\left(1+\frac{\Delta}{8}e^{2J\lambda t}\right),
\end{equation}
where the first approximation relies on $\Delta\ll1$, and in the second we used $\sinh{(J \lambda t)}\approx \exp{(J \lambda t)}/2$, valid at late times.\\

At this point, we want to clarify how we should interpret the consistency of the result \eqref{eq:op_compl_O0}, in view of the assumptions of low-energy and late-time limits used to obtain it.
The late-time limit, in which we could neglect the higher exponential term in \eqref{eq:triple_scaled_Htot_lllr}, requires that we trust \eqref{eq:op_compl_O0} only when we have $l(t)\gg 1$, that is at times $t\gg t_{scr}$. Here we defined the scrambling time $t_{scr}$ as the time at which one observes the onset of the linear behavior in \eqref{eq:op_compl_O0}:
\begin{equation}
    t_{scr}\sim \frac{1}{2J\lambda}\log\biggr(\frac{1}{\Delta}\biggr).
\end{equation}
This quantity becomes large when we consider $\Delta$ parametrically small, as required by the low-energy limit.
Before discussing the interplay of the late-times and low-energy limits, we restate that the picture they paint is totally analogous to the shockwave described in \cref{sec:shock_model_def}. In this gravitational setup, these same limits, underlying the \textit{shockwave approximation}, allowed to build the simplest case for which it was still possible to see some imprints of the energy insertion on the boundary.\\
In order to better understand this double-scaling limit we are performing, let us consider the characteristic regimes of the triple-scaled operator complexity:
\begin{equation}\label{eq:op_compl_regimes}
      l(t)\approx 2\log\left(1+\frac{\Delta}{8}e^{2\lambda J t}\right)\approx 
    \begin{cases}
      \frac{\Delta}{4}e^{2\lambda Jt} & \text{if $t\ll t_{scr}$} \\
     4J\lambda (t-t_{scr}) & \text{if $t\gg t_{scr}$},
    \end{cases}
\end{equation}
The first question we ask ourselves is under what conditions we can trust the first regime of \eqref{eq:op_compl_regimes}, when the late-time approximation we used to obtain it prescribes only the second case $t\gg t_{scr}$ to be rigorous. We know that, when the true exact solution starts to match \eqref{eq:op_compl_regimes} at $t\sim t_{scr}$, the operator complexity is still approximately zero: this means that we need to be agnostic on the precise functional form of the early time profile, but we can state that it will be approximately equal to zero until the scrambling time.
This corresponds to considering the low-energy approximation, where we stop at $\mathcal{O}(\Delta^0)$ order. Notice however, that if we choose $\Delta$ parametrically small we also need to allow the system to evolve for very large times in order to see the second linear regime.
Essentially, we are considering a double-scaled limit where we send $\Delta\to0$, $t_{scr}\to\infty$ keeping, by definition of the scrambling time, $\Delta e^{t_{scr}}\sim O(\Delta^0)$ constant. In practice, this prescription translates into keeping $\Delta$ terms only when they can change the late time regime of the observable of interest. So, for example, $\Delta e^t$ is kept in \eqref{eq:op_compl_regimes} because for very late times $t\gg t_{scr}$ becomes $\propto (\mathrm{const.})\times e^{t}\gg 1$ and gives the leading behavior. We will refer to this double-scaling procedure as the `\textit{shockwave approximation}', to recall its similarity to the one considered in \cref{sec:shockwave_geodesics}. Notice that this is the crudest limit still catching the simplest feature of the light operator insertion, that is still discerning the regimes \eqref{eq:op_compl_regimes}. After these considerations then, we can write \eqref{eq:op_compl_regimes} as:
\begin{equation}\label{eq:op_compl_regimes_approx}
      l(t)\sim2\log\left(1+\frac{\Delta}{8}e^{2\lambda J t}\right)\sim 
    \begin{cases}
       0  & \text{if $t\ll t_{scr}$} \\
      4J\lambda (t-t_{scr}) & \text{if $t\gg t_{scr}$},
    \end{cases}
\end{equation}
and in particular, under the \textit{shockwave approximation}, the calculation in both these regimes will still be reliable. Notice that the details of the operator insertion only feature in the definition of the scrambling time, and this is the reason we claimed it to be the crudest light operator approximation.\\
 In \cref{sec:hol_dict_matt}, we will proceed to identify these regimes with the corresponding ones for the geodesic length \eqref{eq:jt_op_compl_O0} in JT gravity with the shockwave insertion.\\

We conclude this section by noticing that, in the limit discussed above, the result we obtained for the $\mathcal{O}$TFD complexity \eqref{eq:opstate_compl_eom} captures only the linear regime\footnote{in this case we only find the linear regime because its onset for the $\mathcal{O}$TFD complexity $l_+(t)$ happens at times much earlier than scrambling. On the other hand, for $l(t)$, the exact solution is expected to stay small until a very large scrambling time, after which it will coincide with the one we obtained \eqref{eq:op_compl_O0}. So the approximation catching the crudest detail about the light-operator insertion is the one that just discerns the two  pre and post-scrambling characteristic regimes.}. This happens because in the small $\Delta$ limit the leading behavior is given at all times by the $\cosh$:
\begin{equation}\label{eq:otfd_triple_scaled}
    l_+(t)\approx_{\Delta\ll 1}2\log\left(\left(1+\frac{1}{2}\left(\frac{\Delta}{4}\right)^2\right)\cosh2\lambda J t+\frac{\Delta}{4}\right)\approx_{t\gg t_{scr}} 4J\lambda t+O(\Delta^1)
\end{equation}

So, in the low-energy approximation where we neglect the energy inserted by the operator on the boundary, the $\mathcal{O}$TFD state complexity does not encode the operator details, and thus cannot be used to obtain a new entry for matter in the holographic dictionary. Notice, as a consistency check, that we obtained the same late-time result \eqref{eq:otfd_triple_scaled} in \cref{app:details_eom}, where we computed the $\mathcal{O}$TFD complexity by using its total length property.\\

\subsubsection{Switchback effect for a single operator insertion}
In the previous section, we computed the triple-scaled operator complexity and $\mathcal{O}$TFD complexity. By definition, they are obtained by solving the Lanczos algorithm with a seed $\mathcal{O}\ket{TFD}$ and with evolutions respectively $H_R-H_L$ and $H_R+H_L$. Before discussing their holographically dual properties, we wish to verify that these complexities, computed after a single operator insertion, showcase the switchback effect, as introduced in \cref{sec:switch_grav_intro}.\\
In the notation of \cite{Stanford:2014jda}, reviewed in \cref{sec:switch_grav_intro}, the evolution prescriptions for the operator and $\mathcal{O}$TFD complexity, are respectively associated to the following timefolds:
\begin{equation}\label{eq:single_switch_tfolds}
    e^{i(H_L-H_R)t}\mathcal{O}\ket{TFD}\leftrightarrow \raisebox{-45pt}{\includegraphics[scale=0.25]{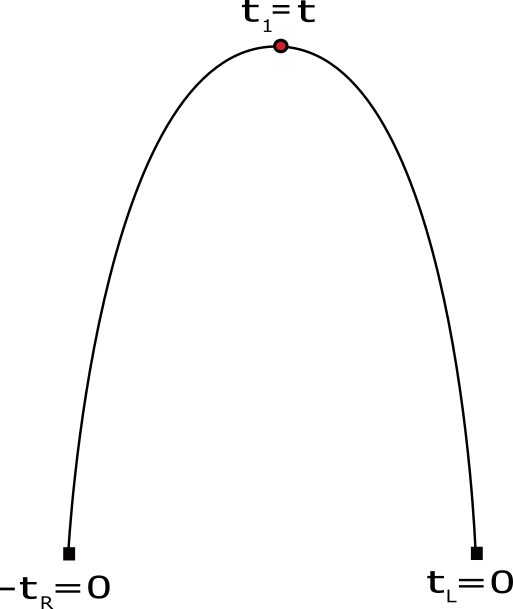}},\quad e^{i(H_L+H_R)t}\mathcal{O}\ket{TFD}\leftrightarrow\quad\raisebox{-50pt}{\includegraphics[scale=0.18]{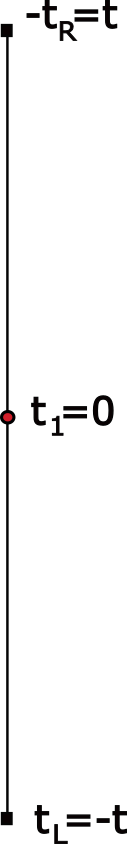}}
\end{equation}
We notice that the prescription to obtain the operator complexity is the insertion of a single switchback perturbation, on a timefold of length $2t$. On the other hand, a through-going insertion, gives the $\mathcal{O}$TFD complexity. The switchback effect for this single insertion prescribes that operator complexity has a delay of $t_{scr}$ after the perturbation, which is precisely the behavior observed in the second regime of \eqref{eq:op_compl_regimes}\footnote{Notice that, as expected, \eqref{eq:op_compl_regimes} coincides with \eqref{eq:precursor_compl}, obtained in \cref{sec:switch_grav_intro} for the gate complexity of the precursor, with a proportionality factor equal to the timescale for triple-scaled lengths $2J\lambda$.}. Consistently, for the $\mathcal{O}$TFD complexity \eqref{eq:otfd_triple_scaled} there is no delay in the linear behavior upon the matter insertion. 

\subsection{Holographic matching}\label{sec:hol_dict_matt}
The holographic parameter identifications we get from the matterless DSSYK/JT gravity duality is \eqref{eq:hol_dict_nomatt}:
\begin{equation}
      2\lambda J =\frac{r_s}{l_{AdS}^2}\;,\quad\quad l_f=l_{AdS} 
\end{equation}
In this section, we describe the holographic duality between DSSYK with light operator insertions and JT gravity with a shockwave. We confirm that the holographic dictionary for the matterless DSSYK-JT gravity remains valid  and clarify the parameter identifications needed for the matter insertions.\\
As explained in \cref{sec:3scale_op_compl} and \cref{sec:shockwave_geodesics}, we have identified a particular ‘low-energy and late-time approximation', referred to as the \textit{shockwave approximation}, in which both the bulk back-reaction description and our simplifying assumptions used to find the operator complexity hold. In this regime we will explicitly establish the duality between the boundary DSSYK with the operator insertion and its gravitational bulk description via the shockwave in JT gravity.\\
We have studied, on both sides of the  duality, two observables corresponding to notions of complexity, namely operator complexity and the $\mathcal{O}$TFD state complexity. On this last one however, as discussed in \cref{sec:3scale_op_compl} and \cref{sec:shockwave_geodesics}, the imprint of the matter insertion is too weak to be detected in the approximation we are considering. Consistently the matterless bulk-boundary correspondence guarantees the matching of the observables \eqref{eq:shock_lgt_+} and \eqref{eq:otfd_triple_scaled}. This is not the case for the triple-scaled total chord number (in DSSYK) and geodesic length (in JT gravity) operators associated to operator complexity. Indeed, we can match the expectation values of these operators \eqref{eq:shock_lgt_-} and \eqref{eq:op_compl_O0}\footnote{this is just a compact way of writing the matching. Remember that rigorously in \textit{shockwave approximation} considered, we are identifying the characteristic regimes of \eqref{eq:op_compl_O0} and \eqref{eq:jt_op_compl_O0}, via the matching of the respective scrambling times.}:\\
\begin{equation}\label{eq:duality_op_compl}
l_{\mathrm{grav}}(t)=2l_{AdS}\log\biggr(1+\frac{E}{8M}e^{\frac{r_s}{l_{AdS}^2}t}\biggr)\quad\leftrightarrow\quad l_{\mathrm{DSSYK}}(t)=2l_f\log\biggr(1+\frac{\Delta}{8}e^{2J\lambda t}\biggr),
\end{equation}
where we reinstated the reference lengths $l_{AdS}$ and $l_f$. Notice that, in the framework of the JT-DSSYK holographic duality, the separate matching between the regimes of \eqref{eq:op_compl_regimes} and \eqref{eq:jt_op_compl_O0} is guaranteed by the matterless holographic dictionary \eqref{eq:hol_dict_nomatt}. This is expected, because, as explained in \cref{sec:3scale_op_compl}, in the \textit{shockwave approximation}, the details regarding the matter insertion are imprinted only in the transition time between regimes, while their functional forms are universal. Then, we can obtain the new entry describing matter in the holographic dictionary, if we identify the scrambling times, at which we have the onset of the linear behavior, in the two theories:
\begin{equation}\label{eq:match_tscr}
   t_{scr}^{\mathrm{JT}}= \frac{l_{AdS}^2}{r_s}\log\frac{M}{E}\quad\longleftrightarrow\quad\frac{1}{2\lambda J}\log\left(\frac{1}{\Delta}\right)=t_{scr}^{\mathrm{DSSYK}}
\end{equation}
This condition suggests that in order to describe this kind of light operator insertions in DSSYK using a shockwave picture, one needs to complement the matterless holographic bulk-to-boundary map with the following entry \cite{Ambrosini:2024sre}:
\begin{eqnarray}\label{eq:new_entry_matt}
    \Delta=\frac{E}{M}
\end{eqnarray}
In other words, we identify $\Delta$, the ratio of the dimensions of operator and Hamiltonian, with the ratio between the energy of the shockwave and the mass of the black hole.\\

With the updated holographic dictionary, we have identified the geodesic length operator anchored at opposite times on the boundary, with the operator K-complexity one defined on DSSYK. At this point, we can write the Hamiltonian of JT gravity \eqref{eq:jt_ham}, whose value on the boundary is unaffected because of the low-energy approximation, using the canonical variable \eqref{eq:shock_lgt_-} and its conjugate momentum. Then, by canonical quantization, we can obtain an Hilbert space for the theory made of eigenstates $\ket{l}$ of the geodesic length operator anchored at $t_L=-t_R$. In this case, the duality \eqref{eq:duality_op_compl}, guaranteed by \eqref{eq:hol_dict_nomatt} and \eqref{eq:new_entry_matt}, is uplifted to an Hilbert space isomorphism between the eigenstates of the length operators whose expectation values are in \eqref{eq:duality_op_compl}:
\begin{equation}
    \lambda\ket{\psi_n^-}\leftrightarrow \ket{l},
\end{equation}
where $\ket{\psi_n^-}$ are the binomial states \eqref{eq:binomial_ansatz}, obtained as the Krylov basis for the evolution $H_R-H_L$ \cite{Ambrosini:2024sre}.\\

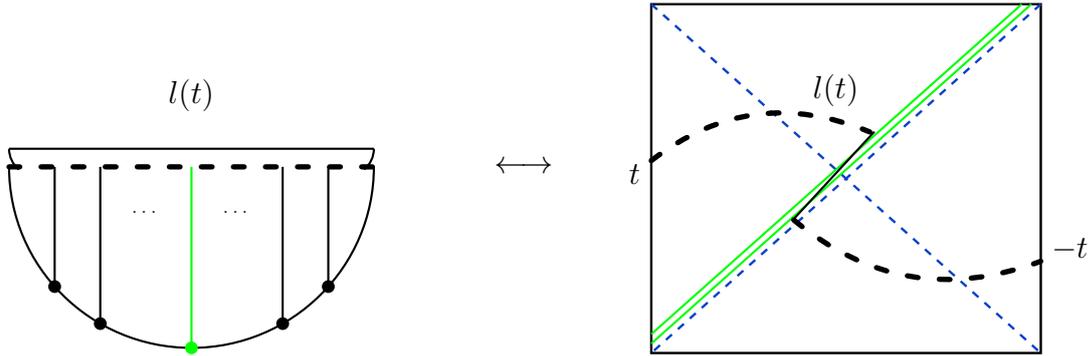
\begin{figure}
\begin{minipage}[b]{0.49\textwidth}
\centering
    \begin{tikzpicture}[scale=1.2]
 
    \draw[thick] (-2,0) arc[start angle=180, end angle=360, radius=2];
    
    \foreach \x in {-1.5,-1.0,1.0, 1.5} {
        \fill[black] (\x,-{sqrt(4-(\x)^2)}) circle (2pt); 
        \draw[thick] (\x,-{sqrt(4-(\x)^2)}) -- (\x,0); 
    }
    \fill[green] (0,-2) circle (2pt); 
    \draw[thick, green] (0,-2) -- (0,0); 
    
    \node at (-0.5,-0.5) {\tiny{$\cdots$}};
    \node at (0.5,-0.5) {\tiny{$\cdots$}};
    \draw [line width=2pt, black, line cap=round, dash pattern=on 4pt off 4\pgflinewidth] (-2.0,0) -- (2.0,0);
    \draw[thick] (-2.0,0.2) -- (2.0,0.2);
    \draw[thick] (-2.0,0.2) arc[start angle=180, end angle=230, radius=0.3];
    \draw[thick] (2.0,0.2) arc[start angle=0, end angle=-50, radius=0.3];
    \node[above] at (0,0.5) {$l(t)$};
    \node[right] at (3.2,0){$\longleftrightarrow$};
\end{tikzpicture}
\end{minipage}
\begin{minipage}[b]{0.49\textwidth}
\centering
\begin{tikzpicture}[scale=0.5]
\draw [ color={rgb,255:red,26; green,95; blue,180} , line width=0.9pt ] (6.25,15.25) rectangle (6.25,15.25);
\draw [ line width=0.9pt ] (6.25,15.5) rectangle (16.5,6.25);
\draw [ color={rgb,255:red,11; green,64; blue,201}, line width=0.9pt, dashed] (6.25,15.5) -- (16.5,6.25);
\draw [ color={rgb,255:red,11; green,64; blue,201}, line width=0.9pt, dashed] (16.5,15.5) -- (6.25,6.25);
\draw [ thick,green] (6.25,6.75) -- (16,15.5);
\draw [ thick,green] (6.25,6.5) -- (16.25,15.5);
    
\node (f) [above left] at (12,12.5){$l(t)$};
\node (l1) [left] at (6.25,11){$t$};
  \node (l2) [right] at (12,12){};
\draw[bend left,line width=2pt,  black, line cap=round, dash pattern=on 4pt off 4\pgflinewidth]  (l1) to node [auto] {} (l2);
\node (l3) [left] at (10,10){};
  \node (l4) [right] at (16.5,9){$-t$};
  \draw[bend right,line width=2pt, black, line cap=round, dash pattern=on 4pt off 4\pgflinewidth]  (l3) to node [auto] {} (l4);
  \draw [thick,black] (12.12,12.12) -- (9.95,9.75);
\end{tikzpicture}
\end{minipage}
\caption{Intuitive representation of the DSSYK-JT gravity duality upon the operator insertion. The re-normalized length (left), obtained by considering the triple-scaled limit of operator K-complexity, is dual to the re-normalized geodesic length (right) anchored at times $t_L=-t_R=t$ on the regularized boundary of AdS$_2$, with some low energy quanta inserted at $t_w=0$. The shockwave backreaction (green, on the right) of these low-energy quanta in the bulk is then associated to the operator chord (green, on the left), in the new updated holographic dictionary \eqref{eq:hol_dict_nomatt}, that contains also the details of the matter insertion.}
\label{fig:duality_shock_matter}
\end{figure}

\section{The switchback effect in K-complexity}\label{sec:switchback_compl}
In this section, we want to describe a particular model of matter insertions that allows us to analyze the switchback effect, purely from the point of view of Krylov complexity. As we explained in \cref{sec:switch_grav_intro}, this is a particular property characteristic of ERB lengths: under a switchback perturbation the growth of the length temporarily slows down and, after a scrambling time, resumes the linear growth behavior. Any complexity notion which is expected to have a bulk dual needs to satisfy this property upon perturbations, as argued in \cite{Belin:2021bga}, and in particular we will check that it is satisfied by operator Krylov complexity: upon the switchback insertion of a second operator, the complexity stays constant for a scrambling time and then resumes the linear growth behavior \cite{Belin:2021bga}. \\

 We consider perturbations given by the insertions of two-sided operators localized on a single time-slice, on top of the single operator background evolution we described in \cref{sec:op_compl_recap}. We now explain this choice of operator insertion: the idea is that it allows us to emulate the effect of the time dependent Hamiltonian, involved in the perturbation caused by the operator insertion,  by modifying the Lanczos procedure at the $n_s$ stage. Indeed, in the semiclassical limit, we can describe perturbations localized in time using insertions localized in chord number, provided we match the parameters of the perturbations using the equations of motion. We provide, in \cref{sec:lanczos_pert}, the definition of a Lanczos evolution operator that includes such perturbations in the evolution, and proceed to analytically solve it in \cref{sec:vertical_multinom_lanczos}, using the triple-scaled and low-energy limit, already introduced in \cref{sec:shock_model_def}, to make contact with the shockwave description in the bulk. In this `\textit{shockwave approximation}', we will show, in \cref{sec:perturbed_3scale_length}, that the complexity associated to the described Lanczos algorithm, which we refer to as the `\textit{perturbed} complexity', showcases the expected switchback effect. \\
 Consistently, this \textit{perturbed} complexity is mapped by the holographic dictionary \eqref{eq:new_entry_matt} to a geodesic length in the bulk shockwave configuration corresponding to the operator insertions in Lanczos algorithm, showing that the duality, outlined in \cref{sec:hol_dict_matt}, between DSSYK and JT gravity, persists upon perturbations.\\
 
\subsection{Perturbations localized in time via localization in chord space}\label{sec:time_to_kbase_pert}
In order to study the properties of the perturbed operator complexity, we need to understand the Lanczos algorithm for the background seed $\mathcal{O}\ket{TFD}$, with an evolution operator that receives a perturbation localized at a certain $t_s$. This perturbation makes the Lanczos evolution operator time-dependent, so that the algorithm becomes considerably harder to solve \cite{Takahashi_2025}. The proposal we describe in the next section is instead to insert in the Lanczos algorithm a perturbation at a certain chord number $n_s$. The idea, as we will momentarily explain, is that we can tune the insertion parameter of such a perturbation so that, in the semiclassical limit, it indeed describes the perturbation at the time $t_s$ we were searching for (see \cref{fig:KrylovOverview} for an intuitive picture).\\

Let us consider the Lanczos algorithm for the seed $\mathcal{O}\ket{TFD}$, with the evolution operator $H_R-H_L$, acting in the chord Hilbert space, but modified by a perturbation at a certain chord number $n_s$, via the insertion of a matter chord with dimension $\Delta_m$. This means that we are adding to the chord diagrams a second flavor of matter-chords, whose intersections with Hamiltonian chords are weighted by the coefficient
\begin{equation}
    \Tilde{q}'=e^{-\lambda \Delta_m}.
\end{equation}
First, let us notice that, as reviewed in \cref{sec:op_compl_recap}, we have identified the total chord number with the position on the Krylov chain when solving the Lanczos algorithm for the background operator before the insertion of the matter perturbation \cite{Ambrosini:2024sre}. This means that the expression for the Lanczos evolution operator localized at chord number $n_s$, can be written by inserting the perturbation matter at the $n_s$-th step of the algorithm. We argue now that such an insertion of matter in the $n_s$-th step of the algorithm, can equivalently describe the perturbation localized in time we want to understand.\\

We consider the generic case of a K-basis $\{\ket{n}\}$ built by solving the Lanczos algorithm with evolution operator $H$ and seed $\ket{0}$: K-complexity is given by the expectation value of the position on the Krylov chain of the evolving wave-packet described by $\phi_n(t)=\bra{n}e^{-iHt}\ket{0}$, that is $C_K(t)=\sum_n n |\phi_n(t)|^2$.
In \cite{Ambrosini:2024sre}, specializing to the evolution $H_R\pm H_L$ and seed $\mathcal{O}\ket{TFD}$, we showed that, in the $\lambda\to 0$ limit, the Lanczos recursion became a wave equation, whose solution is $\phi_n(t)=\delta(n-n(t))$, where $n(t)$ is the of the center of the wave-packet. So, $\lambda$ controls the width of the wave-packet, and, in particular, when we send $\lambda\to 0$, it becomes sharply localized around its peak. This means that, in the triple-scaled and semiclassical limits, in which $\lambda\to 0$, when we solve the equations of motion to find the Krylov complexity, we obtain the expected position $n(t)$ of the localized wave-packet\footnote{this is also the expected number of open chords at time $t$, because of the identification between position on the Krylov basis and total chord number. We notice also that, because of the double-scaling limit in SYK, all finite Krylov chain effects have been driven to infinity. In particular, in the limits we consider in this paper the function $n(t)$ is bijective, so that the condition $n_s=n(t_s)$ is well defined.}. In this context, we can equivalently say that the perturbation, instead of being inserted when the wave-packet is in the position $n_s$, happens at the particular time $t_s$ when $n_s=n(t_s)$ (see \cref{fig:KrylovOverview}).\\

To summarize, we have, in principle, two independent perturbations, one described by a parameter $n_s$, the step at which it enters in the Lanczos algorithm, and the other characterized by its time of insertion $t_s$. We can ask that these perturbations coincide by imposing the relation $n_s=n(t_s)$, between parameters $n_s$ and $t_s$. This condition ensures that the perturbation inserted in the $n_s$-th step of the Lanczos algorithm, or equivalently when we have $n_s$ total open chords, is describing a perturbation at time $t_s$. The prescription to pass from the Lanczos algorithm perturbation to the perturbation in time is by inverting the condition $n_s=n(t_s)$ for $t_s$. Remember that in principle, this identification between perturbations hinges on the fact that the entire wave-packet is localized around $n_s$ when we insert the perturbations. In this paper this is not a problematic request, as we will only be interested in the $\lambda\to 0$ limit, where the \textit{perturbed} operator complexity will assume a geometric meaning (\cref{sec:vertical_multinom_lanczos}), analogously to what happened for the single operator case of \cref{sec:op_compl_recap}.\\
In the next section, we will define the modified Lanczos algorithm described above, and recast it as a time dependent evolution, after explicitly solving the equations of motion.

\subsection{Perturbing the Lanczos algorithm}\label{sec:lanczos_pert}
Now, we describe the modified Lanczos evolution, introduced in the previous section, where we insert some matter perturbation at the $n_s$-th step of the algorithm. We specialize to the specific insertion of matter that is implemented at the level of chord diagrams by a `\textit{time-slice perturbation}'. By this nomenclature, we mean a two-sided insertion upon a state with $n_s$ open chords, which is then associated, from what we discussed in \cref{sec:time_to_kbase_pert}, to a two-sided perturbation on the $t_s$ constant time-slice. There are two heuristic reasons why we are considering this kind of matter insertions: they are diagrammatically tractable in the triple-scaling limit, because we can restrict our attention to a sub-class of diagrams, and they keep our modified Lanczos algorithm manifestly left/right symmetric. The first requirement arises because each matter contribution is characterized by distinct transfer matrix coefficients that enter nonlinearly into the Lanczos coefficients, making the analytical solution considerably harder when several such configurations are present. On the other hand, we need the second left/right symmetric property, because the idea in \cref{sec:perturbed_3scale_length} will be to obtain the triple-scaled \textit{perturbed} complexity by mimicking what we performed in \cref{sec:3scale_op_compl}, where we required the left/right Hamiltonians to be equal.\\

The \textit{time-slice perturbation} we consider is created by the insertion of terms of the form $\Tilde{a}_L \Tilde{a}_R^\dag$, where $\Tilde{a}_{L,R}$ and $\Tilde{a}^\dag_{L,R}$ are annihilation and creation operators associated to an operator insertion $\Tilde{\mathcal{O}}$ of weight $\Delta_m$, inserted on the left/right boundaries, defined with respect to the background matter insertion $\mathcal{O}$ of weight $\Delta$. We consider the Lanczos algorithm for the seed $\mathcal{O}\ket{TFD}$, with the following modified evolution operator:
\begin{equation}\label{eq:pert_Kev}
\begin{aligned}
     T=(a^\dag_R-a^\dag_L)(1-\delta_{n\;n_s})+\frac{(\Tilde{a}_L \Tilde{a}_R^\dag+\Tilde{a}_R \Tilde{a}_L^\dag)}{\sqrt{2}}(a^\dag_R-a^\dag_L)\delta_{n\;n_s}+(a_R-a_L)
\end{aligned}
\end{equation}
We argue that \eqref{eq:pert_Kev} is the perturbed evolution operator associated to the \textit{time-slice perturbation} insertion that we wish to describe. At the $n_s$-th step, when we allow the creation of a new Hamiltonian chord, we also insert the \textit{time-slice perturbation}.
Away from $n_s$, the evolution operator $T$ is the same we used to build the operator K-complexity of $\mathcal{O}$, that is $H_R-H_L$, defined as \eqref{eq:HlHr_def}.\\
A posteriori, at the end of this section, we will recognize that the factor of $\sqrt{2}$ in the denominator normalizes to one the multiplicity of the leading diagram configuration in the triple-scaled limit. We will also notice that, at the level of this leading diagram, the insertion of the two-sided linear combination of $\Tilde{a}_{L,R},\;\Tilde{a}_{L,R}^\dag$ we chose, is equivalent to acting with $\Tilde{\mathcal{O}_L}\Tilde{\mathcal{O}}_R$, modulo a different diagram multiplicity factor.\\Now, we proceed to discuss the features of the Lanczos algorithm with this evolution operator. In particular we will explicitly solve for the Lanczos coefficients in the triple-scaled limit.\\

 Before the perturbation is inserted, for steps $n<n_s$, \eqref{eq:pert_Kev} reduces to the Lanczos algorithm for the background operator studied in \cite{Ambrosini:2024sre}. As summarized in \cref{sec:op_compl_recap}, this results in the Lanczos coefficients $b_n$ and the Krylov basis of binomial states,
\begin{eqnarray}
    b_n=2\sqrt{\frac{1-q^{n/2}}{1-q}(1-\Tilde{q}q^{n/2})},\;\,\ket{\psi_n}=\frac{1}{\prod b_i}\sum_k^n(-1)^k\binom{n}{k}\ket{k,\,n-k}\;\mathrm{for}\;n<n_s\,
\end{eqnarray}
Later in this section, we will make use of the un-normalized version of these binomial states (\cref{eq:binomial_ansatz}), that here we call $\ket{\mathrm{binom}(n_s)}$\footnote{here we use the different nomenclature $\ket{\mathrm{binom}(n_s)}$, instead of $\ket{\chi_{n_s}^-}$, in order to avoid confusion with the Krylov basis $\ket{\chi_n}$, that we are defining for the perturbed evolution when $n>n_s$.}:
\begin{equation}
    (a^\dag_R-a^\dag_L)\ket{\chi_{n_s-1}}=\sum_{k=0}^{n_s}(-1)^k \binom{n}{k}\ket{k,n_s-k}\equiv \ket{\mathrm{binom}(n_s)}
\end{equation}
The novelty of \eqref{eq:pert_Kev} with respect to the un-perturbed evolution enters when we modify the chord creation operator at the $n_s$-th step of the algorithm. The Lanczos recursion for the modified $n_s$-th step reads
\begin{equation}\label{eq:ns_recursion}
    b_{n_s}\ket{\psi_{n_s}}=(a_R-a_L)\ket{\psi_{n_s-1}}+\frac{(\Tilde{a}_L \Tilde{a}_R^\dag+\Tilde{a}_R \Tilde{a}_L^\dag)}{\sqrt{2}}(a^\dag_R-a^\dag_L)\ket{\psi_{n_s-1}}-b_{n_s-1}\ket{\psi_{n_s-2}}\,.
\end{equation}
The Lanczos coefficients $b_n$ for $n<n_s$ are built by solving the unperturbed algorithm. In this case, as summarized in \cref{sec:op_compl_recap}, the binomial states being the Krylov basis for the background operator evolution boils down to the fact that in the $\lambda\to 0$ limit the cancellation identity \eqref{eq:binom_recursion_cancellation_cond} was satisfied
\begin{equation}
    (a_R-a_L)\ket{\psi_{n_s-1}}-b_{n_s-1}\ket{\psi_{n_s-2}}=0.
\end{equation}
Notice that $\lambda\to0$ is the limit we already restricted to in \cref{sec:time_to_kbase_pert} to ensure the localization of wave-packets on the Krylov chain, so this is not a further restriction. By substitution of the above cancellation in the $n_s$ Lanczos recursion \eqref{eq:ns_recursion}, then we obtain:
\begin{equation}
\begin{aligned}
b_{n_s}\ket{\psi_{n_s}}&=\frac{(\Tilde{a}_L \Tilde{a}_R^\dag+\Tilde{a}_R \Tilde{a}_L^\dag)}{\sqrt{2}}(a^\dag_R-a^\dag_L)\ket{\psi_{n_s-1}} \\&\implies \ket{\chi_{n_s}}=\frac{(\Tilde{a}_L \Tilde{a}_R^\dag+\Tilde{a}_R \Tilde{a}_L^\dag)}{\sqrt{2}}(a^\dag_R-a^\dag_L)\ket{\chi_{n_s-1}},
\end{aligned}
\end{equation}
where in the second line we recast the identity using the un-normalized Krylov basis $\{\ket{\chi_n}\}$\footnote{we remind that in order to obtain the usual normalized element of the Krylov basis $\ket{\psi_n}$ from $\ket{\chi_n}$ one divides by the first $n$ Lanczos coefficients: $\{\ket{\psi_n}\}$: $\ket{\chi_n}/\prod_{i=0}^n b_n=\ket{\psi_n}$.}. In the above equation, because of \eqref{eq:adagger_gives_binomial}, the operator $(\Tilde{a}_L \Tilde{a}_R^\dag+\Tilde{a}_R \Tilde{a}_L^\dag)$ inserts the perturbation matter chords on $\ket{\mathrm{binom}(n_s)}$ the binomial state \eqref{eq:binomial_ansatz} with $n_s$ open Hamiltonian chords.
  In particular, the norm of $\ket{\chi_{n_s}}$ will be given by:
  \begin{equation}\label{eq:norm_wick_contraction}
      \begin{aligned}
\braket{\chi_{n_s}|\chi_{n_s}}&=\frac{1}{2}\bra{\mathrm{binom}(n_s)}\Tilde{a}_R \Tilde{a}_L^\dag\Tilde{a}_L \Tilde{a}_R^\dag+\Tilde{a}_L \Tilde{a}_R^\dag\Tilde{a}_L \Tilde{a}_R^\dag+(L\leftrightarrow R)\ket{\mathrm{binom}(n_s)}=\\&=\frac{1}{2}\bra{\mathrm{binom}(n_s)}\wick{\c2{\Tilde{a}_R} \c1{\Tilde{a}_L^\dag}\c1{\Tilde{a}_L} \c2{\Tilde{a}_R^\dag}+\c1{\Tilde{a}_R} \c1{\Tilde{a}_L^\dag}\c1{\Tilde{a}_L} \c1{\Tilde{a}_R^\dag}+\dots+(L\leftrightarrow R)}\ket{\mathrm{binom}(n_s)}
      \end{aligned}
  \end{equation}
  We compute the norm $\braket{\chi_{n_s}|\chi_{n_s}}$ by summing the chord diagrams obtained by all possible contractions of the operators $\Tilde{a}^\dag$'s with $\Tilde{a}$ between the binomial states, with the rules prescribed by the chord algebra of \cite{Lin:2023trc}. We have explicitly indicated the contractions for the first term of \eqref{eq:norm_wick_contraction} in the second line of the equation. By considering all possible configurations, with the correct multiplicity, we obtain the following chord diagrammatic representation for \eqref{eq:norm_wick_contraction}:
  \begin{equation}\label{eq:norm_matter_diagrams}
      \braket{\chi_{n_s}|\chi_{n_s}}=\frac{1}{2}\biggr(2\cdot\raisebox{-10pt}{\includegraphics[scale=0.15]{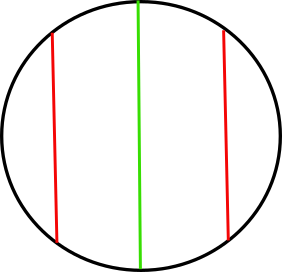} }+4\cdot\raisebox{-10pt}{\includegraphics[scale=0.16]{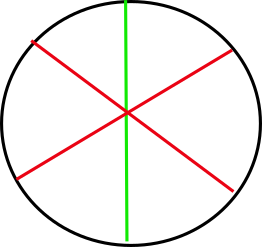} }+2\cdot\raisebox{-10pt}{\includegraphics[scale=0.16]{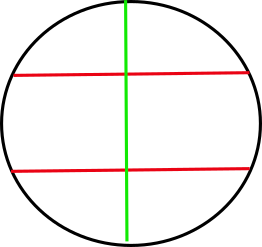} }\biggr),
  \end{equation}
where, in the chord diagrams above, the green chord is the background operator $\mathcal{O}$, and we left implicit the sum over configurations with $n_s$ total Hamiltonian chords before the red chord insertions, which are the ones associated to the perturbation operator $\Tilde{\mathcal{O}}$. We notice that the `time-slice chords' in the last two diagrams are crossing all the $n_s$ open Hamiltonian chords, so they will have an extra penalizing factor of $e^{-\lambda \Delta_m n_s}$, with respect to the first diagram. In particular, this means that, in the triple scaling limit \eqref{eq:triple_scaling_nomatt}, defined by sending $\lambda n_s=l_s\to\infty$, the last two diagrams are exponentially suppressed with the constant $e^{- \Delta_m l_s}$\footnote{our prescription, as in \cref{sec:3scale_op_compl}, is to take the triple-scaling limit, truncate the Hamiltonian to linear order in $\lambda$ and subsequently search for low-energy solutions. The suppression is $e^{-\Delta_ml_s}\propto\lambda^{2\Delta_m}$, and the truncation to $\mathcal{O}(\lambda)$ eliminates the length-dependent terms (appearing at $\mathcal{O}(\lambda)$) coming from the neglected diagrams. This limit procedure however, keeps in principle a sub-leading correction to the ground state energy $E_0\propto 1/\lambda$ to be subtracted off, that we will neglect.}. So, in the triple-scaling limit, the first diagram in \eqref{eq:norm_matter_diagrams}, which we refer to as `vertical matter-chords' configuration, will give the leading contribution to the norm of the state $\ket{\chi_{n_s}}$ and hence to the Lanczos coefficients:
  \begin{equation}\label{eq:norm_matter_diagrams_only_vert}
\braket{\chi_{n_s}|\chi_{n_s}}\approx\raisebox{-9pt}{\includegraphics[scale=0.14]{vertical_m.png} }
  \end{equation}
  Notice that, in retrospect, we chose the normalizing $\sqrt{2}$ factor in \eqref{eq:pert_Kev}, in order to normalize the multiplicity coefficient in front of the `vertical matter-chords' configuration, which is the leading contribution in the triple-scaled limit of interest. We also note that, if, at the $n_s$ step, we chose to perturb with one-sided terms $\Tilde{a}_{R}^\dag+\Tilde{a}_{L}^\dag$, instead of the two-sided $\Tilde{a}_L\Tilde{a}_R^\dag+\Tilde{a}_R\Tilde{a}_L^\dag$, then in this triple-scaled limit we would still have two non-suppressed matter configurations. In this case, then, the transfer matrix entries relative to the two separate configurations combine non-linearly to form the Lanczos coefficients, which become more cumbersome to compute. By contrast, our choice of operator perturbation gives rise to an instance of a switchback effect that is analytically tractable.\\
In the above diagram, on the slice with $n_s$ open chords, the `vertical matter-chords' do not intersect any of them. This means that, in the limit where we can consider only this leading configuration, even though we inserted the matter perturbation at the $n_s$-th step of the algorithm, the norm of the $n_s$-th un-normalized Krylov basis element is given by the norm of the $n_s$-th binomial state. In particular, the $n_s$-th Lanczos coefficient keeps the same form we obtained for the unperturbed operator $\mathcal{O}$ evolution:
\begin{equation}\label{eq:lanczos_ns}
    b_{n_s}=2\sqrt{\frac{1-q^{n_s/2}}{1-q}(1-\Tilde{q}q^{n_s/2})}
\end{equation}\\
After the $n_s$-th step, modified with the insertion of matter, the evolution \eqref{eq:pert_Kev} we chose to describe our modified Lanczos problem, consists again of creation or annihilation operators of Hamiltonian chords only. Let us consider the computation of the norm of a Krylov basis element $\ket{\chi_n}$ with $n>n_s$:
\begin{equation}\label{eq:norm_matter_diagrams1}
\braket{\chi_{n}|\chi_{n}}\supset \bra{\mathrm{binom}(n_s)}\Tilde{a}^\dag\Tilde{a}\underbrace{\;a\dots a^\dag}_{2(n-n_s)}\Tilde{a}\Tilde{a}^\dag\ket{\mathrm{binom}(n_s)}\approx\raisebox{-9pt}{\includegraphics[scale=0.14]{vertical_m.png} }
  \end{equation}
  In this equation, we used $\supset$ to denote that the generic form of a term contributing to the computation of $\braket{\chi_{n}|\chi_{n}}$. Here by generic, we mean that we care only about the number of creation/annihilation operators inserted after $n_s$, without giving details (for the present time) on their precise linear combination creating the $\ket{\chi_n}$ state, or specifying if they are inserted on the left or right of the background operator. Given that we acted with the evolution operator for an additional $n-n_s$ times after the matter perturbation, we will have a string of $a$'s $a^\dag$'s
  separating the $\Tilde{a}\Tilde{a}^\dag$ operators.
  So this norm has the same chord diagram representation as \eqref{eq:norm_matter_diagrams}, but this time we also insert $n-n_s$ additional Hamiltonian chords after the red insertions. In particular, also in this case, the `vertical matter-chords' configuration is the leading contribution, because the other diagrams are again suppressed by the same constant $e^{-l_s\Delta_m}$, due to the intersection with the open $n_s$ Hamiltonian chords. So for all $n>n_s$ we can limit ourselves to solving the Lanczos algorithm considering only the contributions of the `vertical matter-chords'.\\ We remind the Reader that the triple-scaling limit is necessary in order to connect with the bulk shockwave computation of the switchback effect, reviewed in \cref{sec:switch_grav_intro}.\\

  To summarize, in this section we have introduced a modified Lanczos evolution operator \eqref{eq:pert_Kev} that adds a certain matter perturbation at the step $n_s$. We have given an explicit solution of the Lanczos algorithm up to the $n_s$-th step after the matter insertion. In the next section, we will discuss the solution of the algorithm also for $n>n_s$, specializing to the leading configuration of `vertical matter-chords' in the triple-scaled limit.\\

\subsection{Solution in the `vertical' matter-chords configuration}\label{sec:vertical_multinom_lanczos}

Now, we want to solve the Lanczos algorithm with seed $\mathcal{O}\ket{TFD}$ and evolution operator \eqref{eq:pert_Kev} also for $n>n_s$. When $n>n_s$, the evolution \eqref{eq:pert_Kev} goes back to being $H_R-H_L$, as for the unperturbed case. However, the solution of the Lanczos algorithm for $n>n_s$ does not trivially follow from the background operator $\mathcal{O}$ case. This is because the insertion of the matter perturbation modifies the expression of $a_{L,R}$, appearing in the Hamiltonians via \eqref{eq:HlHr_def}, in terms of the chord annihilation operators. For $n<n_s$, we have the relations \eqref{aL_long}:
\begin{equation}\label{eq:trans_matr_oldop}
    \begin{aligned}
       &a_L = \alpha_L \,\frac{1-q^{n_L}}{1-q} +\alpha_R \, \tilde{q} \, q^{n_L} \frac{1-q^{n_R}}{1-q}\\& 
       a_R = \alpha_R \,\frac{1-q^{n_R}}{1-q} +\alpha_L \, \tilde{q} \, q^{n_R} \frac{1-q^{n_L}}{1-q}\,.
    \end{aligned}
\end{equation}

As explained in \cref{sec:op_compl_recap}, these expressions can be obtained by re-summing all possible ways to annihilate an Hamiltonian chord, respectively on the left/right, with the right combinatorial coefficient coming from intersections. When $n>n_s$, the vertical matter-chords split the chord diagrams in four sectors, so that we can describe the system using states $|n_L,\,n_R;n_1,n_2\rangle$, defined as having the following diagrammatic representation: 
\begin{center}
\vspace{5mm}
    \begin{tikzpicture}[scale=1.0]
    \node[left] at (-2.0, -0.75) {$|n_L,\,n_R;n_1,n_2\rangle = $};
    
    \draw[thick] (-2,0) arc[start angle=180, end angle=360, radius=2];
    
    \foreach \x in {-1.85,-1.1,-0.8,0.8,1.1, 1.85} {
        \fill[black] (\x,-{sqrt(4-(\x)^2)}) circle (2pt); 
        \draw[thick] (\x,-{sqrt(4-(\x)^2)}) -- (\x,0); 
    }
    \fill[green] (0,-2) circle (2pt); 
    \draw[thick, green] (0,-2) -- (0,0); 
    \fill[red] (1.4,-{sqrt(4-(1.4)^2)}) circle (2pt); 
    \draw[thick, red] (1.4,-{sqrt(4-(1.4)^2)}) -- (1.4,0);
    \fill[red] (-1.4,-{sqrt(4-(1.4)^2)}) circle (2pt); 
    \draw[thick, red] (-1.4,-{sqrt(4-(1.4)^2)}) -- (-1.4,0);
    \node at (-0.35,-0.5) {\tiny{$\cdots$}};
    \node at (0.35,-0.5) {\tiny{$\cdots$}};
    \node at (-1.6,-0.5) {\tiny{$\cdots$}};
    \node at (1.65,-0.5) {\tiny{$\cdots$}};
    
    \draw[thick] (-2.0,0.2) -- (-1.45,0.2);
    \draw[thick] (-1.3,0.2) -- (-0.1,0.2);
    \draw[thick] (-2.0,0.2) arc[start angle=180, end angle=230, radius=0.2];
    \draw[thick] (-1.3,0.2) arc[start angle=180, end angle=230, radius=0.2];
    \draw[thick] (-1.45,0.2) arc[start angle=0, end angle=-50, radius=0.2];
    \draw[thick] (-0.1,0.2) arc[start angle=0, end angle=-50, radius=0.2];
    \node[above] at (-0.7,0.3) {$n_1$};
    \node[above] at (-1.7,0.3) {$n_L$};

    \draw[thick] (0.1,0.2) -- (1.3,0.2);
    \draw[thick] (1.45,0.2) -- (2,0.2);
    \draw[thick] (0.1,0.2) arc[start angle=180, end angle=230, radius=0.2];
    \draw[thick] (1.45,0.2) arc[start angle=180, end angle=230, radius=0.2];
    \draw[thick] (1.3,0.2) arc[start angle=0, end angle=-50, radius=0.2];
    \draw[thick] (2.0,0.2) arc[start angle=0, end angle=-50, radius=0.2];
    \node[above] at (1.7,0.3) {$n_R$};
    \node[above] at (0.7,0.3) {$n_2$};

\end{tikzpicture}
\end{center}
Now we want to understand how the evolution operator $H_R-H_L$ acts on these states. The left/right creation operators $a_{L,R}^{(II)\,\dag}$ act in the following way\footnote{we denoted the operators with the superscript $(II)$, in order to distinguish these expressions for the vertical matter-chord configuration from the unperturbed case in \eqref{eq:trans_matr_oldop}.}:
\begin{equation}\label{eq:adag_def_pert}
a_L^{(II)\,\dag}\ket{n_L,n_R;n_1,n_2}=\ket{n_L+1,n_R;n_1,n_2},\;a_R^{(II)\,\dag}\ket{n_L,n_R;n_1,n_2}=\ket{n_L,n_R+1;n_1,n_2},
\end{equation} 
However, notice that now an insertion of $a_{L,R}$ can annihilate chords living in any of the sectors. If we define the annihilation operators
\begin{equation}\label{eq:alpha_def_pert}
\begin{aligned}
       &\alpha_L\ket{n_L,n_R;n_1,n_2}=\ket{n_L-1,n_R;n_1,n_2},\quad \;\alpha_R\ket{n_L,n_R;n_1,n_2}=\ket{n_L,n_R-1;n_1,n_2},\\& \alpha_1\ket{n_L,n_R;n_1,n_2}=\ket{n_L,n_R;n_1-1,n_2},\quad\; \alpha_2\ket{n_L,n_R;n_1,n_2}=\ket{n_L,n_R;n_1,n_2-1},
\end{aligned}
\end{equation}
then by summing all possible ways of annihilating the Hamiltonian chords, analogously to \eqref{aL_long}, we obtain the following expressions: 
\begin{equation}\label{eq:vmatt_transm}
\begin{aligned}
    \\&a_L^{(II)}=\frac{1-q^{n_L}}{1-q}\alpha_L+\Tilde{q}'q^{n_L}\frac{1-q^{n_1}}{1-q}\alpha_1+\Tilde{q}'\Tilde{q}q^{n_L+n_1}\frac{1-q^{n_2}}{1-q}\alpha_2+\Tilde{q}'^2\Tilde{q}q^{n_1+n_2+n_L}\frac{1-q^{n_R}}{1-q}\alpha_R
    \\&a_R^{(II)}=\frac{1-q^{n_R}}{1-q}\alpha_R+\Tilde{q}'q^{n_R}\frac{1-q^{n_2}}{1-q}\alpha_2+\Tilde{q}'\Tilde{q}q^{n_R+n_1}\frac{1-q^{n_1}}{1-q}\alpha_1+\Tilde{q}'^2\Tilde{q}q^{n_1+n_2+n_R}\frac{1-q^{n_L}}{1-q}\alpha_L
\end{aligned}
\end{equation}
So, when $n>n_s$, the evolution operator $H_R-H_L=a_R^{(II)\,\dag}-a_L^{(II)\,\dag}+a_R^{(II)}-a_L^{(II)}$ will be modified at the level of the expressions for $a_{L,R}$:
\begin{equation}
\begin{aligned}
     a_R^{(II)}-a_L^{(II)}=&C_R(n_L,n_R;n_1,n_2)\alpha_R+C_L(n_L,n_R;n_1,n_2)\alpha_L+\\&+C_1(n_L,n_R;n_1,n_2)\alpha_1+C_2(n_L,n_R;n_1,n_2)\alpha_2,
\end{aligned}
\end{equation}
where we defined:
\begin{equation}\label{eq:C_coeff}
\begin{aligned}
       &C_R(n_L,n_R;n_1,n_2)=\frac{1-q^{n_R}}{1-q}(1-\Tilde{q}'^2\Tilde{q}q^{n_1+n_2+n_L}),\\&C_L(n_L,n_R;n_1,n_2)=-\frac{1-q^{n_L}}{1-q}(1-\Tilde{q}'^2\Tilde{q}q^{n_1+n_2+n_R}),\\& C_1(n_L,n_R;n_1,n_2)=\Tilde{q}'\frac{1-q^{n_1}}{1-q}(\Tilde{q}q^{n_R+n_1}-q^{n_L}),\\&C_2(n_L,n_R;n_1,n_2)=-\Tilde{q}'\frac{1-q^{n_2}}{1-q}(\Tilde{q}q^{n_L+n_2}-q^{n_R}).
\end{aligned}
\end{equation}
At this point, we give our ansatz for the un-normalized Krylov basis when $n>n_s$:
\begin{equation}  \label{eq:kbase_ansatz_multiop}\ket{\chi_{n>n_s}}=\sum_{k,\,m=0}^{n-n_s,\,n_s}(-1)^{k+m}\binom{n-n_s}{k}\binom{n_s}{m}\ket{k,\,n-n_s-k;\,m,\,n_s-m},
\end{equation}

The heuristic reason to consider \eqref{eq:kbase_ansatz_multiop} as the Krylov basis ansatz is trying to generalize the basis of binomial states \eqref{eq:binomial_ansatz}, found for the single operator case, to the multiple insertions we want to describe now. Even though operators $a_{L,R}$ can annihilate chords in any sector, $a^\dag_{L,R}$ can just create new chords in the left/right sectors respectively. 
The first binomial coefficient in \eqref{eq:kbase_ansatz_multiop} accounts for the new $n-n_s=n_L+n_R$ chords created in the left/right sectors, analogously to what happened in the binomial states \eqref{eq:binomial_ansatz} for the single operator insertion. Instead, the second binomial's dependence on the constant number $n_s=n_1+n_2$ is a reflection of the fact that, after the perturbation is inserted, it is not possible to create new chords between the operators, so that this sector is effectively frozen. We will see that this structure of the chord sectors in \eqref{eq:kbase_ansatz_multiop} is very clearly showcased in the triple-scaled length behavior in \cref{sec:perturbed_3scale_length}.\\ After giving this heuristic motivation, we now proceed to show that our ansatz \eqref{eq:kbase_ansatz_multiop} does indeed solve the Lanczos algorithm when $n>n_s$.\\ 

Now, we prove that \eqref{eq:kbase_ansatz_multiop} is the un-normalized Krylov basis for the modified Lanczos algorithm when $n>n_s$. Let us anticipate that the proof follows in the same way we showed that the binomial states \eqref{eq:binomial_ansatz} were the Krylov basis for the single operator insertion. The outline of the procedure, summarized in \cref{sec:op_compl_recap}, is to show first that (\cref{app:asympt_multiop}):
\begin{equation}\label{eq:adag_chi_multiop}
   (a^{(II)\,\dag}_R-a_L^{(II)\,\dag})\ket{\chi_n}=\ket{\chi_{n+1}}, 
\end{equation}
where the states $\ket{\chi_n}$ are in the form of the ansatz \eqref{eq:kbase_ansatz_multiop}. Then, proving the Lanczos recursion at the $n>n_s$ step reduces to proving the following identity:
\begin{equation}\label{eq:multiop_cancellation}
    (a_R^{(II)}-a_L^{(II)})\ket{\chi_n}\overset{!}{=}b_n^2\ket{\chi_{n-1}}.
\end{equation}
Notice that, even though we are using the basis $\{\ket{\chi_n}\}$ \eqref{eq:kbase_ansatz_multiop} relative to multiple operator insertions, this is precisely the same cancellation condition we had to establish in \cite{Ambrosini:2024sre} to show that binomial states were the Krylov basis for the single operator complexity. In that case, the proof of \eqref{eq:binom_recursion_cancellation_cond}, summarized here in \cref{app:recap_asympt}, hinged on a saddle point approximation, originating from the asymptotic expression of the binomial coefficients \eqref{eq:binom_asymptotics_1p}. Indeed, we will show that \eqref{eq:multiop_cancellation} follows from an analogous saddle point approximation, generalized to the multi-variable case, proving that $\{\ket{\chi_n}\}$ is the un-normalized Krylov basis for $n>n_s$.\\

We proceed by defining a coefficient $c_{k,m}(n)$, analogous to \eqref{eq:ck_def_1p}, for the multi-operator states \eqref{eq:kbase_ansatz_multiop}, as:
\begin{equation}\label{eq:ckm_def_2p}
\begin{aligned}
    (a_R^{(II)}&-a_L^{(II)})\ket{\chi_n}=\\&=\sum_{k=0,m=0}^{n-n_s-1,\,n_s}c_{k,m}(n)(-1)^{k+m}\binom{n-n_s-1}{k}\binom{n_s}{m}\ket{k,\,n-1-n_s-k;\,m,\,n_s-m}
\end{aligned}
\end{equation}
Similarly to its analogue \eqref{eq:ck_def_1p}, the coefficient $c_{k,m}(n)$ parametrizes the deviation in the summands of $(a_R^{(II)}-a_L^{(II)})\ket{\chi_n}$ from those appearing in the definition of $\ket{\chi_{n-1}}$. Let us define:
\begin{equation}
    C_{L,R,1,2}(k,m)\equiv C_{L,R,1,2}(k,n-n_s-k;m,n_s-m),
\end{equation}
where $C_{L,R,1,2}(k,n-n_s-k;m,n_s-m)$ are the coefficients appearing in \eqref{eq:C_coeff}, obtained from \eqref{eq:vmatt_transm}. Then, after some algebra, we obtain
\begin{equation}
    \begin{aligned}
        (&a_R^{(II)}-a_L^{(II)})\ket{\chi_n}=\sum_{k,m}\binom{n-n_s-1}{k}\binom{n_s}{m}\biggr(C_R(k,m)\frac{n-n_s}{n-n_s-k}-C_L(k,m)\frac{n-n_s}{k+1}+\\&-C_1(k,m)\frac{n_s+1}{m+1}+C_2(k,m)\frac{n_s+1}{n_s+1-m}\biggr)(-1)^{k+m}\ket{k,\,n-1-n_s-k;\,m,\,n_s-m}\,.
    \end{aligned}
\end{equation}
So, by isolating the definition \eqref{eq:ckm_def_2p} in the equation above, we find:
\begin{equation}\label{eq:ckm_2p_expression}
\begin{aligned}
    c_{k,m}(n)=C_R(k,m)\frac{n-n_s}{n-n_s-k}&-C_L(k,m)\frac{n-n_s}{k+1}+\\&-C_1(k,m)\frac{n_s+1}{m+1}+C_2(k,m)\frac{n_s+1}{n_s+1-m}
\end{aligned}
\end{equation}
Our main interest lies in taking the triple-scaling limit, in order to make contact with gravity and observe the switchback effect. We leave the discussion of this triple-scaling for the next section, as, for the present asymptotic analysis, it is sufficient to work in the following semiclassical limit:
\begin{equation}\label{eq:semiclass_limit_multisector}
 \lambda\to0, \;{n_L,n_1,n_2,n_R\to\infty}\quad\mathrm{with}\quad l_{L,R,1,2}=\lambda n_{L,R,1,2}
\quad\mathrm{fixed}\end{equation}
Now, let us define $l_s=l_1+l_2$, $l'=l-l_s$, $x=l_L-l'/2$ and $y=l_1-l_s/2$. Then, as in \eqref{eq:binom_asymptotics_1p}, we can obtain the following asymptotic expressions for the binomial coefficients $\binom{n-n_s}{k}$ and $\binom{n_s}{m}$ near $\lambda\sim 0$:
\begin{align}
    &\qquad \qquad \qquad \qquad \qquad \qquad \qquad\qquad\binom{n-n_s}{k}\overset{\text{semicl.}}{=} \binom{\frac{l-l_s}{\lambda}}{l_L/\lambda} \nonumber \\
     &\underset{\lambda\sim 0}{\sim} \sqrt{\frac{2l'}{\pi\lambda(l'^2-4x^2)}} ~ \text{exp}\left\{ \frac{1}{\lambda} \Big( l'\log(2l')-(l'-2x)\log(l'-2x)-(l'+2x)\log(l'+2x) \Big) \right\}~, \label{eq:binom_asymptotics_2p}
\end{align}
\begin{align}
    &\qquad \qquad \qquad \qquad \qquad \qquad \qquad\qquad\binom{n_s}{m}\overset{\text{semicl.}}{=} \binom{l_s/\lambda}{l_1/\lambda} \nonumber \\
     &\underset{\lambda\sim 0}{\sim} \sqrt{\frac{2l_s}{\pi\lambda(l_s^2-4y^2)}} ~ \text{exp}\left\{ \frac{1}{\lambda} \Big( l_s\log(2l_s)-(l_s-2y)\log(l_s-2y)-(l_s+2y)\log(l_s+2y) \Big) \right\}~.
     \label{eq:binom_asymptotics_2p(1)}
\end{align}

At this point, the idea to prove \eqref{eq:multiop_cancellation} is the same of the one used for \eqref{eq:binom_recursion_cancellation_cond} in \cite{Ambrosini:2024sre}. In the semiclassical limit \eqref{eq:semiclass_limit_multisector}, we can approximate the sum over chord numbers, appearing for example in \eqref{eq:ckm_def_2p}, as integrals over the associated semiclassical lengths. Then, by virtue of the asymptotic expansion of the binomial coefficients \eqref{eq:binom_asymptotics_2p} and \eqref{eq:binom_asymptotics_2p(1)}, we can see the $\lambda\to0$ limit in \eqref{eq:semiclass_limit_multisector} as a saddle point approximation around $x=0$, $y=0$. Intuitively, in the semiclassical limit \eqref{eq:semiclass_limit_multisector}, the binomial coefficients $\binom{n-n_s-1}{k}$ and $\binom{n_s}{m}$ get squeezed and localize the sum \eqref{eq:ckm_def_2p}, by acting as nascent Dirac delta functions, respectively around $2\lambda k=\lambda(n-n_s)$ and $2\lambda m=\lambda n_s$. For more details on this asymptotic analysis you can refer to \cref{app:asympt_analysis} and appendix (D) in \cite{Ambrosini:2024sre}. An important technical detail we report here is that, because of the alternating signs in \eqref{eq:kbase_ansatz_multiop}, for the advent of these nascent delta functions, we need the multi-particle inner product between states $\ket{n_L,n_R;n_1,n_2}$ to be, even slightly, asymmetric around the centers of the summation domains of each variable. This is manifestly true for the single particle inner product \eqref{eq:scal_prod_semiclass_1p}, and we discuss this assumption for the multi-particle case at hand in \cref{app:inner_prod_multipart}. \\
So, we proved that \eqref{eq:multiop_cancellation} holds, by virtue of the $\lambda\to 0$ limit behaving as a saddle point approximation. Then, analogously to \eqref{eq:b_nplus1_induction_proof}, we can find the Lanczos coefficients
\begin{equation}\label{eq:perturbed_lanczos-}
\begin{aligned}
    b_n&=\sqrt{c_{k=\frac{n-n_s}{2},m=n_s/2}(n)}=\\&=\sqrt{2\biggr(C_R\biggr(\frac{n-n_s}{2},\frac{n_s}{2}\biggr)-C_L\biggr(\frac{n-n_s}{2},\frac{n_s}{2}\biggr)-C_1\biggr(\frac{n-n_s}{2},\frac{n_s}{2}\biggr)+C_2\biggr(\frac{n-n_s}{2},\frac{n_s}{2}\biggr)\biggr)}=\\&=2\sqrt{\frac{1-q^{\frac{n-n_s}{2}}}{1-q}\biggr(1-\Tilde{q}'^2\Tilde{q}q^{\frac{n_s+n}{2}}\biggr)+\Tilde{q}'\frac{1-q^{n_s/2}}{1-q}\biggr(q^{\frac{n-n_s}{2}}-\Tilde{q}q^{n/2}\biggr)},
\end{aligned}
\end{equation}
where in the last line we inserted the expressions for $C_{L,R,1,2}((n-n_s)/2,n_s/2))$ from \eqref{eq:C_coeff}. Notice that if our perturbation operator is the identity, that is for $\Tilde{q}'=1$, \eqref{eq:perturbed_lanczos-} coincides, when $n=n_s$, with the Lanczos coefficients for the single unperturbed operator insertion.\\

At this point, we have the solution of the Lanczos algorithm up to the step $n_s$ from \cref{sec:lanczos_pert}, and in this section we solved it for $n>n_s$. The last condition we need to check is the smoothness of the interface between the two solutions we found, that is we need to check that \eqref{eq:kbase_ansatz_multiop} solves the Lanczos recursion step just after the operator insertion. The Lanczos recursion at the $n_s+1$ step, after inserting \eqref{eq:adag_chi_multiop}, reduces to:
\begin{equation}
      (a_R^{(II)}-a_L^{(II)})\ket{\chi_{n_s}}-b_{n_s}^2\ket{\chi_{n_s-1}}=0
\end{equation}
Notice that the result \eqref{eq:multiop_cancellation}, that we proved in this section, concerns states \eqref{eq:kbase_ansatz_multiop} and Lanczos coefficients of the form \eqref{eq:perturbed_lanczos-}. However, in the condition we are examining now we have $\ket{\chi_{n_s}}$ in the form \eqref{eq:kbase_ansatz_multiop}, but the Lanczos coefficient $b_{n_s}$ is \eqref{eq:lanczos_ns}, which differs from \eqref{eq:perturbed_lanczos-}, evaluated at $n_s$, by a $\Tilde{q}'$ factor. So we have that $(a_R^{(II)}-a_L^{(II)})\ket{\chi_{n_s}}-b_{n_s}^2\ket{\chi_{n_s-1}}\propto (\sqrt{\Tilde{q}'}-1)\ket{\chi_{n_s-1}}$. This tells us that the interface between the unperturbed solution of the Lanczos algorithm \eqref{eq:binomial_ansatz} and the perturbed one \eqref{eq:kbase_ansatz_multiop} is smooth when the perturbation is given by a low-energy operator. Notice that, when we take $\lambda\to 0$, this condition translates in considering $\Delta_m$ finite, and, in particular, is satisfied in the limit $\Delta_m\to0$, that we will take in \cref{sec:perturbed_3scale_length}. As for the triple-scaled limit, considered in \cref{sec:vertical_multinom_lanczos}, we note that this is not a further restriction, as we need the low-energy limit to make contact with the gravitational shockwave picture where this property is showcased on the gravity side.\\

This concludes our proof of the identity \eqref{eq:multiop_cancellation}, and thus of the fact that the full solution of the Lanczos algorithm with seed $\mathcal{O}\ket{TFD}$ and evolution operator \eqref{eq:pert_Kev} is given by
\begin{equation}\label{eq:kbase_2op_full}
\begin{aligned}
    \ket{\chi_{n}}=\begin{cases}\sum_{k}^{n}(-1)^{k}\binom{n}{k}\ket{k,\,n-k}&\mathrm{when}\quad n\leq n_s\\
    \sum_{k,\,m=0}^{n-n_s,\,n_s}(-1)^{k+m}\binom{n-n_s}{k}\binom{n_s}{m}\ket{k,\,n-n_s-k;\,m,\,n_s-m}&\mathrm{when}\quad n> n_s\end{cases}
\end{aligned}
\end{equation}
\begin{equation}\label{eq:lanczos_2op_full}
\begin{aligned}
b_n=\begin{cases}2\sqrt{\frac{1-q^{\frac{n}{2}}}{1-q}\bigr(1-\Tilde{q}q^{n/2}\bigr)}&\mathrm{when}\quad n\leq n_s\\
    2\sqrt{\frac{1-q^{\frac{n-n_s}{2}}}{1-q}\bigr(1-\Tilde{q}'^2\Tilde{q}q^{\frac{n_s+n}{2}}\bigr)+\Tilde{q}'\frac{1-q^{n_s/2}}{1-q}\bigr(q^{\frac{n-n_s}{2}}-\Tilde{q}q^{n/2}\bigr)}&\mathrm{when}\quad n> n_s\end{cases}  
\end{aligned}
\end{equation}
 This is an approximate solution, hinging on the following facts: other matter configurations are suppressed with respect to the `vertical chords', the $\lambda\to 0$ limit promotes the binomial coefficients in \eqref{eq:kbase_2op_full} to nascent delta functions, and we have smoothness between the cases in \eqref{eq:lanczos_2op_full}. As we discussed, these approximations hold, and \eqref{eq:kbase_2op_full}\eqref{eq:lanczos_2op_full} give a reliable solution to the Lanczos algorithm with evolution \eqref{eq:pert_Kev}, in the cases of interest of this paper, where we take a low-energy and triple-scaled limit, in order to make contact with the gravitational description.\\ Notice that the expression of \eqref{eq:kbase_ansatz_multiop} confirms that, in the multi-operator case, when $\lambda\to 0$, the Krylov basis is a linear combination of states characterized by the same total chord number $n$. So, analogously to the way the single operator complexity was a \textit{total length} $l=l_L+l_R$, in this case the perturbed Krylov complexity will be a \textit{total length} $l=l_L+l_R+l_1+l_2=l_s+l_L+l_R$. We will confirm in \cref{sec:3scale_op_compl}, that, as we showed for the single operator complexity in \cref{sec:bulk_dual}, this feature is associated to the complexity having a geometrical nature.\\

\subsection{Lanczos coefficients for the perturbed $\mathcal{O}$TFD complexity}\label{sec:pert_otfd_compl}
We re-state that our objective is to compute the triple-scaled length, identified with the perturbed operator complexity, stemming from the modified evolution \eqref{eq:pert_Kev}, and verify that it showcases the switchback effect. However, as noticed in \cref{sec:bulk_dual}, the leading order contribution picked up by the saddle point approximation for the $H_R-H_L$ evolution does not capture the operator imprint in this low-energy triple-scaled limit. So, the procedure we will follow to compute the triple-scaled \textit{perturbed} operator complexity will be analogous to the one of \cref{sec:3scale_op_compl}, leveraging its \textit{total length} property. This will result in a triple-scaled complexity evolution that exhibits the switchback effect for Krylov complexity.\\
In this section, we compute the Lanczos coefficients pertaining the perturbed $\mathcal{O}$TFD complexity, from which we can find the triple-scaled expression of $H=H_L+H_R$ after the perturbation, which is the first ingredient of the procedure described above, performed in \cref{sec:perturbed_3scale_length}.
This will mostly be a repetition of \cref{sec:lanczos_pert} and \cref{sec:vertical_multinom_lanczos}, apart from sign differences, so we will proceed invoking less intermediate steps while, highlighting the important differences when necessary.\\

We wish to solve the modified Lanczos algorithm with seed $\mathcal{O}\ket{TFD}$ and evolution operator:
\begin{equation}\label{eq:pert_Kev_otfd}
\begin{aligned}
     (a^\dag_R+a^\dag_L)(1-\delta_{n\;n_s})+\frac{(\Tilde{a}_L^\dag \Tilde{a}_R+\Tilde{a}_R^\dag \Tilde{a}_L)}{\sqrt{2}}(a^\dag_R+a^\dag_L)\delta_{n\;n_s}+(a_R+a_L)
\end{aligned}
\end{equation}
Notice that \eqref{eq:pert_Kev_otfd} is perturbed in the same way as \eqref{eq:pert_Kev}, with the difference being that the evolution operator outside of the $n_s$ step is $H_L+H_R$ instead of $H_L-H_R$. If we are interested in taking a low-energy triple-scaled limit, we can approximate the full solution of the Lanczos recursion with: 
\begin{equation}\label{eq:kbase_2otfd_full}
\begin{aligned}
    \ket{\chi_{n}^+}=\begin{cases}\sum_{k}^{n}\binom{n}{k}\ket{k,\,n-k}&\mathrm{when}\quad n\leq n_s\\
    \sum_{k,\,m=0}^{n-n_s,\,n_s}\binom{n-n_s}{k}\binom{n_s}{m}\ket{k,\,n-n_s-k;\,m,\,n_s-m}&\mathrm{when}\quad n> n_s\end{cases}
\end{aligned}
\end{equation}
\begin{equation}\label{eq:lanczos_2otfd_full}
\begin{aligned}
b_n^+=\begin{cases}2\sqrt{\frac{1-q^{\frac{n}{2}}}{1-q}\bigr(1+\Tilde{q}q^{n/2}\bigr)}&\mathrm{when}\quad n\leq n_s\\
    2\sqrt{\frac{1-q^{\frac{n-n_s}{2}}}{1-q}\bigr(1+\Tilde{q}'^2\Tilde{q}q^{\frac{n_s+n}{2}}\bigr)+\Tilde{q}'\frac{1-q^{n_s/2}}{1-q}\bigr(q^{\frac{n-n_s}{2}}+\Tilde{q}q^{n/2}\bigr)}&\mathrm{when}\quad n> n_s\end{cases}  
\end{aligned}
\end{equation}
As anticipated, this result can be shown by reproducing the steps of \cref{sec:vertical_multinom_lanczos}, so here we just give an outline of the procedure.\\
The solution \eqref{eq:kbase_2otfd_full}\eqref{eq:lanczos_2otfd_full} when $n<n_s$, is the one obtained for the unperturbed $\mathcal{O}$TFD complexity, that we summarized in \cref{sec:op_compl_recap}.
The matter perturbation at the $n_s$-th step in \eqref{eq:pert_Kev_otfd}, creates additional matter chords in the same configurations of \eqref{eq:norm_matter_diagrams}. Similarly, if we are eventually interested in considering the triple-scaled limit, we can restrict our attention to the `vertical matter-chords' configuration. After specializing to these chord diagrams, we can prove that:
\begin{equation}
    (a_R^{(II)\,\dag}+a_L^{(II)\,\dag})\ket{\chi^+_n}=\ket{\chi^+_{n+1}}
\end{equation}
So, again, proving that $\{\ket{\chi^+_n}\}$ is the (un-normalized) Krylov basis for the evolution \eqref{eq:pert_Kev_otfd} when $n>n_s$ reduces to proving:
\begin{equation}\label{eq:cancellation_otfd_pert}
    (a_R^{(II)}+a_L^{(II)})\ket{\chi^+_n}\overset{!}{=}(b_n^+)^2\ket{\chi^+_{n-1}},
\end{equation}
where the coefficients $a_{R,L}^{(II)}$ are the ones in \eqref{eq:vmatt_transm}. Now, analogously to \eqref{eq:ckm_def_2p}, we can define a coefficient $c_{k,m}^+(n)$ via
\begin{equation}\label{eq:ckm_def_2p_otfd}
    (a_R^{(II)}+a_L^{(II)})\ket{\chi_n^+}=\sum_{k=0,m=0}^{n-n_s-1,\,n_s}c_{k,m}^+(n)\binom{n-n_s-1}{k}\binom{n_s}{m}\ket{k,\,n-1-n_s-k;\,m,\,n_s-m},
\end{equation}
that again has the role of parameterizing the difference between the summands in \eqref{eq:ckm_def_2p_otfd} and in the definition of $\ket{\chi_{n-1}^+}$. At this point, \eqref{eq:cancellation_otfd_pert} follows by the same saddle point approximation\footnote{notice that in this case we don't have alternating signs in the definition of the Krylov basis \eqref{eq:kbase_2otfd_full}, so in general the sums appearing are better controlled. This translates into the fact that we do not need the extra condition of the inner product asymmetry in order for the binomial coefficients to behave like nascent delta functions (see \cref{app:asympt_analysis} and appendix D in \cite{Ambrosini:2024sre}).} we leveraged in \cref{sec:vertical_multinom_lanczos}, so that, after some algebra, the Lanczos coefficients turn out to be:
\begin{equation}\label{eq:perturbed_lanczos+}
    b_n^+=\sqrt{c^+_{k=\frac{n-n_s}{2},m=n_s/2}}=2\sqrt{\frac{1-q^{\frac{n-n_s}{2}}}{1-q}\biggr(1+\Tilde{q}'^2\Tilde{q}q^{\frac{n_s+n}{2}}\biggr)+\Tilde{q}'\frac{1-q^{n_s/2}}{1-q}\biggr(\Tilde{q}q^{n/2}+q^{\frac{n-n_s}{2}}\biggr)},
\end{equation}
As in \cref{sec:vertical_multinom_lanczos}, continuity between the cases of \eqref{eq:lanczos_2otfd_full} is needed in order to ensure that the $n_s+1$ recursion is satisfied, and this holds when $\Tilde{q}'\sim 1$. \\

In the next section, we will recast explicitly \eqref{eq:perturbed_lanczos+} in the low-energy triple-scaled limit we assumed we would eventually take when we derived this result in the fixed `vertical matter-chords' configuration. Before doing this however, we ask ourselves, if, in order to compute the triple-scaled lengths, it would have been sufficient to just compute these perturbed $\mathcal{O}$TFD Lanczos coefficients, without addressing the evolution \eqref{eq:pert_Kev} pertaining the \textit{perturbed} operator complexity. Indeed, it is true that in \cref{sec:3scale_op_compl} we built the triple-scaled lengths by summing solutions to the equations of motion relative to the low-energy Hamiltonian that can be obtained from the limit of the $\mathcal{O}$TFD Lanczos coefficients. However, we could compute operator complexity as a sum of such solutions because we leveraged its property of being a \textit{total length}. In view of the analysis we will perform in the next section, the purpose of \cref{sec:vertical_multinom_lanczos} was precisely showing that the \textit{perturbed} operator complexity is also a \textit{total length}, so that it can be computed by generalizing the procedure of \cref{sec:3scale_op_compl} to the multi-insertion case.\\

\subsection{The triple-scaled perturbed operator complexity}\label{sec:perturbed_3scale_length}
In this section, we want to find the triple-scaled length associated to the perturbed operator complexity. Notice that, because we have chosen the perturbation in the evolution operator \eqref{eq:pert_Kev_otfd} to be explicitly left/right symmetric, at this point, the idea is to repeat the same procedure of \cref{sec:bulk_dual}.\\
We want to build the total Hamiltonian $H=H_L+H_R$ from the triple-scaled Lanczos coefficients associated to the $\mathcal{O}$TFD complexity found in the previous section:
\begin{equation}\label{eq:perturbed_lanczos+_normalt}
    b_n^+=2\frac{J}{\sqrt{\lambda}}\sqrt{\biggr[\frac{n-n_s}{2}\biggr]_q\biggr(1+\Tilde{q}'^2\Tilde{q}q^{\frac{n_s+n}{2}}\biggr)+\biggr[\frac{n_s}{2}\biggr]_q\Tilde{q}'\biggr(\Tilde{q}q^{n/2}+q^{\frac{n-n_s}{2}}\biggr)},
\end{equation}
where, comparing to equation \eqref{eq:perturbed_lanczos+}, we multiplied by $J/\sqrt{\lambda}$ to pass to the normalization of \cite{Lin:2022rbf}. Let us now consider the following triple-scaling limit:
\begin{equation}
    \lambda\to0,\;l=\lambda n \to\infty\quad\mathrm{with}\quad \frac{e^{-l}}{(2\lambda)^2}=e^{-\Tilde{l}}\;\mathrm{fixed}
\end{equation}
Notice that, in the above limit we are not renormalizing all lengths appearing in \eqref{eq:perturbed_lanczos+_normalt} and, crucially, that $l_s$ stays finite, $l_s=\lambda n_s \not\to\infty$. \footnote{this new prescription for the triple-scaled limit is implemented after the matter perturbation insertion, that is after the splitting in the two extra sectors detaches the length $l_s$ from the boundary. Notice, in particular, that when we require all matter configurations, except for the `vertical' one, to be suppressed in \eqref{eq:norm_matter_diagrams}, coherently to what we requested, we still have $l_s\to\infty$ by courtesy of triple scaling. As we will see in this section, after this step, there is no more need of triple-scaling $l_s$, and this length remains frozen at its renormalized value upon the perturbation insertion.
}. The general prescription, proposed by \cite{Lin:2022rbf}, is that only lengths anchored to the boundary are in need of being renormalized, and hence triple-scaled. This makes sense from a bulk point of view, as only geodesics anchored on the regularized boundary have a potentially divergent contribution that needs to be subtracted. So, if we write the Lanczos coefficients \eqref{eq:perturbed_lanczos+_normalt} in terms of $l_p\equiv \Tilde{l}-l_s$, we obtain the following triple-scaled expression $b^{+,TS}$:
\begin{equation}
    b^{+,TS}(l_p)=b_0(\lambda)-2 \lambda  \left(J e^{-l_p-l_s} \left(e^{l_p/2} \left(\Delta_m+\Delta  e^{l_s/2}+\Delta_m e^{l_s}\right)+2\right)\right)+\mathcal{O}(\lambda^2),
\end{equation}
where, as in \cref{sec:op_compl_recap}, $b_0(\lambda)=\frac{2 J}{\lambda }+O(\lambda^0)=E_0(\lambda)/2$.
The Lanczos coefficients \eqref{eq:perturbed_lanczos+_normalt} are obtained considering the evolution $H_L+H_R$, so we can write this operator as
\begin{equation}\label{eq:htot_pert_3scaled_2}
\begin{aligned}
    H_R&+H_L = e^{i\lambda k_p} b^{TS}(l_p) + b^{TS}(l_p) e^{-i\lambda k_p}=\\&=E_0(\lambda)-4J \lambda   \left(\frac{k_p^2}{2}+ \Delta_m e^{-l_s-l_p/2}+\Delta_m e^{-l_p/2}+ \Delta  e^{-\frac{l_p+l_s}{2}}+2e^{-l_p-l_s}\right)+\mathcal{O}(\lambda^2), 
\end{aligned}
\end{equation}
where we introduced the momentum $k_p$, conjugate to $l_p$. As we did in \cref{sec:op_compl_recap}, we subtract the ground state energy $E_0(\lambda)$ and change the sign of the remaining term in order to ensure boundedness from below. Also, similarly to \cref{sec:3scale_op_compl}, we will divide \eqref{eq:htot_pert_3scaled_2} by an additional factor of $2$, in order to avoid doubling the relevant timescales with respect to the non-perturbed case. So the Hamiltonian whose equation of motion we wish to solve is the following:

\begin{equation}\label{eq:ham_perturbed}
\frac{H_L+H_R}{2}=2J \lambda   \left(\frac{k_p^2}{2}+ \Delta_m e^{-l_s-l_p/2}+\Delta_m e^{-l_p/2}+ \Delta  e^{-\frac{l_p+l_s}{2}}+2e^{-l_p-l_s}\right)+\mathcal{O}(\lambda^2)
\end{equation}
Notice that, upon the perturbation insertion, the Hamiltonian \eqref{eq:triple_scaled_Htot} is modified to \eqref{eq:ham_perturbed}. In particular, not only they have different coefficients appearing in their functional forms, but they are expressed in terms of distinct canonical variables. This means that, upon the time of insertion $t_s$, we will have the onset of a new dynamical length, whose values are obtained by solving the equation of motion of \eqref{eq:ham_perturbed}. Before solving the equation of motion of the Hamiltonian \eqref{eq:ham_perturbed}, we note that this is the time-dependent evolution we set ourselves to find at the beginning of \cref{sec:time_to_kbase_pert}. In this section, we will be interested just in the post-perturbation physics at $t>t_s$ contained in \eqref{eq:ham_perturbed}, because we already discussed in \cref{sec:3scale_op_compl} what happens for $t<t_s$. However, for the purpose of highlighting the time-dependence of the system, we can write the following Hamiltonian, encapsulating both the pre/post-perturbation physics:
\begin{equation}
\begin{aligned}
    \frac{H_L+H_R}{2}&=2J \lambda   \left(\frac{k^2}{2}+ \Delta  e^{-l/2}+2e^{-l}\right)(1-\Theta(t-t_s))+\\&+2J \lambda  \Theta(t-t_s) \left(\frac{k_p^2}{2}+ \Delta_m e^{-l_s-l_p/2}+\Delta_m e^{-l_p/2}+ \Delta  e^{-\frac{l_p+l_s}{2}}+2e^{-l_p-l_s}\right),
\end{aligned} 
\end{equation}
where we can exhibit the full dependence on $t_s$ explicitly, by inserting back $l_s$, computed as a solution at time $t_s$ of the unperturbed Hamiltonian.\\

\begin{figure}
\begin{minipage}{0.49\textwidth}
 \includegraphics[width= 0.85\textwidth]{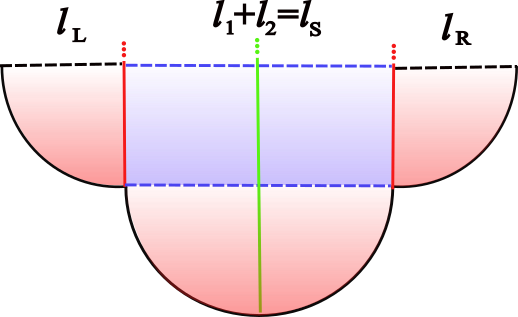}    
\end{minipage}
\begin{minipage}{0.50\textwidth}
\begin{tikzpicture}[scale=0.85]
\begin{axis}[clip=false,
xmin=0,xmax=200, ymin=0,xtick={50,100,150,200},ylabel near ticks,
xlabel near ticks,
xlabel={$2J\lambda t$},
ylabel={$l(t)$}]
\addplot[domain=0:100,black]{2*ln(1+.000000001*exp(x)/8)};
\addplot[domain=100:200,black]{2*ln(1+0.0000001*exp(x-100)/8)+2*ln(1+0.000000001*exp(100)/8)};
\addplot[domain=0:200,blue,dashed]{+2*ln(1+0.000000001*exp(100)/8)}
node[pos=0.1,above] {$l_s$};
\draw[red,dashed] ({axis cs:100,0}|-{rel axis cs:0,0}) -- ({axis cs:100,0}|-{rel axis cs:0,1})
node[pos=0.1,right] {$t_s$};
\end{axis}
\end{tikzpicture}
\end{minipage}
\caption{\textbf{On the left:} Intuitive representation of the DSSYK disk with an operator insertion $\mathcal{O}$ (green chord), receiving a two-sided $\Tilde{\mathcal{O}}$ perturbation (red chords) after a time $t_s$. The length is dynamical in the regions shaded in red, where geodesic lengths are anchored to the boundary. Upon the perturbation insertion, the chord length, in the blue-shaded region, freezes to the value it had at time $t_s$, and we have new dynamical variables. \textbf{On the right:} Plot of the perturbed operator complexity $l(t)$ \eqref{eq:perturbed_opcompl_lgt_full}, corresponding to the intuitive representation on the left, with $\Delta=10^{-9}$, $\Delta_m=10^{-7}$ and $2J\lambda t_s=100$. Upon the perturbation insertion in $t_s$, the length freezes to $l_s=l(t_s)$, however, due to the appearance of the two new dynamical variables, it resumes the linear growth after a scrambling time.}
    \label{fig:freezing}
\end{figure}
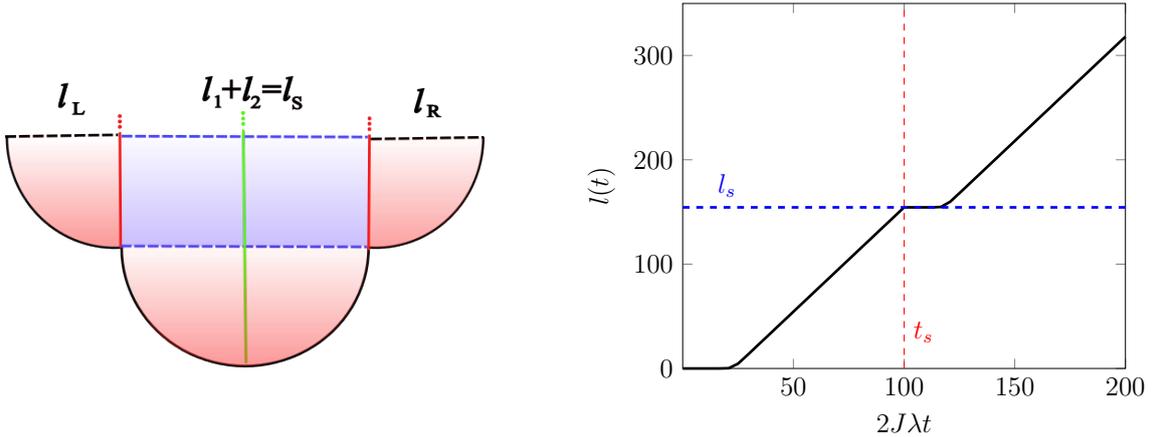

We can find the total lengths associated to the perturbed $\mathcal{O}$TFD and operator complexities in the \textit{shockwave approximation}, in a totally analogous way to \cref{sec:bulk_dual}. The change in dynamical variables ensures that we can set the initial value of the new length $l_p(t_s)=0$, while the low-energy perturbation assumption tells us that we should impose the wanted solution to have the same energy of the matterless case \eqref{eq:matterless_energy}. In particular, these conditions are precisely the ones used to find the total lengths $l_L+l_R$ in \cref{app:details_eom}, with $x_0=0$, $t_0=t_s$ and $C=\Delta e^{-l_s/2}+\Delta_m(1+e^{-l_s})$. The $\mathcal{O}$TFD complexity is again equal to the matterless case, while, from \eqref{eq:tot_lgt_minus_app}, we obtain the following perturbed operator complexity:
\begin{equation}\label{eq:lp_genericC}
l_p(t>t_s)=2\log\left(1+\frac{\Delta e^{-l_s/2}+\Delta_m(1+e^{-l_s})}{8}e^{2\lambda J (t-t_s)}\right)
\end{equation}
We define $t_{*}'\sim -1/(2\lambda J)\log(C)$ as the time-scale associated to the advent of the linear regime of the length $l_p(t)$, and $t_{scr}'\sim -1/(2\lambda J)\log(\Delta_m)$ as the scrambling time associated to an operator of dimension $\Delta_m$. Notice that, even though we have chosen $t_s\gg t_{scr}$ when inserting the perturbation, we may have a parametrically small $\Delta_m$ such that $t_s\lessapprox t_{scr}'$. In this case, we have a non-negligible contribution to $t_{*}'$ coming from the backreacted background operator chord, which becomes leading if $t_s\ll t_{scr}'$, so that $C\approx \Delta e^{-l_s/2}+\Delta_m$, and the delay does not depend only on the details of the perturbation, but also on all previous times of insertion.\\
In further discussions, we will be interested instead in the case where $t_s\gg t_{scr}, \;t_{scr}'$, so that we have $C\approx \Delta_m$ and $t_{*}'\approx t_{scr}'$, that is the delay upon the insertion only depends on the details of the perturbation operator. In this parameter region, \eqref{eq:lp_genericC} becomes: 

\begin{equation}
l_p(t>t_s)\approx2\log\left(1+\frac{\Delta_m}{8}e^{2\lambda J (t-t_s)}\right)
\end{equation}
The equation above, which consists of the dynamical part of the length $l_L+l_R$, anchored to the boundary, coincides with the triple-scaled operator complexity of the perturbation $\Tilde{\mathcal{O}}$, translated by the time of insertion $t_s$.
In order to obtain the total length, that is the full perturbed operator complexity, we need to also add $l_1+l_2=l_s$, so that the complete complexity profile is (\cref{fig:freezing}):
\begin{equation}\label{eq:perturbed_opcompl_lgt_full}
    \begin{aligned}
&l(t)=\\&2\log\biggr(1+\frac{\Delta}{8}e^{2\lambda J t}\biggr)(1-\Theta(t-t_s))+\Theta(t-t_s)\left(l_s+2\log\left(1+\frac{\Delta_m}{8}e^{2\lambda J (t-t_s)}\right)\right),\\&\quad\quad\quad\quad\quad\quad\quad\quad\quad\quad\quad\quad\quad\quad\quad\quad\mathrm{where}\quad l_s=2\log\biggr(1+\frac{\Delta}{8}e^{2\lambda J t_s}\biggr)
    \end{aligned}
\end{equation}
Now we want to study, after the perturbation, the regimes of \eqref{eq:perturbed_opcompl_lgt_full} in the `\textit{shockwave approximation}', as described in \cref{sec:shockwave_geodesics} and \cref{sec:3scale_op_compl}, used to derive it. When $t>t_s$, we have:
\begin{equation}\label{eq:pert_op_compl_regimes_approx}
      l(t)\sim l_s+2\log\left(1+\frac{\Delta_m}{8}e^{2\lambda J (t-t_s)}\right)\sim 
    \begin{cases}
       l_s  & \text{if $t-t_s\ll t'_{scr}$} \\
      l_s+4J\lambda (t-t_s-t'_{scr}) & \text{if $t-t_s\gg t'_{scr}$},
    \end{cases}
\end{equation}
At this point, in order to make contact with \eqref{eq:switchback_erb_def}, we restrict to the specific low-energy regime where the dimensions $\Delta,\; \Delta_m\ll 1$ define parametrically equivalent scrambling times $t'_{scr}\sim t_{scr}$. In this parameter region, using that for $t_s\gg t_{scr}$, from \eqref{eq:op_compl_regimes_approx}, we have $l_s\sim 4J\lambda (t_s-t_{scr})$, we observe the following regimes for the perturbed length:
\begin{equation}\label{eq:pert_op_compl_regimes_approx_sametscramble}
      l(t)\sim  
    \begin{cases}
    0& \text{if $t\ll t_{scr}$}\\
       4J\lambda (t-t_{scr})  & \text{if $t_{scr}\ll t< t_s$} \\
       4J\lambda (t_s-t_{scr})& \text{if $t_s<t\ll t_s+t_{scr}$} \\
      4J\lambda (t-2t_{scr}) & \text{if $t\gg t_s+t_{scr}$},
    \end{cases}
\end{equation}
From this analysis, we understand that ultimately the effect of the perturbation is a further delay of a scrambling time upon the restoration of the linear regime. The regimes in \eqref{eq:pert_op_compl_regimes_approx_sametscramble} coincide with those of the dual bulk length \eqref{eq:app_2switch} identified using the holographic dictionary we defined in \cref{sec:hol_dict_matt} (see \cref{app:switch_erb} for the explicit bulk dual computation). Then, consistently, \eqref{eq:pert_op_compl_regimes_approx_sametscramble} also has the same behavior expected of the switchback effect for the insertion configuration we are considering, as we will discuss in \cref{sec:switch_explicit}.\\

From this procedure, it seems like the main mechanism behind the switchback effect in DSSYK is the change in dynamical lengths, that is in the lengths anchored to the boundary. By this we mean that if we insert a two-sided perturbation operator at time $t_s$, then we can split the system in four regions. We can define a chord number, and hence a length, in each of these regions, and, as we discussed in \cref{sec:perturbed_3scale_length}, Krylov complexity is a total length, and so is computed by their sum. The length-region correspondence is the following: $l_L$ is the length from the left boundary to the perturbation chord, $l_1$ and $l_2$ go from perturbation to background operator and viceversa, and finally $l_R$ goes from the second perturbation to the right boundary. Notice that $l_L$ and $l_R$ are new variables defined upon the matter insertion, while the value of the complexity before the perturbation is encoded in $l_1+l_2=l_s=l(t_s)$. Upon the perturbation insertion, the lengths $l_1$ and $l_2$ are detached from the boundary and get frozen to their renormalized value at this time. Simultaneously, the new lengths $l_L$ and $l_R$ are attached to the boundary, but will show the same scrambling dynamics that we observed at early times for $l_1$ and $l_2$\footnote{to avoid confusion, notice that, upon the perturbation insertion, we are changing the name of our lengths variables. The convention is that $l_L$ and $l_R$ are always the dynamical lengths anchored to the boundary, while $l_{1,2}$ are frozen. So when we improperly talk about `early time behavior' of $l_{1,2}$, we rather mean, the early time behavior of the dynamical variables whose value at $t_s$ is frozen in $l_{1,2}$.}. This intuitively justifies what we proved rigorously above: we obtain a length $l_p(t)=(l_L+l_R)(t)$ equal to the (unperturbed) operator complexity of $\mathcal{\Tilde{O}}$, but translated by $t_s$ in the past. In order to obtain the total length of the wormhole we also need to sum $l_s$, and this gives the fundamental picture of the scrambling time delay upon insertion expected from the switchback effect.\\
Notice that we could have inferred this characteristic behavior directly from the structure of the Krylov basis \eqref{eq:kbase_ansatz_multiop} after the perturbation. From the form of these states, we know that operator complexity is going to be a total length $l=l_L+l_1+l_2+l_R$, but also that the sector of $l_1$ and $l_2$ has a fixed number of open chords. The fact that growth is relegated to the left/right regions anchored to the boundary is a consequence of the operators $a^\dag_{L,R}$ appearing in the evolution \eqref{eq:pert_Kev} being able to create chords only in these sectors\footnote{remember that we are using the convention of \cite{Berkooz:2018jqr}, where chords are created below all other open chords, and annihilated arbitrarily.}.\\

\subsubsection{The switchback effect of operator K-complexity}\label{sec:switch_explicit}

Now we want to understand the evolution procedure defined in \cref{sec:lanczos_pert} from the gravitational point of view, in particular the boundary anchoring prescriptions of the relevant bulk geodesics. Our purpose is to understand what timefold is associated to our case, and check that the K-complexity result obtained in \cref{sec:perturbed_3scale_length} showcases the switchback effect.\\ 

The evolution procedure we underlined in \cref{sec:lanczos_pert} consists in evolving $\mathcal{O}\ket{TFD}$ with $H_R-H_L$ until the time $t_s$, upon which we insert additional matter. We perform this perturbation by acting with the two-sided operator $\Tilde{\mathcal{O}}_L\Tilde{\mathcal{O}}_R$, inserted at time $t_s$, and then we continue the evolution for a time $t-t_s$:
\begin{equation}\label{eq:op_switch_evolution}
\begin{aligned}    \mathcal{O}\ket{TFD}\rightarrow_{t_s}&\Tilde{\mathcal{O}}_L\Tilde{\mathcal{O}}_Re^{i(H_L-H_R)t_s}\mathcal{O}\ket{TFD}\rightarrow_{t-t_s}\\&\rightarrow e^{i(H_L-H_R)(t-t_s)}\Tilde{\mathcal{O}}_L\Tilde{\mathcal{O}}_Re^{i(H_L-H_R)t_s}\mathcal{O}\ket{TFD},
\end{aligned}
\end{equation}
where above the arrows are denoting the evolution for the time indicated in the subscript. The operators $\Tilde{\mathcal{O}}_L$, $\mathcal{O}$, inserted on the left boundary, and $\Tilde{\mathcal{O}}_R$, on the right, evolve respectively with the Hamiltonians $H_L$ and $H_R$. So using the fact that the TFD state is symmetric under $H_R-H_L$, we can write the final state obtained from the evolution scheme \eqref{eq:op_switch_evolution} as:
\begin{equation}\label{eq:timefold_switch}
    e^{iH_L t}\Tilde{\mathcal{O}}_R(-t+t_s)\Tilde{\mathcal{O}}_L(-t_s)\mathcal{O}e^{-iH_Lt}\ket{TFD}\quad\leftrightarrow \quad\raisebox{-50pt}{\includegraphics[scale=0.26]{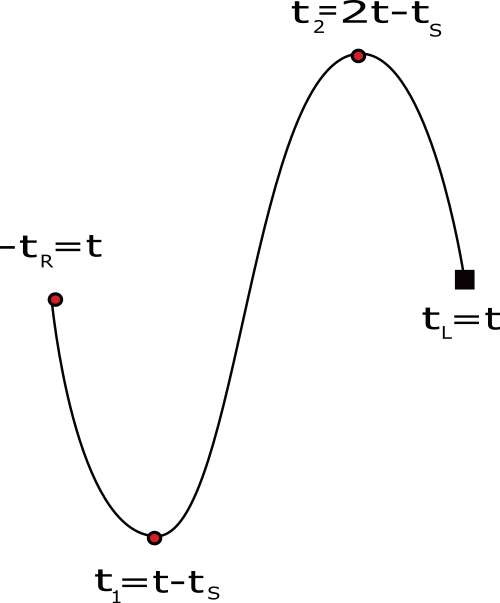}}
\end{equation}
By using \eqref{eq:reflection_princ}, we can bring this evolution prescription in the form \eqref{eq:evolution_example}, with three operator insertions at times $t_0=t$, $t_1=t-t_s$ and $t_2=2t-t_s$. So the timefold contour in \eqref{eq:timefold_switch}, outlined, with the notation of \cref{sec:switch_grav_intro}, by the sequence of times $t_0=-t_R=t$, $t_1=t-t_s$, $t_2=2t-t_s$ and $t_3=t_L=t$, has length $2t$ and contains two switchbacks. So, we expect that the \textit{perturbed} operator complexity showcases the characteristic switchback effect in the form of \cref{eq:switchback_erb_def} for $n_{sb}=2$ and $t_f=2t$:
\begin{equation}\label{eq:switch_2sw}
    l(t)\propto 2(t-2t_{scr}),
\end{equation}
which is indeed precisely the behavior observed in \cref{eq:pert_op_compl_regimes_approx_sametscramble}.\\
The characteristic regime \eqref{eq:switch_2sw}, as presented here, is the expected behavior of an ERB length under two switchback perturbations. At the level of the \textit{shockwave approximation}, this matching is already sufficient to determine that the bulk dual of the perturbation presented in \eqref{eq:pert_Kev}, is a shockwave corresponding to the operator insertion via the dictionary \eqref{eq:new_entry_matt}. In \cref{app:switch_erb}, we will explicitly exhibit the dual shockwave configuration to the DSSYK perturbation scheme considered here, and compute the geodesic length associated to the perturbed operator complexity of \cref{sec:perturbed_3scale_length}. As expected, these two observables match under the holographic dictionary established in \cref{sec:bulk_dual}, and in particular they showcase the same characteristic asymptotic switchback behavior.\\

To summarize, in this section we showed that indeed the operator complexity perturbed using the evolution scheme defined in \cref{sec:lanczos_pert} showcases the characteristic behavior associated to two switchback insertions. Analogously, we can consider instead the triple-scaled length associated to the perturbed $\mathcal{O}$TFD complexity, underlined by the evolution $H_R+H_L$. As discussed, at least at the level of the \textit{shockwave approximation}, this complexity is linear at all times and does not encode any detail on the matter insertion. In particular, with the evolution prescription $H_L+H_R$, we are considering timefolds with only through-going insertions, so it is expected that the associated complexity always presents the linear behavior without any scrambling time delays.\\
The procedure we used, at this point, is easily applicable to the case of multiple insertions similar to \eqref{eq:pert_Kev}, and the associated \textit{perturbed} operator complexity will show the behavior associated to arbitrary $n$-switchbacks. In the next section, we will briefly go over the changes needed to generalize the discussion above to this case with many operator insertions.

\subsection{Generalization to many switchback insertions}\label{sec:multiop_switch_gen}
In the previous sections, we understood how the perturbation defined in \cref{sec:lanczos_pert} introduces a switchback insertion to the evolution of the background operator complexity. At this point, we briefly discuss how to generalize the arguments in order to insert multiple such switchbacks.\\

The idea for building a Lanczos evolution containing arbitrary $m$ operator insertions at times $t_1,\,\dots t_{m}$ is the same as \cref{sec:time_to_kbase_pert}. We will consider the Lanczos algorithm with seed $\mathcal{O}\ket{TFD}$ and evolution operator modified at steps $n_s^{(1)},\,\dots n_s^{(m)}$ as:
\begin{equation}\label{eq:pert_Kev_multi}
\begin{aligned}
     (a^\dag_R-a^\dag_L)\biggr(1-\sum_{i=1}^m\delta_{n\;n_s^{(i)}}\biggr)+\frac{(\Tilde{a}_L \Tilde{a}_R^\dag+\Tilde{a}_R \Tilde{a}_L^\dag)}{\sqrt{2}}(a^\dag_R-a^\dag_L)\sum_{i=1}^m\delta_{n\;n_s^{(i)}}+(a_R-a_L),
\end{aligned}
\end{equation}
and again we can associate these chord numbers of insertion to the corresponding times by inverting the semiclassical relation $n(t_i)=n_s^{(i)}$. Notice that, for notational simplicity, we chose to act with the same $\Tilde{a}_{L,R},\,\Tilde{a}^\dag_{L,R}$, meaning that at each $n_s^{(i)}$ we are inserting the same operator $\Tilde{\mathcal{O}}$. \\
After each operator insertion, we can repeat analogously the analysis performed in \cref{sec:vertical_multinom_lanczos} and \ref{sec:perturbed_3scale_length}. In particular the norms of the Krylov states, that we will define momentarily, are still dominated, in the triple-scaling limit, by the `vertical matter-chords' configuration. Notice that the triple-scaling limit also suppresses the contribution originating from the possibility of a matter chord generated at $n_s^{(i)}$ being annihilated by a chord of the same kind in a previous insertion in $n_s^{(j)}$.\\
After $m$ insertions of the kind of \eqref{eq:pert_Kev_multi}, the perturbation matter chords, in the leading `vertical' configuration, will have split the system in $2(m+1)$ sectors, so that we can describe the states in the Hilbert space of interest as:
\begin{center}
\vspace{5mm}
    \begin{tikzpicture}[scale=1.0]
    \node[left] at (-2.0, -0.75) {$|n_L,\,n_R;n_1,n_{2m};\dots\rangle = $};
    
    \draw[thick] (-2,0) arc[start angle=180, end angle=360, radius=2];
    
    \foreach \x in {-0.8,0.8} {
        \fill[red] (\x,-{sqrt(4-(\x)^2)}) circle (2pt); 
        \draw[thick][red] (\x,-{sqrt(4-(\x)^2)}) -- (\x,0); 
    }
    \fill[green] (0,-2) circle (2pt); 
    \draw[thick, green] (0,-2) -- (0,0); 
    \fill[red] (1.4,-{sqrt(4-(1.4)^2)}) circle (2pt); 
    \draw[thick, red] (1.4,-{sqrt(4-(1.4)^2)}) -- (1.4,0);
    \fill[red] (-1.4,-{sqrt(4-(1.4)^2)}) circle (2pt); 
    \draw[thick, red] (-1.4,-{sqrt(4-(1.4)^2)}) -- (-1.4,0);
    \node at (-0.35,-0.5) {\tiny{$\cdots$}};
    \node at (0.35,-0.5) {\tiny{$\cdots$}};
    \node at (-1.6,-0.5) {\tiny{$\cdots$}};
    \node at (1.65,-0.5) {\tiny{$\cdots$}};
     \node at (1.1,-0.5) {\tiny{$\cdots$}};
      \node at (1.1,-0.5) {\tiny{$\cdots$}};
    \draw[thick] (-2.0,0.2) -- (-1.45,0.2);
    \draw[thick] (-1.3,0.2) -- (-0.85,0.2);
    \draw[thick] (-2.0,0.2) arc[start angle=180, end angle=230, radius=0.2];
    \draw[thick] (-1.3,0.2) arc[start angle=180, end angle=230, radius=0.2];
    \draw[thick] (-1.45,0.2) arc[start angle=0, end angle=-50, radius=0.2];
    \draw[thick] (-0.85,0.2) arc[start angle=0, end angle=-50, radius=0.2];
    \node[above] at (-1.05,0.3) {$n_1$};
    \node[above] at (-1.7,0.3) {$n_L$};

    \draw[thick] (1.45,0.2) -- (2,0.2);
     \node[above] at (0.3,0.4) {$\dots$};
    \draw[thick] (1.45,0.2) arc[start angle=180, end angle=230, radius=0.2];
    \draw[thick] (2.0,0.2) arc[start angle=0, end angle=-50, radius=0.2];
    \node[above] at (1.7,0.3) {$n_R$};

\end{tikzpicture}
\end{center}
Analogously to \cref{sec:vertical_multinom_lanczos}, we can show that the Krylov basis after the $m$-th insertion is:
\begin{equation}  \label{eq:kbase_ansatz_multiop_mult}
\begin{aligned}
\ket{\chi_{n>n_s^{(m)}}}=&\sum_{k_1,\dots\,k_m=0}^{n-n_s^{(m)},\,n_s^{(m)}-n_s^{(m-1)},\,\dots\,n_s^{(1)}}(-1)^{\sum k_i}\binom{n-n_s^{(m)}}{k_m}\binom{n_s^{(m)}-n_s^{(m-1)}}{k_{m-1}}\dots\binom{n_s^{(1)}}{k_1}\times\\&\times\ket{k_m,\,n-n_s^{(m)}-k_m;\,k_{m-1}, n_s^{(m)}-n_s^{(m-1)}-k_{m-1},\dots;k_1, n_s^{(1)}-k_1}.
\end{aligned}
\end{equation}
Notice that, in this case, the states above are the Krylov basis, when $\lambda\to0$, by virtue of the usual saddle point approximation of \cref{sec:op_compl_recap} and \ref{sec:perturbed_3scale_length}: each binomial coefficient localizes one of the integrals, coming from the continuum approximation of the sums, to the center of its domain via saddle point approximation\footnote{again when considering the evolution with $H_R-H_L$, we are assuming that $m$ particle inner product is asymmetric around the center of each integration domain, so that we can rigorously treat the binomials as nascent delta functions (\cref{app:asympt_multiop}).}. Crucially, this shows that, even under many perturbation insertions, operator complexity still has the property, encoding its geometrical nature, of being a total length $l=l_L+l_R+l_1+\dots+l_{2m}$, where we defined $l_i=\lambda n_i$ for $i=1,\dots \,m$. Notice also that the structure of \eqref{eq:kbase_ansatz_multiop_mult} suggests to us that the same freezing effect of \cref{sec:perturbed_3scale_length} happens here. All chord sectors are frozen to the chord number value at which they have been disconnected from the boundary of the disk, and only $n_L+n_R$ can grow upon evolution with the two-sided Hamiltonian.\\
At this point, the procedure to find the Lanczos coefficients after the $m$-th perturbation at $n_s^{(m)}$ is analogous to \cref{sec:vertical_multinom_lanczos}: they can be computed from the coefficients of the transition matrix, analogous to \eqref{eq:vmatt_transm}, evaluated, as in \cref{sec:vertical_multinom_lanczos}, in the center of the summation domain of each variable appearing in \eqref{eq:kbase_ansatz_multiop_mult}. \\

Similarly, we can obtain the Lanczos coefficients associated to the seed $\mathcal{O}\ket{TFD}$ and evolution operator 
\begin{equation}\label{eq:pert_Kev_multi_plus}
\begin{aligned}
     (a^\dag_R+a^\dag_L)\biggr(1-\sum_{i=1}^m\delta_{n\;n_s^{(i)}}\biggr)+\frac{(\Tilde{a}_L \Tilde{a}_R^\dag+\Tilde{a}_R \Tilde{a}_L^\dag)}{\sqrt{2}}(a^\dag_R+a^\dag_L)\sum_{i=1}^m\delta_{n\;n_s^{(i)}}+(a_R+a_L),
\end{aligned}
\end{equation}
which pertains the analysis of the $\mathcal{O}$TFD complexity, perturbed by $m$ operator insertions at $n_s^{(1)},\dots,\,n_s^{(m)}$.\\
Armed with the Lanczos coefficients associated to the perturbed $H_L+H_R$ evolution, we can find the triple-scaled total lengths repeating the procedure of \cref{sec:3scale_op_compl} and \ref{sec:perturbed_3scale_length}.
First, let us describe the triple scaling limit we perform to link our matter perturbed DSSYK to a bulk picture. Remember that a single operator insertion splits the theory in two left and right sectors, so that $H_L$ and $H_R$ can be written as a function of canonical variables $l_{L,R}=\lambda n_{L,R}$ and their conjugates. Further operator insertions split the theory in many sectors, with associated lengths $l_L$, $l_1\dots,\,l_{2m},\, l_{R}$ and the total length, that is our would-be complexity, is just the total length given by their sum. We follow the triple scaling prescription presented in \cite{Lin:2022rbf}, so that $l_L, l_R\to\infty$ and $\lambda\to0$ such that $e^{-l_{L,R}}/(2\lambda)^2$ is kept constant. This is consistent from the bulk point of view, as $l_{L,R}$ are the only lengths anchored on the boundary, and thus the only ones in need of a renormalization.\\

In order to find $l_L+l_R$, that is the dynamical length, we need to repeat the procedure of \cref{app:details_eom} for the Hamiltonian $H_L+H_R$ after the $m$-th insertion. Of course, the expression of this evolution operator will depend on the Lanczos coefficients, and hence on the specific form of the analogue of \eqref{eq:vmatt_transm}. However, we expect the Hamiltonian to be in the same form of \eqref{eq:htot_generalizedC_app}, with $C\to C^{(m)}=C^{(m-1)}e^{-(l_s^{(m)}-l_s^{(m-1)})/2}+\Delta_m(1+e^{-l_s^{(m)}})$, where we defined $l_s^{(m)}=\lambda n_s^{(m)}$\footnote{the base step of the recursion is given by the case analyzed in \cref{sec:perturbed_3scale_length}, where, in this notation, we considered $m=1$, $C^{(0)}=\Delta$ and $l_s^{(0)}=l_s$.}.
Analogously to what happened for the single perturbation in \cref{sec:perturbed_3scale_length}, this constant acts like an effective operator dimension, that takes into account not only the $\Delta_m$ of the last insertion, but also, in the general case, the backreacted contributions of all previously inserted matter perturbations. This effective dimension $C^{(m)}$ is the parameter controlling the scrambling time associated to the dynamics of the new lengths anchored to the boundary, after the $m$-th insertion. However, in the limit where we have $t_i\gg t_{scr},\;t_{scr}' $ for all $i\leq m$, we have that $C^{(m)}\approx \Delta_m$, and after each insertion the scrambling time delay is controlled by the dimension of the last operator inserted. So in the end, we expect the picture of \eqref{eq:pert_op_compl_regimes_approx} to repeat, picking up an extra scrambling time delay after each perturbation insertion. In particular, the operator complexity perturbed by $m$ insertions defining parametrically equivalent scrambling times $t_{scr}\sim t_{scr}'$ is expected to be at asymptotically late times:\\
\begin{equation}
    l(t>t_m)=4\lambda J(t-mt_{scr}),
\end{equation}
which is precisely the characteristic behavior of $m$ switchbacks insertions in \eqref{eq:switchback_erb_def}.
Let us notice that, also in this more general case, the switchback effect is driven by the change in dynamical region of the DSSYK disk (\cref{fig:freezing}). When a perturbation matter is inserted, we perform a length redefinition where the previously dynamical variables $l_{L,R}$ get frozen at their re-normalized value and sent into $l_{1,2m}$ respectively (and concurrently all other $l_i\to l_{i+1}$). Then, we define a new pair of dynamical variables, which we call again $l_{L,R}$, anchored to the boundary, and encoding the scrambling details after the $m$-th perturbation.\\
The bulk dual of these matter insertions of DSSYK will be given by a shockwave configuration in JT gravity, generalizing the discussion of \cref{app:geod_details}. Let us briefly notice that, in the limit of many matter insertions $m\to\infty$, the evolution \eqref{eq:pert_Kev_multi} gives a prescription to recursively build a microscopic configuration related, after taking the triple-scaled limit in the `\textit{shockwave approximation}', to the `accordion' geometry introduced in \cite{Shenker_2014_manyshock}.\\

\section{Discussion}\label{sec:final_discussion}
In this paper, building on \cite{Ambrosini:2024sre}, we showed that the geometric nature of operator Krylov complexity can be seen as a direct consequence of the fact that its associated Krylov basis is a certain combination of states with fixed total chord number.
By leveraging this \textit{total chord length} property, we computed the triple-scaled operator complexity and identified its dual with a geodesic anchored at times $t_L=-t_R=t$ in the shockwave-backreacted bulk of JT gravity:
\begin{equation}
l_{\mathrm{grav}}(t)=2l_{AdS}\log\biggr(1+\frac{E}{8M}e^{\frac{r_s}{l_{AdS}^2}t}\biggr)\quad\leftrightarrow\quad l_{\mathrm{DSSYK}}(t)=2l_f\log\biggr(1+\frac{\Delta}{8}e^{2J\lambda t}\biggr),
\end{equation}
This relation identifies a new entry in the bulk-to-boundary dictionary containing the details of the matter insertion:
\begin{equation}\label{eq:hol_dict_compl}
    2\lambda J=\frac{r_s}{l_{AdS}^2},\quad l_f=l_{AdS},\quad\Delta=\frac{E}{M}
\end{equation}
As discussed, the bulk-to-boundary correspondence implied by the above parameter correspondence holds in the late-time limit of light insertions, as these are the assumptions underlying the \textit{shockwave approximation}. It would be interesting to find a similar duality in the presence of matter, that also captures the early time details of the complexity profile. A further avenue of study would be to compute the operator Krylov complexity outside of the $\lambda\to 0$ limit, where we staged our analysis both here and in \cite{Ambrosini:2024sre}. In particular, as for example in \cite{Heller:2024ldz}, a comparison with the bulk sine-dilaton gravity model of \cite{Blommaert:2024ymv}, could be useful in understanding the geometrization, or lack thereof, of operator K-complexity in this parameter regime.\\

Then, we showed how to perturb the Lanczos algorithm in order to describe the operator insertion at the level of the evolution of the system. In the triple-scaled limit, we find that, also in this multi-operator case, the Krylov basis is a certain linear combination of states with fixed total chord number. Even though the diagrams are split into several sectors by the matter insertions, we argue that the geometric nature of this \textit{perturbed} operator Krylov complexity can be seen as a consequence of this \textit{total chord length} property, in an analogous manner to the single-operator case. The key mechanism driving the appearance of the switchback effect in DSSYK is the following: the insertion of the new matter chords creates new dynamical variables anchored to the boundary, while the chord numbers in the region between the operators freeze to the value they had at the moment of the perturbation. The behavior of this \textit{perturbed} operator K-complexity exactly matches, in the \textit{shockwave approximation}, the switchback effect of the ERB lengths, computed in the shockwave configuration in JT gravity dual to the operators in DSSYK:
    \begin{equation}
      l(t)\sim  
    \begin{cases}
    0& \text{if $t\ll t_{scr}$}\\
       4J\lambda (t-t_{scr})  & \text{if $t_{scr}\ll t< t_s$} \\
       4J\lambda (t_s-t_{scr})& \text{if $t_s<t\ll t_s+t_{scr}$} \\
      4J\lambda (t-2t_{scr}) & \text{if $t\gg t_s+t_{scr}$},
    \end{cases}
\end{equation}
This result, generalizable to $m$ insertions, shows that the bulk-to-boundary map defined in the single shockwave case persists upon perturbations, and, more crucially, that operator K-complexity indeed describes a geometric quantity in the bulk.\\ 
 The analysis in this paper was limited to the insertion of a specific class of two-sided operators, for which we could solve analytically the Lanczos algorithm and repeat the triple-scaled computation we performed for the single-operator case, which required symmetry between the left and right boundaries. It would be interesting to study more general types of matter insertion, in particular the one of single-sided operators. As we discussed, this would require the additional complication of considering multiple contributing matter-chords configuration in the steps of the Lanczos algorithm. More crucially however, we note that in the two-sided case we considered, we could describe switchbacks by virtue of the two-sided insertions. If we wish to consider single-sided operators, in order to have switchbacks, we also need to understand how to reverse time in the evolution after the insertion of the perturbation.\\
 Additionally, it would be interesting to understand if it is possible to use our model of modified Lanczos algorithm to study other time-dependent evolutions, for example quenches in more general models.\\

Finally, it would be rewarding to understand if the Lanczos algorithm is the basis mechanism for which an emergent notion of length arises also in more general models and parameter regimes of DSSYK. Some cases of interest would be DSSYK in the spectral region where it describes de Sitter \cite{Narovlansky:2023lfz, Susskind:2022bia, Okuyama:2025hsd, heller2025sitterholographiccomplexitykrylov}, as well as supersymmetric extensions of the model and their integrable deformations \cite{Berkooz:2024ofm}, in particular to study K-complexity as a probe of chaos in the recent fortuitous/monotone states discussion \cite{Chang:2024lxt,Chen:2024oqv,Aguilar-Gutierrez:2025sqh}.

\section*{Acknowledgments}
We thank Adrián Sanchez-Garrido and Ruth Shir for initial collaboration, and numerous enlightening discussions. We also thank Brian Swingle and Lenny Susskind for early email correspondence regarding the switchback effect in Krylov complexity. ER acknowledges NYU for their hospitality during the period where some of the work was done. This research is supported in part by the Fonds National Suisse de la Recherche Scientifique (Schweizerischer Nationalfonds zur Förderung der wissenschaftlichen Forschung) through the Project Grant 200021\_215300 and the NCCR51NF40-141869 The Mathematics of Physics (SwissMAP). ER  acknowledges partial support from Israel’s Council for Higher Education grant.
\appendix

\section{Details on solutions of the equations of motion}\label{app:details_eom}

In the left/right symmetric configuration we are interested in studying in \cref{sec:bulk_dual}, the Hamiltonians $H_{L,R}$ are
\begin{equation}        
\begin{aligned}
     H= H_{L,R}=2J\lambda\left(e^{-l_{L,R}}\Delta+\frac{k_{L,R}^2}{8}+2 e^{-2l_{L,R}}\right),
\end{aligned}
\end{equation}
written as a function of the canonical variables $l_{L,R}=l/2$ and their conjugate momenta $k_{L,R}$. The equation of motion for the Hamiltonian above is:
\begin{equation}
   (J\lambda)^2(  \Delta e^{-l_{L,R}(t)}+4e^{-2l_{L,R}(t)}  )=l''_{L,R}(t)
\end{equation}
As discussed in \cref{sec:shock_model_def}, the shockwave we want to match the operator chord with, is created by inserting some low energy quanta very far into the past. For the purpose of matching with these geodesics we can make the simplifying assumption of searching for solutions at late times by solving instead the simplified differential equation\footnote{after finding a solution $f(t)$, by substituting it back into the equation of motion, it is possible to check that the neglected term $\propto e^{-2f(t)}$ is exponentially suppressed at late times for which $f(t)\gtrsim 1$. This means that the solution of the simplified Hamiltonian will agree with the full one from the scrambling time onward. Notice that the same happens in the \textit{shockwave approximation}: only after the scrambling time we can trust the geodesic length expression we compute, because only after the scrambling time the shockwave crossing is described by a null shift \cite{Shenker_2014}.}:
\begin{equation}
    (J\lambda)^2\Delta e^{-l_{L,R}(t)}=l''_{L,R}(t),
\end{equation}
which can be obtained as the equation of motion for the approximated late-time Hamiltonian $H_{L,R}\sim2J\lambda\left(e^{-l_{L,R}}\Delta+\frac{k_{L,R}^2}{8}\right)$.\\

In the remaining part of this section, we will compute the solutions to the slightly more general differential equation
\begin{equation}\label{eq:diffeq_gen}
    (J\lambda)^2 C e^{-l_{L,R}(t)}=l''_{L,R}(t),
\end{equation}
which can be obtained as the equation of motion for the late-time limit of the Hamiltonian:
\begin{equation}\label{eq:htot_generalizedC_app}
    H=2J\lambda\left(e^{-l_{L,R}}C+\frac{k_{L,R}^2}{8}+2 e^{-2l_{L,R}}\right)
\end{equation}
For the appropriate choice of the constant $C$, solving this generalized case will give us what we need to compute the triple-scaled lengths in the late-time limit both in \cref{sec:3scale_op_compl} and \cref{sec:perturbed_3scale_length}. \\

A class of solutions for \eqref{eq:diffeq_gen} is given by the functions:
\begin{equation}\label{eq:solution_diffeq_general}
f(t)=\log\left(\frac{-CJ^2\lambda^2}{c_1}(-1+\cosh{\sqrt{c_1}(t+c_2)})\right)
\end{equation}
Now we explain what conditions we wish to impose while searching for solutions in the above form. The functional forms of physical two-sided observables, that is the total bulk lengths we want to build via sums of \eqref{eq:solution_diffeq_general}, are uniquely determined if we fix the value of the length and the momentum at a certain time $t_0$. However, what we are searching in the form \eqref{eq:solution_diffeq_general} are the one-sided lengths that need to be summed together, and thus need weaker conditions. We choose to impose that \eqref{eq:solution_diffeq_general} has a certain value $x_0/2$ at a time $t_0$ and has boundary energy equal to the matterless case. We choose these conditions to ensure respectively the left/right symmetry and consistency with low-energy insertion approximation used to derive \eqref{eq:diffeq_gen} from the explicit expressions of $H_{L,R}=H$. As it turns out, imposing these conditions already singles out only two possible solutions. All two-sided lengths that are equal to $x_0$ at $t_0$, and in particular the two objects we are looking for\footnote{as a reminder, these are the operator and $\mathcal{O}$TFD complexity, which are both total lengths and thus computable via this procedure.}, are be built by summing such solutions. \\

Now we impose the aforementioned conditions on the general class of solutions \eqref{eq:solution_diffeq_general}, in order to find the appropriate constants $c_{1,\,2}$.
We start by searching for constants $c_2$ so that \eqref{eq:solution_diffeq_general} is equal to $x_0/2$ at a certain time $t_0$, and we find two possible values, for which we obtain respectively:
\begin{equation}\label{eq:f1f2_sol}
    \begin{aligned}
        &f_1(t)=\log \biggr(-\frac{1}{2}C e^{x_0}\sinh \biggr(J\lambda e^{-x_0/2} (t-t_0)-\frac{1}{2}\cosh^{-1}\biggr({1-\frac{4e^{-x_0/2}}{C}\biggr)}\biggr)^2\biggr),\\&
        f_2(t)=\log \biggr(-\frac{1}{2}C e^{x_0}\sinh \biggr(J\lambda e^{-x_0/2} (t-t_0)+\frac{1}{2}\cosh^{-1}\biggr({1-\frac{4e^{-x_0/2}}{C}\biggr)}\biggr)^2\biggr),
    \end{aligned}
\end{equation}
where we chose $c_1=4\lambda^2 J^2 e^{-x_0}$ to set the boundary energy of these solutions equal to $4J\lambda e^{-x_0}$, as for the matterless case initialized to $x_0$ in $t_0$ \cite{Rabinovici:2023yex}\footnote{in the main text of this paper, we always consider $x_0=0$, so that here we have the usual boundary energy \eqref{eq:matterless_energy}, dual to \eqref{eq:bdry_energy}.}.\\

Now in the small $C$ limit these functions become:
\begin{equation}\label{eq:f1f2}
    \begin{aligned}
    &f_1(t)\approx\log \left(\frac{1}{4} Ce^{x_0}\left(1-e^{2 J \lambda  (t_0-t)e^{-x_0/2}}\right)+e^{2 J \lambda  e^{-x_0/2}(t_0-t)+\frac{x_0}{2}}\right)=f(-t)\\&
        f_2(t)\approx \log \left(\frac{1}{4} Ce^{x_0}\left(1-e^{2 J \lambda  (t-t_0)e^{-x_0/2}}\right)+e^{2 J \lambda  e^{-x_0/2}(t-t_0)+\frac{x_0}{2}}\right)=f(t)
    \end{aligned}
\end{equation}
So if we now want to construct a total length, we need to sum two solutions chosen between $f_1$ and $f_2$ that we respectively assign to $l_L$ and $l_R$. It is clear then that we can construct two objects: one by choosing the same solution twice (the $\mathcal{O}$TFD state complexity), and one by choosing different solutions (the operator complexity). Notice that this choice, given that $f_1(t)$ and $f_2(t)$ are given by the same function $t$ but evaluated at times $t$ and $-t$ respectively, correctly mirrors the anchoring points prescription for the evolutions underlying the $\mathcal{O}$TFD state and operator complexity computations.\\
If we choose $l_L(t)=f_1(t)$ and $l_R(t)=f_2(t)$ in \eqref{eq:f1f2_sol}, for small $C$, we get the (two-sided) total length associated to a boundary evolution with $H_R-H_L$ (that is the operator complexity):
\begin{eqnarray}\label{eq:tot_lgt_minus_app}
    l(t)\approx x_0+2 \log \left(1+\frac{1}{2} C e^{x_0/2} \sinh ^2(J \lambda e^{-x_0/2}  (t-t_0))\right).
\end{eqnarray}
Instead, choosing $l_L(t)=l_R(t)=f_1(t)$ gives the total length associated to the evolution $H_R+H_L$ (the $\mathcal{O}$TFD complexity):
\begin{eqnarray}\label{eq:tot_lgt_plus_app}
    l_+(t)\approx x_0+4J\lambda(t-t_0),
\end{eqnarray}
which, being independent of $C$, is not showcasing any details of the matter insertion, at least in the late-time and low-energy approximation used here.\\
Let us now address what is the effect of neglecting for simplicity the $\propto e^{-2l_{L,R}}$ term while solving the Hamiltonian. As discussed, we know that the solutions we obtained here will rigorously match the full one from the scrambling time $\sim\log\left(1/\Delta\right)$ onward. On the other hand, the small energy $E$ quanta inserted on the JT gravity boundary need a scrambling time $\sim \log(1/E)$ to form the shockwave \cite{Shenker_2014}. So, as anticipated, we can match the behavior of a certain geodesic before/after the scrambling time with the ones of operator complexity, and for this purpose the simplified solution we found suffices.\\

In the discussion above, we imposed that the solutions we identified with the left/right lengths, when inserted back into the Hamiltonian, have boundary energy $4 J\lambda$, equal to the matterless case. Before concluding this appendix, we show that neglecting the small amount of energy inserted, say on the left boundary, is consistent with the shockwave approximation, and in particular does not spoil the left/right symmetry we leveraged to find our results. If we consider again \cref{eq:solution_diffeq_general}, with $c_2$ fixed by imposing that the length is zero at time $t_0=0$, we have that its boundary energy is $c_1/(J\lambda)$. Now, imagine that, due to the operator insertion, the left solution has some amount of extra energy proportional to $\Delta$, that is $c_1^L=4J^2\lambda^2+c_3\Delta$, where $c_3$ is a constant we keep for generality, while we still choose $c_1^R=4J^2\lambda^2$ for the right solution. Repeating the procedure above, we find the following operator complexity: $l(t)\approx \log((1+\Delta \sinh{J\lambda t}^2/2)^2+c_3\Delta t/(8J\lambda)+\dots)$. Notice, in particular, that the extra term obtained, becomes of $\mathcal{O}(1)$ at much later times $\sim 1/\Delta$ than scrambling $\sim \log(1/\Delta)$. In particular, it does not modify either the pre or post-scrambling characteristic regimes of the complexity in the shockwave approximation \cref{eq:op_compl_regimes_approx}. This provides a consistency check that the left/right symmetry leveraged to obtain our results is broken at a higher order than what is captured by the shockwave approximation.

\subsection{Solution for the Morse potential}\label{app:morse_sol}
We can obtain the triple-scaled total Hamiltonian from the Lanczos coefficients associated to the $\mathcal{O}TFD$ complexity:
\begin{equation} \label{eq:H_tot_triple_scaled} 
      \frac{H_L+H_R}{2}=2J\lambda\left(e^{-\Tilde{l}/2}\Delta+\frac{k^2}{2}+2 e^{-\Tilde{l}}\right)
\end{equation}
We start with a closely related Hamiltonian:
\begin{equation}\label{eq:Ham_eow}
     H=\frac{2}{A}\left(\mu e^{-\Tilde{l'}}+\frac{k'^2}{4}+ e^{-2\Tilde{l'}}\right),
\end{equation}
that was obtained in \cite{Gao_2022} for JT gravity with an insertion of an end-of-the-world brane with tension $\mu$. This suggests that the DSSYK matter chord could be equivalently interpreted as the insertion in the bulk of an end-of-the-world brane with tension related to the operator dimension. Given $A$ and $B$, values of the dilaton at the boundary and horizon in this system, the particular solution with energy $2B^2/A$ is known to be \cite{Gao_2022}:
\begin{equation}\label{eq:eow_solution}
     \Tilde{l'}(t)=\log\left(\frac{2\mu+2\sqrt{\mu^2+B^2}\cosh{\frac{B}{A}t}}{B^2}\right)
\end{equation}
If we perform the change of variable $\Tilde{l'}= \Tilde{l}/2-\log 2$ we can bring \eqref{eq:Ham_eow} in the form of \eqref{eq:H_tot_triple_scaled}, with the identification of $A=\frac{2}{J\lambda}$ and $\mu=\Delta$. To first approximation in $\lambda$ and $\Delta$ we can search for a solution of \eqref{eq:H_tot_triple_scaled} with the same energy as the matterless one, that is $4J\lambda=2B^2/A$. So the solution of \eqref{eq:H_tot_triple_scaled} in terms of our original length variable, which is the triple-scaled $\mathcal{O}$TFD state complexity, is:
\begin{equation}\label{eq:sol_Htot}
    \Tilde{l}(t)=2\log\left(\sqrt{1+\left(\frac{\Delta}{4}\right)^2}\cosh{2J\lambda t}+\frac{\Delta}{4}\right)
\end{equation}
If we substitute it back in the Hamiltonian \eqref{eq:H_tot_triple_scaled} we can check that indeed the energy of this solution is equal to $4 J\lambda$, as obtained for the matterless case.\\

\section{Details on asymptotic analysis for multiple operator insertions}\label{app:asympt_analysis}
In this appendix, we first give a summary of the asymptotic analysis of appendix D in \cite{Ambrosini:2024sre} (\cref{app:recap_asympt}), that allowed us to identify the Krylov bases \eqref{eq:binomial_ansatz} relative to the $\mathcal{O}$TFD and operator complexities for a single-matter insertion. Then, in \cref{app:asympt_multiop}, we discuss how to generalize the discussion to the multi-operator case we introduced in \cref{sec:vertical_multinom_lanczos} and \ref{sec:pert_otfd_compl}. As we remind in the following sections, both for the single and multi-operator cases, identifying the Krylov basis for the Lanczos evolution $H_R-H_L$ requires a particular asymmetry in the inner product, which is an assumption we discuss in \cref{app:inner_prod_multipart}.

\subsection{Asymptotic analysis for the single-operator insertion}\label{app:recap_asympt}
In this section, we briefly summarize some of the more technical contents of appendix D of \cite{Ambrosini:2024sre}. In particular, we remind the reader why a certain inner product asymmetry is needed in order to show that the binomial states are the Krylov basis in relation to evolution $H_R-H_L$. Then, assuming the same asymmetry is present in the multi-particle inner product, the same procedure can be translated analogously to the many-insertions binomial states \eqref{eq:kbase_ansatz_multiop}, discussed in \cref{sec:vertical_multinom_lanczos}.\\

We wish to consider a two-sided Hamiltonian $H_R+zH_L$, for some $z\in \mathbb{R}$ (required for hermiticity). Applying the Lanczos algorithm with this evolution operator and seed state $|0,0\rangle$ would lead us to generalized binomial states, defined as\footnote{After a slight change of notaton, the states $\ket{\chi_n^{(+1)}}\equiv\ket{\chi_n^{(+)}}$ are recognized as the Krylov basis of the $\mathcal{O}$TFD complexity, while $\ket{\chi_n^{(-1)}}\equiv\ket{\chi_n^{(-)}}$ as the ones of operator complexity, discussed in \cref{sec:op_compl_recap}.}
\begin{equation}
    \label{Generalized_binomial_state}
    |\chi^{(z)}_n\rangle := (a_R^\dagger + z a_L^\dagger)^n~|0,0\rangle = \sum_{k=0}^{n}z^k\binom{n}{k}|k,n-k\rangle~.
\end{equation}
If we are interested in computing the Lanczos coefficients associated to the recursion at hand, we need to study the norm of these states, given by:
\begin{equation}
    \label{Norm_binom_generalized}
    \langle\chi^{(z)}_{n+1} | \chi^{(z)}_{n+1} \rangle = \sum_{k^\prime,k=0}^{n}c^{(z)}_k(n+1)~z^{k^\prime + k}\binom{n}{k^\prime}\binom{n}{k}\langle k^\prime,n-k^\prime | k,n-k\rangle=:\mathcal{F}_{n+1}(z;\lambda,\Delta)~,
\end{equation}
where we have introduced a generalized $c$-coefficient analogous to \eqref{eq:ck_def_1p}
\begin{equation}
    \label{ckn_generalized}
    c^{(z)}_k(n) := n\frac{[n-k]_q}{n-k}\left( 1+z\widetilde{q}q^k \right) + z^2 n\frac{[k+1]_q}{k+1}\left( 1+\frac{1}{z}\widetilde{q}q^{n-1-k} \right)~.
\end{equation}
If we consider the semiclassical limit where $\lambda\to0$ and $k,\,k',\,n\to\infty$, keeping $l_L=\lambda k,\,l_L'=\lambda k',\,l=\lambda n$ fixed, the above coefficient has a smooth expression as a function of $l$ and $l_L$ regardless of $z$:
\begin{equation}
    \label{ckn_generalized_semicl}
    c^{(z)}_k(n)\underset{\lambda\sim 0}{\sim} c^{(z)}(l,l_L;\lambda) \equiv \frac{l}{l-l_L} \frac{1-e^{-(l-l_L)}}{\lambda}(1+z\widetilde{q}e^{-l_L}) +z^2 \frac{l}{l_L}\frac{1-e^{-l_L}}{\lambda}\biggr(1+\frac{1}{z}\widetilde{q}e^{-(l-l_L)}\biggr)~.
\end{equation}

Notice however, that for $z<0$ the function $z^{l_L/\lambda}$ is wildly oscillating, and thus does not admit such smooth limiting form in the semiclassical limit. So, in particular for the case of operator complexity $z=-1$, the localization to the center of the summation domain of \eqref{Norm_binom_generalized}, by virtue of the asymptotic expression of $\binom{n}{k}$, is not directly applicable. Indeed, if we consider the asymptotic expression $\eta_\lambda (l,l_L;z)$ of $z^k\binom{n}{k}/\lambda$ as a whole, when $\lambda\to0$, we have:
\begin{equation}
    \label{nascent_generalized_asymptotic}
    \eta_\lambda (l,l_L;z)\underset{\lambda\sim 0}{\sim} \sqrt{\frac{l}{2\pi\lambda l_L (l-l_L)}} \text{exp}\left\{\frac{1}{\lambda}\Big( l\log(l) -l_L\log(l_L/z)-(l-l_L)\log(l-l_L) \Big)\right\}~.
\end{equation}
This function, for $\lambda\sim 0$, is sharply localized at the extremal point $l_L=l_L^{(*)}(z;l)$  of the exponent:
\begin{equation}
    \label{z_dependent_center}
    l_L^{(*)}(z;l)=\frac{lz}{z+1}~.
\end{equation}
This shows that, in the semiclassical limit, the expression $z^k{\lambda}\binom{n}{k}/\lambda$ becomes asymptotically proportional to a (nascent) Dirac delta function $\delta\bigr(l_L - l_L^{(*)}(z;l) \bigr)$. However, the Lanczos evolution pertaining the computation of operator complexity is associated to $z=-1$: the localization point is $l_L^{(*)}\equiv \infty$ in this case. This is a signal that, because of the alternating phases, \eqref{Norm_binom_generalized} might, in principle, pick up contributions from the edges of the summation domain. Now, we wish to show that the asymmetry of the inner product nevertheless allows us to consider the semiclassical limit of $z^k\binom{n}{k}/\lambda$, as if it were a nascent delta function at the center of the summation domain.\\

In the semiclassical limit, the summation appearing in \eqref{Norm_binom_generalized} can be computed via the following integral over the semiclassical variables\footnote{this integral representation is derived when $z>0$, that is when $\mathcal{F}(z;\lambda,\Delta)$ is well-defined in the semiclassical limit. We can consider other values in the full complex plane, including the point of interest $z=-1$, via analytical continuation of this expression.}:
\begin{align}
    &\mathcal{F}_{n+1}(z;\lambda,\Delta) \nonumber\\
    &\underset{\lambda\sim 0}{\sim} [n]_q! \int_{\mathbb{R}^2} dl_L dl_L^\prime c^{(z)}(l,l_L;\lambda)\eta_{\lambda}(l,l_L;z)\eta_\lambda(l,l_L^\prime;z) \left( \frac{\left(1-e^{-l}\right)/2}{\cosh\left(\frac{l_L-l_L^\prime}{2}\right)-e^{-l/2}\cosh\left( \frac{l-l_L-l_L^\prime}{2} \right)} \right)^{2\Delta}~, \label{Norm_binom_generalized_integral_line2}
\end{align}
where we have defined
\begin{equation}
    \label{Nascent_generalized}
    \eta_\lambda(l,l_L;z) := \frac{1}{\lambda}\eta\left(\frac{l}{\lambda},\frac{l_L}{\lambda}\right)~,\quad \text{for} \quad \eta(l,l_L)=\left\{ \begin{aligned}
        & \frac{z^{l_L} ~\Gamma(l +1)}{\Gamma(l_L +1) \Gamma(l-l_L+1)}~,\quad 0\leq l_L\leq l~, \\
        &0~,\qquad\qquad\qquad \qquad\qquad~\text{else}~.
    \end{aligned} \right.  
\end{equation}

Now let us specialize to the case $z=-1$ pertaining the evolution originating the operator complexity. The expression $(-1)^{(l_L/\lambda)}\binom{l/\lambda}{l_L/\lambda}$, for $\lambda\to0$, gets squeezed and become sharply peaked around $l_L=l/2$ in absolute value, but presents a wildly oscillating behavior with frequency $\propto1/\lambda$. In the semiclassical limiting procedure, there always exists a limiting scheme such that this oscillating function is odd around the center of the integration domain. Now, let us imagine to integrate such oscillating $\eta_\lambda(l,l_L;-1)$ against a test function $g(l_L)$:
\begin{equation}
    \int_{\mathbb{R}}dl_L ~\eta_\lambda(l,l_L,-1) g(l_L)=\int_{\mathbb{R}}dl_L~ \eta(l/\lambda,l_L) g(\lambda l_L),
\end{equation}
we want to understand under which condition we have that:
\begin{equation}\label{eq:integral_delta_condition}
    \int_{\mathbb{R}}dl_L ~\eta_\lambda(l,l_L,-1) g(l_L)\underset{\lambda\sim 0}{\overset{?}{\sim}} g(l/2)~\eta\left(\frac{l}{\lambda},\frac{l}{2\lambda}\right)~.
\end{equation}
In particular, in the case we are examining here, the test function we are integrating again comes from the semiclassical limit expressions of the coefficient $c_k(n)$ \eqref{ckn_generalized_semicl} and the inner product \eqref{eq:scal_prod_semiclass_1p}.\\
Notice that, the integration \eqref{eq:integral_delta_condition} will yield exactly zero (up to surface terms) if the smooth test function $g(l_L)$ is even around the center of the integration domain $l_L=l/2$, because the contributions to the integral coming from the right of $l/2$ will get exactly canceled out by those to the left. In particular, there are various ways to take the semiclassical limit of the sum \eqref{Norm_binom_generalized}, from which we obtain different results if the test function is symmetric around the middle of the summation domain (hence formally the limit does not exist). Nevertheless, we can still make the following point: if the test function integrated against $\eta_\lambda(l,l_L;z=-1)$ is even slightly asymmetric around $l_L=l/2$, the integrand contributions in any neighborhood to the right of $l_L=l/2$ will not exactly cancel out against those coming from the left, and in fact this non-zero difference will get amplified the smaller $\lambda$ gets. So for the set of test functions that are not symmetric around $l_L=\frac{l}{2}$, we have that \eqref{eq:integral_delta_condition} holds, and in particular $\eta_\lambda (l,l_L;z=-1)$ is indeed acceptable as a delta function at $l_L=l/2$. Notice that this procedure works also for parametrically small $\Delta$, as long as we perform the low energy limit after we already restricted to the $\lambda\to0$ one. Conversely, if we have strictly $\Delta=0$, the inner product becomes a constant, and is thus even around $l_L=l/2$. Therefore, by the argument above, in this case $\eta_\lambda (l,l_L;z=-1)$ does not localize to a delta function for any $\lambda$, and indeed vanishes because of the alternating phases appearing in the integrand \cite{Ambrosini:2024sre}. \\

\subsection{Generalization of the asymptotic analysis to multi-operator insertions}\label{app:asympt_multiop}
As we mentioned in the previous section, the asymptotic analysis needed to prove that the states \eqref{eq:kbase_ansatz_multiop} constitute the Krylov basis relative to the perturbed operator complexity, is a straightforward generalization of the single-insertion case. Here we give a proof of some identities we used in \cref{sec:vertical_multinom_lanczos} and give a brief outline on how the asymptotic analysis of the previous section generalizes to this case.\\

Let us start by showing that the identity $(a_R^{(II)\,\dag}-a_L^{(II)\,\dag})\ket{\chi_n}= \ket{\chi_{n+1}}$ \eqref{eq:adag_chi_multiop} holds for states $\ket{\chi_n}$ defined in \eqref{eq:kbase_ansatz_multiop} for $n>n_s$:
\begin{equation}\label{eq:adagg_gives_multiop1}
\begin{aligned}
     (a_R^{(II)\,\dag}-a_L^{(II)\,\dag})\ket{\chi_n}&=\sum_{k,\,m=0}^{n-n_s,\,n_s}(-1)^{k+m}\binom{n-n_s}{k}\binom{n_s}{m}\ket{k,\,n-n_s-k+1;\,m,\,n_s-m}+\\&-\sum_{k,\,m=0}^{n-n_s,\,n_s}(-1)^{k+m}\binom{n-n_s}{k}\binom{n_s}{m}\ket{k+1,\,n-n_s-k;\,m,\,n_s-m}
\end{aligned}
\end{equation}
In the second term we can change summation variable $k\to k+1$ and use the binomial identity $\binom{n-n_s}{k+1}=\binom{n-n_s+1}{k}-\binom{n-n_s}{k}$ to obtain
\begin{equation}\label{eq:adagg_gives_multiop2}
    \begin{aligned}
      &(a_R^{(II)\,\dag}-a_L^{(II)\,\dag})\ket{\chi_n}=\\&= \sum_m(-1)^{m}\binom{n_s}{m}\biggr(\ket{0,\,n-n_s+1;\,m,\,n_s-m}+(-1)^{n-n_s}\ket{n-n_s+1,\,0;\,m,\,n_s-m}\biggr)\\&+\sum_{m,k=1}^{k=n-n_s}(-1)^{k+m}\binom{n-n_s+1}{k'}\binom{n_s}{m}\ket{k,\,n+1-n_s-k;\,m,\,n_s-m}=\ket{\chi_{n+1}}\,.
    \end{aligned}
\end{equation}
Notice that the first two terms, coming respectively from the first and second terms of \eqref{eq:adagg_gives_multiop1}, give the $k=0$ and $k=n+1$ summands appearing in the definition of $\ket{\chi_{n+1}}$ of \eqref{eq:kbase_ansatz_multiop}. So by re-summing \eqref{eq:adagg_gives_multiop2} we obtain by definition \eqref{eq:kbase_ansatz_multiop} the state $\ket{\chi_{n+1}}$ and we showed \eqref{eq:adag_chi_multiop}:
\begin{equation}
    (a_R^{(II)\,\dag}-a_L^{(II)\,\dag})\ket{\chi_n}= \ket{\chi_{n+1}}
\end{equation}\\
As we explained in \cref{sec:vertical_multinom_lanczos}, the Lanczos basis for $n>n_s$ is built by the recursion relation:
\begin{equation}\ket{\chi_{n+1}}=(a_R^{(II)\,\dag}-a_L^{(II)\,\dag})\ket{\chi_n}+(a_R^{(II)}-a_L^{(II)})\ket{\chi_n}-b_n^2\ket{\chi_{n-1}}
\end{equation}
So if we insert \eqref{eq:adag_chi_multiop} in the $n$-th Lanczos step we obtain:
\begin{equation}
\cancel{\ket{\chi_{n+1}}}=\cancel{(a_R^{(II)\,\dag}-a_L^{(II)\,\dag})\ket{\chi_n}}+(a_R^{(II)}-a_L^{(II)})\ket{\chi_n}-b_n^2\ket{\chi_{n-1}}
\end{equation}

As anticipated, given \eqref{eq:adag_chi_multiop}, showing that $\{\ket{\chi_n}\}$ in \eqref{eq:kbase_ansatz_multiop} is the (un-normalized) Krylov basis reduces to proving the following identity
\begin{equation}\label{eq:multiop_cancellation_app}
    (a_R^{(II)}-a_L^{(II)})\ket{\chi_n}\overset{!}{=}b_n^2\ket{\chi_{n-1}}.
\end{equation}
This cancellation is totally analogous to \eqref{eq:binom_recursion_cancellation_cond}, found for the single insertion case. Indeed, it does follow from the same argument: in the semiclassical limit we can treat the binomial coefficients as nascent Dirac delta functions at the center of the summation domain. In the remaining part of the section we will give an outline on how to repeat the analysis of \cref{app:recap_asympt} for the multi-operator case.\\

Let us consider the following generalization of the states \eqref{eq:kbase_ansatz_multiop}, arising, as in \cref{app:recap_asympt}, from the modified Lanczos algorithm with evolution $H_R+zH_L$:
\begin{equation}\label{eq:kbase_multiop_genz}
    \ket{\chi^{(z)}_{n>n_s}}=\sum_{k,\,m=0}^{n-n_s,\,n_s}(z)^{k+m}\binom{n-n_s}{k}\binom{n_s}{m}\ket{k,\,n-n_s-k;\,m,\,n_s-m}
\end{equation}
The norm of these states, from which we find the Lanczos coefficients, can be written as:
\begin{equation}
    \label{eq:norm_multikbase_gen}
    \begin{aligned}
        &\langle\chi^{(z)}_{n+1} | \chi^{(z)}_{n+1} \rangle = \sum_{k^\prime,k,m,m^\prime=0}^{n-n_s,n_s}c^{(z)}_{k,m}(n+1)~z^{k^\prime + k+m+m^\prime}\binom{n-n_s}{k^\prime}\binom{n-n_s}{k}\binom{n_s}{m}\binom{n_s}{m^\prime}\times\\&\times\braket{k^\prime,n-n_s-k^\prime;m^\prime,n_s-m^\prime| k,n-n_s-k;m,n_s-m} =:\mathcal{F}_{n+1}(z;\lambda,\Delta,n_s,\Delta_m)~,
    \end{aligned}
\end{equation}
where the coefficient $c^{(z)}_{k,m}(n)$ is given by the analogue of \eqref{eq:ckm_2p_expression}, adapted for the evolution with $H_R+zH_L$. In particular, notice that for $z=1$ we will obtain the asymptotic analysis relative to the computation of the perturbed $\mathcal{O}$TFD complexity of \cref{sec:pert_otfd_compl}, while $z=-1$ gives the operator complexity of \cref{sec:vertical_multinom_lanczos}. Notice that, in contrast to \cref{app:recap_asympt}, the function $\mathcal{F}_{n}$ depends not only on $z$, $\lambda$ and $\Delta$, but also on the parameters $n_s$ and $\Delta_m$, encoding the details of the operator perturbation.\\

If we consider the semiclassical limit where $\lambda\to0$ and $k,\,k',\,m,\,m',\,n_s,\,n\to\infty$, keeping $l_L=\lambda k,\,l_L'=\lambda k',\,l_1=\lambda m,\,l_1'=\lambda m', \,l_s=\lambda n_s,\,l=\lambda n$ fixed, the coefficient $c^{(z)}_{k,m}(n)$ has again a smooth expression $c^{(z)}(l,l_L,l_1)$ as a function of $l$, $l_L$ and $l_1$ regardless of $z$. In this limit, we can write \eqref{eq:norm_multikbase_gen} via the following integral representation over the semiclassical variables: 
\begin{align}
    &\mathcal{F}_{n+1}(z;\lambda,\Delta,n_s,\Delta_m) \underset{\lambda\sim 0}{\sim}\nonumber \\
    & \int_{0}^{l_s}dl_1dl_1'\int_{0}^{l-l_s} dl_L dl_L^\prime c^{(z)}(l,l_L,l_1)G(l_L,l_L',l_1,l_1',l,l_s;z)F(l_L,l_L',l_1,l_1')~, \nonumber\\& \mathrm{with}\qquad G(l_L,l_L',l_1,l_1',l,l_s;z)\equiv\eta_{\lambda}(l-l_s,l_L;z)\eta_\lambda(l-l_s,l_L^\prime;z)\eta_\lambda(l_s,l_1;z)\eta_\lambda(l_s,l_1^\prime;z)\label{eq:norm_integral_multiop},
\end{align}
where again we have
\begin{equation}
    \eta_\lambda(l,l_L;z) := \frac{1}{\lambda}\eta\left(\frac{l}{\lambda},\frac{l_L}{\lambda}\right)~,\quad \text{for} \quad \eta(l,l_L)=\left\{ \begin{aligned}
        & \frac{z^{l_L} ~\Gamma(l +1)}{\Gamma(l_L +1) \Gamma(l-l_L+1)}~,\quad 0\leq l_L\leq l~, \\
        &0~,\qquad\qquad\qquad \qquad\qquad~\text{else}~,
    \end{aligned} \right.  
\end{equation}
and we denoted with $F(l_L,l_L',l_1,l_1')$ the semiclassical limit of the multi-particle inner product $\braket{k^\prime,n-n_s-k^\prime;m^\prime,n_s-m^\prime| k,n-n_s-k;m,n_s-m}$. \\
At this point the same considerations of \cref{app:recap_asympt} apply to \eqref{eq:norm_integral_multiop}. In particular, the functions $\eta_\lambda$ act as nascent Dirac delta functions when $\lambda\to 0$, and localize the contribution of each integration variable to the center of its integration domain. Similarly to \cref{app:recap_asympt}, \eqref{eq:norm_integral_multiop} is well defined for $z>0$, and if we wish to consider $z<0$ we need to perform an analytical continuation. In particular, repeating the considerations of \cref{app:recap_asympt}, for the operator complexity evolution $z=-1$, the functions $\eta_\lambda$, when $\lambda\to0$, localize the contributions to the integral around the center of each variable's integration domain if the function $F(l_L,l_L',l_1,l_1')$, encoding the inner product, is not even around these points. In the next section, we will give some intuition on why we expect this property to hold in the general multi-particle case.\\
The many-insertions case discussed in \cref{sec:multiop_switch_gen} is again a generalization of the discussion presented here for the basis \eqref{eq:kbase_ansatz_multiop_mult}. In practice, the presence in \eqref{eq:kbase_ansatz_multiop_mult} of many binomials, allows us to go over the same localization arguments of \cref{app:recap_asympt}, in norm computations such as \eqref{eq:norm_integral_multiop}, in the case where we have more integration variables.

\subsection{Properties of the multi-particle inner product}\label{app:inner_prod_multipart}

In this section, we want to discuss the inner product between states $\ket{n_0,\dots, n_l}$, living in the generic $l$-particle chord Hilbert space.
In particular, we are interested in the semiclassical limit of scalar products appearing in the Lanczos coefficients computations \eqref{eq:norm_integral_multiop}:
\begin{equation}\label{eq:inner_prod_3p_def}
    \braket{k,n-n_s-k;m,n_s-m|k',n-n_s-k';m',n_s-m'}\underset{\mathrm{semicl.\;lim.}}{\to} F(l_L,l_L',l_1,l_1')
\end{equation}
We want to understand if this inner product is not even around the center of the integration domain of each variable in \eqref{eq:norm_integral_multiop}. If this property holds, then, as reviewed in \cref{app:asympt_analysis}, we can consider the $\lambda\to0$ limit of the binomial coefficients as nascent delta functions around the point where their absolute value is maximal (that is, the center of the integration domain). \\
In this section, we explain how to rigorously construct the three-particle inner product, and show its limiting analytical expressions coming from the two and one-particle cases computed in \cite{lin2023symmetry}. Then, we turn to the insertion configuration where \cite{Stanford:2014jda} stages the discussion of the switchback effect, with an arbitrary number of insertions $m$ with the same weight $\Delta_m$, whose analytical inner product expression was computed in \cite{Xu:2024gfm}. All these analytical formulas presented have the wanted asymmetry, heuristically showing that it is a good assumption to make also in more general configurations of weights.\\

In the three-particle case pertaining the discussion of \cref{sec:vertical_multinom_lanczos} and \ref{sec:pert_otfd_compl}, we can write from \eqref{eq:vmatt_transm} the following recursion relation:
\begin{equation}\label{eq:recursion_inner_prod_3p}
    \begin{aligned} &\braket{n_L',n_R';n_1',n_2'|n_L,n_R;n_1,n_2}=\bra{n_L'-1,n_R';n_1',n_2'}a_L^{(II)}\ket{n_L,n_R;n_1,n_2}=\\&
   = [n_L]\braket{n_L'-1,n_R';n_1',n_2'|n_L-1,n_R;n_1,n_2}
    +\braket{n_L'-1,n_R';n_1',n_2'|n_L,n_R;n_1-1,n_2}\times\\& \times\Tilde{q}'q^{n_L}[n_1]+\Tilde{q}'\Tilde{q}q^{n_L+n_1}[n_2]\braket{n_L'-1,n_R';n_1',n_2'|n_L,n_R;n_1,n_2-1}+\Tilde{q}'^2\Tilde{q}q^{n_1+n_2+n_L}[n_R]\times\\&\qquad\qquad\qquad\qquad\qquad\qquad\qquad\qquad\qquad\quad\;\times\braket{n_L'-1,n_R';n_1',n_2'|n_L,n_R-1;n_1,n_2},
    \end{aligned}
\end{equation}
where $[m]=(1-q^m)/(1-q)$ is the $q$-integer.
 Analogously, we can write the recursion originating from acting with $a_R^{(II)}$ in the first line above. These recursions can be used to lower the chord numbers in each sectors, until the inner product is determined by the 0-particle boundary condition:
\begin{equation}\label{eq:bdry_cond_3p}
\braket{0,n_m,0,0|0,n_m,0,0}=\prod_{k=0}^{n_m}\frac{1-q^{k}}{1-q}
\end{equation}
In practice however, solving \eqref{eq:recursion_inner_prod_3p}\eqref{eq:bdry_cond_3p} proves quite cumbersome already in this three-particle case, and we will not perform it here. Instead, we will give a justification stemming from the known explicit expression, computed in \cite{Lin:2023trc}, of the two-particle semiclassical inner product:
\begin{equation}\label{eq:inner_prod_semiclass_2p}
    \begin{aligned}
& \left\langle n_L^{\prime}, n_I', n_R^{\prime} \mid n_L, n_I, n_R\right\rangle \\
& \quad=[n]!\left[\frac{\left(1-c^2\right) / 2}{\cosh \frac{x_V-x_V^{\prime}}{2}-c \cosh \frac{x_V+x_V^{\prime}}{2}}\right]^{2 \Delta_V}\left[\frac{\left(1-c^2\right) / 2}{\cosh \frac{x_W+x_W^{\prime}}{2}-c \cosh \frac{x_W-x_W^{\prime}}{2}}\right]^{2 \Delta_W},
\end{aligned}
\end{equation}
where we defined:
\begin{equation}
    \begin{aligned}
n_L  =\frac{1}{2} n+y_V,&\quad
n_I =y_W-y_V,\quad
n_R =\frac{1}{2} n-y_W \\
x_{V / W} & =\lambda y_{V / W},\quad
\lambda n =-2 \log c=l
\end{aligned}
\end{equation}
Now we want to use this result to analyze the case of interest \eqref{eq:inner_prod_3p_def} for the discussion of \cref{sec:vertical_multinom_lanczos} and \ref{sec:pert_otfd_compl}. In particular, the three-particle Hilbert space sector collapses on the two-particle sector if we set $\Delta=0$, so that, using \eqref{eq:inner_prod_semiclass_2p} with $\Delta_V=\Delta_W=\Delta_m$, we can compute \eqref{eq:inner_prod_3p_def}, when $\Delta=0$, as:
\begin{equation}\label{eq:inner_prod_3pto2p}
     \begin{aligned}
& \left\langle k', n-n_s-k';n_s,0 \mid k,n-n_s-k; n_s,0\right\rangle= \left\langle k',n_s; n-n_s-k' \mid k, n_s; n-n_s-k\right\rangle\\
& \quad=[n]!\left[\frac{\left(1-c^2\right) / 2}{\cosh \frac{l_L-l_L'}{2}-c \cosh \frac{l-l_L-l_L'}{2}}\right]^{2 \Delta_m}\left[\frac{\left(1-c^2\right) / 2}{\cosh \frac{l-2l_s-l_L-l_L'}{2}-c \cosh \frac{l_L-l_L'}{2}}\right]^{2 \Delta_m},
\end{aligned}
\end{equation}
Notice that this limiting form is the one giving the leading contribution to the inner product \eqref{eq:inner_prod_3p_def} in the parameter region with $1\gg\Delta_m\gg\Delta$. In particular, \eqref{eq:inner_prod_3pto2p} presents the wanted asymmetry, as it is not even around $l_L\to l-l_s-l_L$. Analogously if we consider the case $\Delta_m\ll\Delta$, the leading contribution to \eqref{eq:scal_prod_semiclass_1p} is given by the collapse to the one-particle sector expression \eqref{eq:scal_prod_semiclass_1p}, which presents the requested asymmetry.\\
Lastly, we have the following semiclassical expression obtained in \cite{Xu:2024hoc} for arbitrary number of matter chords with same weight $\Delta_m$\footnote{the expression obtained in \cite{Xu:2024hoc} contains also a sum over permutations of matter-chords of the same kind, with the associated intersection penalty factors. Here we suppressed this summation, because in this paper we specialized to chord states in a single matter configuration, the `vertical chords' one of \cref{sec:vertical_multinom_lanczos}, with no intersections between operator chords.}:
\begin{equation}\label{eq:inn_prod_multip_symm}
    \left.\left\langle n_0, \cdots, n_l \mid m_0, \cdots, m_l\right\rangle\right|_{q \rightarrow 1}= \widetilde{\sum_{k_{i j}}}  q^{\Delta_m\operatorname{tr}\left(\mathbb{D}^\pi_{il} k_{ls}\right)} \cdot \frac{\prod_{i=0}^l n_{i}!m_{i}!}{\prod_{i, j=0} k_{i j}!},
\end{equation}
where the $\widetilde{{\sum_{k_{ij}}}}$ is a sum over variables $k_{ij}$ with the constraints
\begin{equation}
    \sum_{j=0}^l k_{i j}=n_i, \quad \sum_{i=0}^l k_{i j}=m_j, \quad i, j=0,1, \cdots, l,
\end{equation}
and $\mathbb{D}^\pi_{il}$ is the distance matrix $\mathbb{D}_{i j}^\pi=|i-j|$ for $i, j=0,1, \cdots, l $. So the chord numbers in each sector appear at the level of the constraints, and an appropriate permutation of rows/columns of the matrix $(k_{ij})$ implements inversions around the center of the domain of integration of such variables. In particular, the trace in \eqref{eq:inn_prod_multip_symm} is manifestly not invariant around these transformations, so that the full multi-particle inner product picks up some non-trivial $\Tilde{q}'$ factors.\\

We consider the fact that the inner product shows the wanted lack of symmetry in all the parameter regions where we have analytical control, as a good heuristic indication that this will be the case also for arbitrarily many generic weights. Notice however that, if our goal is just to describe the parameter region where we can make contact with the timefold arguments of \cite{Stanford:2014jda} and obtain \cref{eq:pert_op_compl_regimes_approx_sametscramble}, then observing the asymmetry in \cref{eq:inn_prod_multip_symm} rigorously suffices.

\section{Details on geodesic lengths computation}\label{app:geod_details}

\subsection{ERB lengths in single-shockwave geometries}\label{app:geod_details_single}
In this section, we give details on how to compute geodesic lengths anchored on the boundary and crossing a shockwave interface, reproducing the result \eqref{eq:shock_geod_leftright} from \cite{Shenker_2014}\footnote{in \cite{Shenker_2014} they compute geodesic lengths in $AdS_3$, while here we adapt the same computation to the $AdS_2$ case. We obtain the same result, in view of the dimensional reduction we discussed in \cref{sec:jt_to_btz_dimred}.}.\\
We can parametrize Lorentzian $AdS_{d+1}$ as a sub-manifold of $\mathbb{R}^{2,\,d}$ via the constraint:
\begin{equation}
    \vec{X} \cdot \vec{X}=-\left(X_0\right)^2-\left(X_{d+1}\right)^2+\left(X_1\right)^2+\cdots+\left(X_{d}\right)^2=-l_{AdS}^2,\end{equation}
where $\vec{X}\in \mathbb{R}^{2,\,d}$. In these embedding coordinates of $AdS_{d+1}$, we know how to compute the geodesic length between points $A$ and $B$\footnote{ a reference on how to compute this known result can be found, for example, in appendix A of \url{https://www.marcosmarino.net/uploads/1/3/3/5/133535336/polycopie-qfcs.pdf}.}: 
\begin{equation}\label{eq:geod_lgt_ads_general}
d(A,\,B)=l_{AdS}\,\mathrm{cosh}^{-1}\biggr(-\frac{\vec{X}_A\cdot\vec{X}_B}{l_{AdS}^2}\biggr),
\end{equation}
where $\vec{X}_{A,B}$ are respectively the coordinates of $A,B$ in this embedding formalism.\\
We now specialize to the $d=1$ case for $AdS_2$. Given the embedding coordinates of the anchoring points $\vec{X}\equiv(T_1,\,X_1,\,T_2)$ and $\vec{X}'\equiv(T_1',\,X_1',\,T_2')$, from \eqref{eq:geod_lgt_ads_general} we have:
\begin{equation}
    l_{AdS}^2\cosh \frac{d(\vec{X},\,\vec{X}')}{l_{AdS}}=T_1 T_1^{\prime}+T_2 T_2^{\prime}-X_1 X_1^{\prime}.
\end{equation}
We can write these embedding coordinates as a function of the Schwarzschild coordinates in the black hole patch appearing in \eqref{eq:metric_schw}, and of the Kruskal coordinates of \eqref{eq:metric_kruskal}\cite{Shenker_2014}:
\begin{equation}\label{eq:embed_coord_to_schw_to_krusk}
    \begin{aligned}
T_1/l_{AdS} & =\frac{1-U V}{1+U V} =\frac{r}{r_s} \\
T_2/l_{AdS} & =\frac{V+U}{1+U V}=\frac{1}{r_s} \sqrt{r^2-r_s^2} \sinh \frac{r_s t}{l_{AdS}^2} \\
X_1/l_{AdS} & =\frac{V-U}{1+U V}=\frac{1}{r_s} \sqrt{r^2-r_s^2} \cosh \frac{r_s t}{l_{AdS}^2}
\end{aligned}
\end{equation}
Before proceeding, let us consider what happens when we try to extend the coordinates of \cref{eq:metric_schw}, defined in the right patch, first to the interior of the black hole, and then to the left patch. \\
If we want to extend the coordinates to the black hole interior region $r<r_s$ we can imagine analytically continuing $t\to t+i\pi l_{AdS}^2/2r_s$, which effectively sets $T_2/l_{AdS}=\sqrt{1-r^2/r_s^2} \cosh r_s t/l_{AdS}^2$ and $X_1/l_{AdS}=\sqrt{1-r^2/r_s^2} \sinh r_s t/l_{AdS}^2$. We can just analogously repeat $t\to t+i\pi l_{AdS}^2/2r_s$ when we cross the horizon again to access the left patch. Now let us note that crucially the future light cone is inverted, meaning that $t_L$, the time on the left boundary, will flow oppositely with respect to $t_R$ on the right. Analogously for the coordinates of \cref{eq:kruskal_coord_def} in the right patch, the access to the left sector is effectively implemented by $U\to -V$ and $V\to -U$. This is the notation considered in \cite{Shenker_2014}, while for the results in this paper we chose to perform the additional change $t_L\to -t_L$.  \\

Now we use \eqref{eq:geod_lgt_ads_general} to find the lengths $l_L$ and $l_R$ respectively from the left/right boundaries to the crossing point on the shockwave interface, where we need to impose the appropriate shift conditions \eqref{eq:shift_sol}. In the late-time limit underlying the \textit{shockwave approximation}, the position of the backreacted quanta is exponentially close to the horizon. This means that we can approximately describe the position of the shockwave via the condition $U\approx0$, so we have:
\begin{equation}
    \begin{aligned}
\cosh \frac{l_L}{l_{AdS}} & =\frac{r}{r_s}+\frac{1}{r_s} \sqrt{r^2-r_s^2} e^{r_s t_L / l_{AdS}^2} (V_w+\alpha) \\
\cosh \frac{l_R}{l_{AdS}} & =\frac{r}{r_s}-\frac{1}{r_s} \sqrt{r^2-r_s^2} e^{-r_s t_R / l_{AdS}^2}V_w\,, 
\end{aligned}
\end{equation}
where $V_w$ is the value of the coordinate $V$ where we cross the $U=0$ shockwave and $r$ is the radius of the geodesic anchoring point. Notice that for generic $V_w$, which at this stage is just a free parameter for the crossing, there is no guarantee that the total length $l=l_L+l_R$ is a geodesic, albeit $l_L$ and $l_R$ are. The procedure we perform, in order to ensure that $l$ is a geodesic length, is to choose the crossing point $V_w$ so that the derivative of $l$ with respect to this parameter is zero.\\ 
The lengths $l_{L,R}$ are anchored to the regularized boundary, so that we have $r=l_{AdS}/\epsilon$ with small cutoff parameter $\epsilon$. The total length, as a function of the crossing point, will then be:
\begin{equation}\label{eq:lgt_non_extremized}
   l_{AdS} \biggr(\cosh^{-1}\biggr(\frac{r}{r_s}+\sqrt{\frac{r^2}{r_s^2}-1}\, e^{r_s t_L / l_{AdS}^2} (V_w+\alpha)\biggr)+\cosh^{-1}\biggr(\frac{r}{r_s}-\sqrt{\frac{r^2}{r_s^2}-1}\, e^{-r_s t_R / l_{AdS}^2}V_w\biggr)\biggr)
\end{equation}
Now we proceed to extremize this with respect to the shockwave crossing point $V_w$. By imposing that the derivative of \eqref{eq:lgt_non_extremized} with respect to $V_w$ vanishes to leading order in $r/r_s\gg1$, we obtain:
\begin{equation}\label{eq:vw_minimized}
    V_w=\frac{1}{2}\biggr(-\alpha-e^{-\frac{r_s t_L}{l_{AdS}^2}}+e^{\frac{r_s t_R}{l_{AdS}^2}}\biggr)
\end{equation}
If we substitute this result back into \eqref{eq:lgt_non_extremized}, and expand for $r/r_s\gg 1$ we obtain the total geodesic length $l$:
\begin{equation}\label{eq:lgt_single_ren}
    l(t_L,t_R)\approx2l_{AdS}\log 2r/r_s+2l_{AdS}\log\biggr(\cosh\frac{r_s}{2l_{AdS}^2}(t_R+t_L)+\frac{\alpha}{2}e^{-\frac{r_s(t_R-t_L)}{2l_{AdS}^2}}\biggr)
\end{equation}
The first term is just an $\epsilon$ divergent term due to anchoring on the regularized boundary. Analogously to what we performed in \cref{sec:jt_dual_recap} for \eqref{eq:jt_geodesic}, we can define the re-normalized version of this length $\Tilde{l}$ by subtracting the $\epsilon$ divergence, thus obtaining \eqref{eq:shock_geod_leftright}. 

\subsection{ERB lengths with multiple shockwaves}\label{app:switch_erb}
In this section, we repeat the computation of the geodesic length performed in the previous section, but in the presence of multiple shockwaves, and, in particular, in the bulk configuration dual to the boundary operator perturbations introduced in \cref{sec:lanczos_pert}. We will essentially follow \cite{Stanford:2014jda}, where an analogous computation was performed for extremal surfaces in the BTZ black hole, and show that ERB lengths indeed exhibit the switchback effect. \\

\begin{figure}
\centering
    \begin{tikzpicture}[scale=0.7]
\draw [ line width=0.9pt ] (6.25,15.5) -- (16.5,15.5);            
\draw [ line width=0.9pt ] (6.25,15.5) -- (6.25,7.85);            
\draw [ line width=0.9pt ] (16.5,15.5) -- (16.5,7.85);            

\draw [ line width=0.9pt, dotted ] (6.25,7.85) -- (6.25,7);       
\draw [ line width=0.9pt, dotted ] (16.5,7.85) -- (16.5,7);       

\draw [ line width=0.9pt, dotted ] (6.25,7) -- (16.5,7);          

\draw [ color={rgb,255:red,11; green,64; blue,201}, line width=0.9pt, dashed ]
      (6.25,15.5) -- (13.5,8.25);
\draw [ color={rgb,255:red,11; green,64; blue,201}, line width=0.9pt, dashed ]
      (16.5,15.5) -- (9.25,8.25);

\draw [ thick,red ] (10.25,9.8) -- (16,15.5);
\draw [ thick,red ] (10.25,9.6) -- (16.25,15.5);
\draw [ thick,red ] (6.55,15.5) -- (12.0,10.1);
\draw [ thick,red ] (6.8,15.55) -- (12.25,10.1);

\draw [ thick,red, dotted ] 
    (10.25,9.8) -- ({10.25 + 0.1*(10.25 - 16)}, {9.8 + 0.1*(9.8 - 15.5)});
\draw [ thick,red, dotted ] 
    (10.25,9.6) -- ({10.25 + 0.1*(10.25 - 16.25)}, {9.6 + 0.1*(9.6 - 15.5)});
\draw [ thick,red, dotted ] 
    (12.0,10.1) -- ({12.0 + 0.1*(12.0 - 6.55)}, {10.1 + 0.1*(10.1 - 15.5)});
\draw [ thick,red, dotted ] 
    (12.25,10.1) -- ({12.25 + 0.1*(12.25 - 6.8)}, {10.1 + 0.1*(10.1 - 15.55)});
\draw [ thick,green ] (8.8,8.9) -- (15.6,15.5);
\draw [ thick,green ] (9,8.9) -- (15.85,15.5);
\draw [ thick,green, dotted ] 
    (8.8,8.9) -- ({8.8 + 0.1*(8.8 - 15.6)}, {8.9 + 0.1*(8.9 - 15.5)});
\draw [ thick,green, dotted ] 
    (9,8.9) -- ({9 + 0.1*(9 - 15.85)}, {8.9 + 0.1*(8.9 - 15.5)});
\node (lsr) [right] at (16.5,12){$t_R$};
\node (lssr) [right] at (11.6,10.9){};

\node (l1) [left] at (6.25,12){$t_L$};
\node (l2) [left] at (9.2,13){};

\node (l3) [left] at (8.7,13.5){};
\node (l4) [left] at (13.5,13){};
\node (l5) [left] at (8.9,14){};
\node (f) [above left] at (12,12.5){$l(t)$};

\draw[bend left,line width=2pt, black, line cap=round] (6.3,12) to node [auto] {} (9.2,13);
\draw[bend right,line width=2pt, black, line cap=round] (l4) to node [auto] {} (l3);
\draw [thick,black] (13.15,13.18) -- (12.55,12.0);
\draw [thick,black] (8.7,13.6) -- (9.2,13);
\draw[bend right,line width=2pt, black, line cap=round] (12.6,12.0) to node [auto] {} (lsr);
\end{tikzpicture}
    \caption{A shockwave (green) background, perturbed by a two-sided matter insertion (red), representing the bulk dual of the perturbation described in \cref{sec:lanczos_pert}. Crossing each shockwave amounts to a null shift, in opposite directions for right/left moving shocks, along the interface, so that the length of a geodesic anchored at points $t_L$, $t_R$ on the regularized boundary will need to be extremized with respect to two crossing parameters.} 
    \label{fig:JT_2shockwave}
\end{figure}
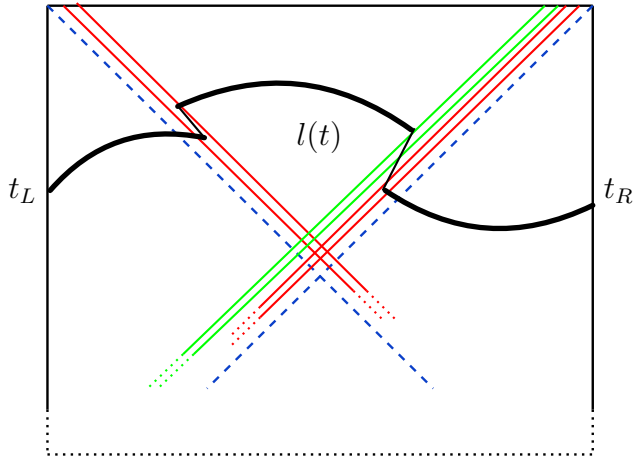

We consider multiple shockwave perturbations, characterized by shift parameters:
\begin{equation}\label{eq:multi_alpha_def}
    \alpha_i=\frac{E}{4M}e^{\pm r_s t_i/l_{AdS}^2},
\end{equation}
where the sign, appearing before the times of insertion $t_i$, is chosen negative for a shock on $U\approx 0$, and positive for one on $V\approx 0$.  Computing lengths in this multi-shockwave background is analogous to the calculation of the previous section: we need to sum the distances between the bulk portions split by the shockwaves, where at each crossing point we implement a shift of $\pm \alpha_i$, respectively if the shock is on $U\approx 0$ or $V\approx 0$ (see for example \cref{fig:JT_2shockwave}). Then, we impose that the derivative of this length with respect to each crossing parameter is null, in order to obtain the boundary-anchored geodesic. In the following part of this section, we proceed to perform this procedure in the shockwave configuration corresponding to the DSSYK operator insertions introduced in \cref{sec:lanczos_pert}.\\

As anticipated, we want to consider the shockwave configuration whose boundary dual is the two-sided perturbation inserted in the Lanczos algorithm in \cref{sec:lanczos_pert}. The computation of the \textit{perturbed} complexity, carried out in \cref{sec:perturbed_3scale_length}, corresponds to considering the geodesic length anchored at $t_R=-t_L=t$, in a background where we inserted shockwaves at $t_0=t$,  $t_1=t-t_s$ and $t_2=2t-t_s$. Notice that we are describing the second insertion, originally in the past of the right boundary, in the future of the left. This is allowed as they originate, when $t-t_s\gg t_{scr}$, the same $V\approx0$ shockwave. This is the same procedure performed on the boundary in \cref{sec:switch_explicit}, so that we can apply the usual timefold considerations for one-sided insertions. The low-energy quanta inserted in $t_0$ and $t_1$ backreact into, exponentially close, $U\approx 0$ shockwaves, while the ones in $t_2$ localize on $V\approx 0$ at late times. These shockwaves are then characterized respectively by the following shift parameters:
\begin{equation}
    \alpha_0=\frac{E}{4M}e^{-r_s t/l_{AdS}^2},\quad \alpha_1=\frac{E}{4M}e^{-(t-t_s) r_s/l_{AdS}^2},\quad \alpha_2=\frac{E}{4M}e^{(2t-t_s) r_s/l_{AdS}^2}
\end{equation}
For simplicity we are considering here perturbation of energy $E'=E$. This assumption can be in principle relaxed by requiring that $E'$ and $E$ just define the same parametrically large scrambling time-scale $t_{scr}$, which is a requirement we need in order to find \eqref{eq:switchback_erb_def}. As noted in \cite{Shenker_2014}, this condition results in considering $E,\;E'$ of the order of the energy of a quantum at the Hawking temperature, so that $t_{scr}\propto \log S$. \\
The insertion at $t_0$ is through-going, while the ones in $t_1$ and $t_2$ are switchbacks, correspondingly to the timefold picture in DSSYK, that we gave in \cref{sec:switch_explicit}. Given that the shocks inserted at $t_0$ and $t_1$ are localized on exponentially close null surfaces at very late times, we neglect the distance between them, and just sum their shift parameters upon crossing of both. However, notice that, if we choose $t_s\gg t_{scr}$, we have $\alpha_0\ll\alpha_1$ so that the total shift $\alpha_0+\alpha_1\approx \alpha_1$.\\

 In this configuration, we want to compute the geodesic length, anchored at times $t_L$ and $t_R$ on the regularized boundaries, and in particular show that it has the same characteristic switchback effect behavior of \eqref{eq:pert_op_compl_regimes_approx}.
The system of shockwaves effectively splits boundary anchored lengths in three sections: $l_{L,R}$ going respectively from the left/right boundary to the closest null interface, and $l_1$ going between the two horizons (\cref{fig:JT_2shockwave}). Repeating the procedure used in the previous section, we can compute the total length by summing the distances \eqref{eq:geod_lgt_ads_general}, expressed with the maximally extended coordinates \eqref{eq:embed_coord_to_schw_to_krusk}, between the points on the regularized boundaries and the shockwaves, whose crossing points are determined by the parameters $U_2$ and $V_1$. We obtain the following distance
\begin{equation}\label{eq:lgt_non_extremized_2shock}
\begin{aligned}
   l_{AdS} \cosh^{-1}\biggr(\frac{r}{r_s}-\sqrt{\frac{r^2}{r_s^2}-1}\, &e^{r_s t_L / l_{AdS}^2} (U_2-\alpha_2)\biggr)+l_{AdS}\cosh^{-1}\biggr(1+2U_2 (V_1+\alpha_1)\biggr)+\\&+l_{AdS}\cosh^{-1}\biggr(\frac{r}{r_s}-\sqrt{\frac{r^2}{r_s^2}-1}\, e^{-r_s t_R / l_{AdS}^2}V_1\biggr)=l(t),
\end{aligned}
\end{equation}
which still needs to be extremized with respect to $U_2$ and $V_1$ in order to obtain a geodesic length. This procedure is, in full generality, quite cumbersome to carry out. However, as for the DSSYK insertion, the situation simplifies if we consider the late-time limit $t-t_s\gg t_{scr}$ together with $t_s\gg t_{scr}$. In this approximation we obtain:
\begin{equation}\label{eq:lgt_non_extremized_2shock_log}
\begin{aligned}
   l_{AdS} \biggr(\log\biggr(-e^{r_s t_L / l_{AdS}^2} (U_2-\alpha_2)\biggr)+\log\biggr( -e^{-r_s t_R / l_{AdS}^2}V_1\biggr)+\log\biggr(U_2 (V_1+\alpha_1
   )\biggr)\biggr)=\Tilde{l}(t),
\end{aligned}
\end{equation}
where, as in \eqref{eq:lgt_single_ren}, we renormalized the length by subtracting the $\epsilon$ divergent term $2\log2r/r_s$. Now, we can solve the equations that set to zero the derivatives of \eqref{eq:lgt_non_extremized_2shock_log} with respect to $U_2$ and $V_1$, and obtain $U_2=\alpha_2/2$ and $V_1=-\alpha_1/2$. 
If we plug these extremized crossing parameters back into \eqref{eq:lgt_non_extremized_2shock_log}, we obtain the geodesic length:
\begin{equation}\label{eq:app_2switch}
    l(t)\approx \frac{r_s}{l_{AdS}^2}(-t_R-2t_1+2t_2-t_L-4t_{scr})=\frac{2r_s}{l_{AdS}^2}(t-2t_{scr}),
\end{equation}
which is precisely the late time regime observed on the DSSYK side of the duality in \cref{eq:pert_op_compl_regimes_approx_sametscramble}.\\

We can generalize the above procedure to an arbitrary number of shockwaves and obtain, for example, the following, in the case with an odd number of switchbacks \cite{Stanford:2014jda}:

\begin{equation}\label{eq:switch_erb}
\begin{aligned}
     l(t_L,t_R)\approx&\log(-V_1e^{-r_st_R/l_{AdS}^2})+\log(U_2(V_1+\alpha_1))+\dots+\log(V_n(U_{n-1}-\alpha_{n-1}))+\\&+\log((V_n+\alpha_n)e^{r_st_L/l_{AdS}^2})\biggr|_{U_i=\alpha_i/2,\;V_i=-\alpha_i/2}=\frac{2r_s}{l_{AdS}^2}(t_f/2-n_{sb}t_{scr}),
\end{aligned}
\end{equation}
where $t_f$ is the length of the timefold and $n_{sb}$ is the number of switchbacks inserted.
We can obtain the same result if we have an even number of switchbacks, and through-going insertion do not significantly change the ERB length, as we have seen above. So, from \eqref{eq:switch_erb}, it is proven that ERB lengths indeed show the same switchback effect behavior expected for computational complexity, as reviewed in \cref{sec:switch_grav_intro}. This is also the same late-time regime observed in the \textit{perturbed} operator complexity with arbitrary $m=n_{sb}$ perturbations, discussed in \cref{sec:multiop_switch_gen}, whose dual is then built from such shockwave configurations.
\newpage

\section{Glossary of notation}\label{app:glossary}
\begin{itemize}
    \item $N$: the number of SYK Majorana fermions in \eqref{H_SYK}.
    \item $p$: number of fermions participating in each Hamiltonian interaction \eqref{H_SYK}.
    \item $\Tilde{p}$: same as above, but referred to an operator $\mathcal{O}$ \eqref{eq:random_operator_def} 
    \item $\lambda\equiv 2p^2/N$: the parameter kept fixed in the double-scaling limit of SYK.
    \item $J$: coupling strength entering in the normalization of the probability distribution from which we extract the random coupling appearing in \eqref{H_SYK}.
    \item $q\equiv e^{-\lambda}$: the penalty factor associated to an intersection of Hamiltonian chords.
    \item $[n]_q\equiv\frac{1-q^n}{1-q}$: is the $q$- integer.
    \item $\Delta\equiv\frac{\Tilde{p}}{p}$: parameter characterizing the relative dimension, with respect to \eqref{H_SYK}, of the operator $\mathcal{O}$.
     \item $\Delta_m$: as above, but for the perturbation operator $\Tilde{\mathcal{O}}$, introduced in \cref{sec:lanczos_pert}.
    \item $\Tilde{q}\equiv q^\Delta$: the penalty factor for an intersection between a Hamiltonian chord and a $\mathcal{O}-\mathcal{O}$ chord.
   \item $\Tilde{q}'\equiv q^{\Delta_m}$: same as above, but for the perturbation operator chord $\Tilde{\mathcal{O}}-\Tilde{\mathcal{O}}$.
    \item $\ket{0}\equiv\ket{TFD}$: the $0$-chord state, identified with the infinite-temperature thermofield double state $\ket{TFD}$.
    \item $\ket{n_L,n_R}$: state in the one-particle sector of the Hilbert space, with $n_L$ open Hamiltonian chords to the left of the operator $\mathcal{O}$ insertion, and $n_R$ to the right (as defined in \cref{sec:dssyk_matter_recap}).
    \item $\ket{n_L,n_R;n_1,n_2}$: state living in the three-particle sector of the Hilbert space, with fixed open Hamiltonian chord number in each sector of the chord diagram split by a `vertical matter-chord' (as defined in \cref{sec:vertical_multinom_lanczos}).
    \item $H_{L,R}$: Hamiltonians, in the effective disorder-averaged theory, describing the evolution to the left/right of the operator $\mathcal{O}$, as defined in \eqref{eq:HlHr_def}.
    \item $H=\frac{H_L+H_R}{2}$: total Hamiltonian considered for the evolution prescription for $\mathcal{O}$TFD complexity.
    \item $a^\dag_{L,R}$: creation operators of Hamiltonian chords, respectively on the left/right of the operator insertion \eqref{eq:adag_leftright_def}.
    \item $a^{(II)\;\dag}_{L,R}$: same as above, but acting in the multi-particle Hilbert space that we have after the insertion of a perturbation operator in the `vertical matter-chords' configuration of \cref{sec:vertical_multinom_lanczos} \eqref{eq:adag_def_pert}.
    \item $a_{L,R}$: Hermitian conjugate, with respect to the inner product \eqref{eq:scal_prod_semiclass_1p}, of $a^\dag_{L,R}$ \eqref{aL_long}.
    \item $a^{(II)}_{L,R}$: as above, but relative to $a^{(II)\;\dag}_{L,R}$, after the perturbation insertion of \cref{sec:vertical_multinom_lanczos} \eqref{eq:vmatt_transm}.
    \item $\alpha_{L,R}$: left inverse of $a^\dag_{L,R}$, respectively annihilating a Hamiltonian chord on the left/right of the operator $\mathcal{O}$ insertion \eqref{eq:alpha_lr_def}.
    \item $\alpha^{(II)}_{L,R}$: as above, but relative to $a^{(II)\;\dag}_{L,R}$, acting in the multi-particle Hilbert space obtained after the perturbation insertion \eqref{eq:alpha_def_pert}. 

    \item $\Tilde{a}^\dag_{L,R},\; \Tilde{a}_{L,R}$: creation/annihilation operators of a $\Tilde{\mathcal{O}}-\Tilde{\mathcal{O}}$ matter chord, respectively to the left/right of the background $\mathcal{O}-\mathcal{O}$ chord.
    \item $\ket{\psi_n},\;\ket{\chi_n}$: the normalized and un-normalized Krylov basis elements, pertaining the evolution with $H_R-H_L$. For the single-operator case, we called them $\ket{\psi_n^-},\;\ket{\chi_n^-}$.
    \item $\ket{\psi^+_n},\;\ket{\chi^+_n}$: same as above, but for evolution with $H=(H_L+H_R)/2$.
    \item $b_n,\; b_n^+$: Lanczos coefficients associated to evolution with $H_R-H_L$ and $H$ respectively.
    \item $n$: the sum of chord numbers in each sector of a diagram split by matter chords.
    \item $l\equiv \lambda n$: total length whose anchoring points are evolved with $H_L-H_R$.
    \item $\Tilde{l}$: re-normalized version of $l$ \eqref{eq:triple_scaling_nomatt}, from \cref{sec:bulk_dual} onward we suppress the $\Tilde{\cdot}$ notation.
    \item $l_+$: total length obtained with boundary evolution prescription $H$
    \item $n_s$: number of the Krylov basis element upon which we insert the perturbation, coinciding with the number of open chords on the chord diagram slice prescribed by the Lanczos evolution.
    \item $l_s=\lambda n_s$: semiclassical length of the slice upon the perturbation insertion.
    \item $t_s$: the time of insertion of the perturbation, related to $n_s$ via the inversion of the semiclassical condition $n(t_s)=n_s$.
    \item $l_p\equiv l_L+l_R$: is the dynamical part of the total length that exhibits growth upon time evolution.
    \item $r_s$: the radius of the event horizon in JT gravity. 
    \item $l_{AdS}$: the characteristic dimension of the $AdS_2$ background of JT gravity.
    \item $M$: mass of the black hole in JT gravity, inherited from the dimensional reduction of the BTZ metric (\cref{sec:jt_to_btz_dimred}).
    \item $E$: energy of the shockwave perturbation inserted on the boundary of JT gravity.
    \item $t_{scr}\sim \frac{l_{AdS}^2}{r_s}\log(\frac{M}{E})$: scrambling time for shockwave propagation in JT gravity.
    \item $t_{scr}\sim \frac{1}{2J\lambda}\log(\frac{1}{\Delta})$: scrambling time associated to operator insertions in DSSYK.
\end{itemize}
\newpage

\bibliography{references}
\end{document}